\documentstyle[12pt,epsf,epsfig,rotating]{article}
\input axodraw.sty

\begin{document}

\title{Search for new physics  at LHC}
\author{N.V.Krasnikov and V.A.Matveev \\
INR RAS, Moscow 117312}

\date{September 2003}
\maketitle
\begin{abstract}
We review the search for new physics to be performed  at the Large 
Hadron Collider(LHC). Namely, we review the expectations for the Higgs boson, 
supersymmetry and exotica detection at LHC. 
We also describe the main parameters of the CMS and ATLAS 
detectors.  

\end{abstract}

\newpage

\section{Introduction} 

\begin{flushright}
\end{flushright}
 
The SM (Standard Model) \cite{SM}  which describes within an 
unprecedental scale of 
energies and distances the strong and electroweak interactions of elementary 
particles relays on a few basic principles - the renormalizability, the 
gauge invariance and the spontaneous breaking of the underlying gauge 
 symmetry. The principle of the 
renormalizability \cite{BOG} which is considered 
often as something beyond the limits of experimental test is one of the 
most important (if not the major) ingredients of the quantum field theory.
The SM gauge group $SU_c(3)\otimes SU_L(2) \otimes U(1)$ is spontaneously 
broken to  $SU_c(3) \otimes U_{em}(1)$ by the existence  of scalar field 
with nonzero vacuum expectation value, leading to massive vector bosons - the 
$W^{\pm}$ and $Z$ - which mediate the weak interactions; the photon remains 
massless. One  physical degree of freedom remains in  the scalar sector, a 
neutral scalar boson (Higgs boson) $H$, which is the last nondiscovered 
particle of the SM. It should be noted that the existence of the Higgs 
boson is direct consequence of the renormalizability of the SM model. 
The $SU_c(3)$ gauge group describes the strong interactions 
(quantum chromodynamics or QCD). The eight vector gluons carry colour 
charges and are selfinteracting. Due to the property of asymptotic freedom 
the effective QCD coupling constant $\alpha_{s}$ is small for large 
momentum transfers that allows to calculate reliably deep inelastic 
cross sections. The fundamental fermions in the SM are leptons and quarks; 
the left-handed states  are doublets under $SU_L(2)$ gauge group, 
while the right-handed states are singlets. There are three generations 
of fermions, each generation identical except for mass. 

Despite the apparent striking success of the SM, there are a lot 
of reasons why it is not the ultimate theory. In the SM 
the neutrinos are massless and hence there are no neutrino oscillations.  
However there is strong evidence for neutrino oscillations 
\cite{NEUT1} coming from 
measurements of neutrinos produced in the atmosphere and from a 
deficit in the flux of electron neutrinos from sun. It is 
easy to extend the SM to include neutrino masses, however 
the natural explanation of small neutrino masses is rather untrivial 
and  probably it requires qualitatively new physics. 
In the SM an elementary Higgs field generates masses for the $W$, $Z$ and 
fermions. For the SM  to be consistent the  Higgs boson mass
should be relatively light $M_H \leq 1~TeV$. The tree-level 
Higgs boson mass receives quadratically 
divergent corrections at quantum level:
 $\delta M^2_H \sim \Lambda^2$, where $ \Lambda$ is some 
ultraviolet cutoff. The natural ultraviolet cutoff in particle 
physics is usually assumed to be 
the Planck scale $M_{PL} \sim 10^{19}~GeV$ or 
grand unification scale $M_{GUT} \sim 10^{16}~GeV$. Hence  the natural 
scale for the Higgs boson mass is $O(\Lambda)$. To explain 
the smallness of the Higgs boson mass  some delicate cancellation 
 is required that is rather untrivial  ``fine tuning'' or gauge 
hierarchy problem . At present the supersymmetric solution 
\cite{GOLDFAND}, \cite{SUSY}
of the 
gauge hierarchy problem is the most fashionable one. It predicts that 
the masses of supersymmetric particles have to be lighter than 
$O(1)~TeV$. Other possible explanation is 
based on models with ``technicolour'' \cite{TECH}. 
Also we can't exclude the possibility 
that the natural scale of the nature is $\Lambda \sim O(1)~TeV$. 
At any rate all solutions of the gauge hierarchy problem predict 
the existence of new physics at TeV scale \footnote{There is crucial 
difference between the Higgs boson prediction and the prediction of 
new physics at TeV scale. Really, electroweak models without Higgs boson 
are nonrenormalizable ones and we simply can't make quantitative 
radiative calculations for such models. The SM with small 
Higgs boson mass is consistent renormalizable quantum field theory, 
however within the SM we can't explain naturally the smallness of the 
Higgs boson mass (the smallness of electroweak scale) in comparison 
with Planck scale.}.    
Other untrivial problem is that the SM can't predict the fermion masses, 
which vary over at least five orders of magnitude (fermion problem).

The scientific programme at the LHC (Large Hadron Collider) 
\cite{LHC} which will be 
the biggest particle accelerator complex ever built in the world consists in 
many goals. Among them there are two supergoals:

a. Higgs boson discovery,

b. supersymmetry discovery.

LHC \cite{LHC} will accelerate mainly two proton beams with the total energy 
$\sqrt{s} = 14~TeV$. At low luminosity stage (first two-three years of the 
operation) the luminosity is planned to be $L_{low} = 10^{33}cm^{-2}
s^{-1}$ with total luminosity $L_{t} = 10~fb^{-1}$ per year. 
At high luminosity stage the luminosity is planned to be $L_{high} = 
10^{34}cm^{-2}s^{-1}$ with total luminosity $L_{t} = 100~fb^{-1}$ 
per year. Also the LHC will accelerate heavy ions, for example, 
Pb-Pb ions at 1150 TeV in the centre of mass and luminosity up to 
$10^{27}~cm^{-2}s^{-1}$.
Bunches of protons will intersect at four points where detectors are 
placed. There are planned to be two big detectors  at the LHC: 
the CMS (Compact Muon Solenoid) \cite{CMS} and ATLAS (A Toroidal LHC Apparatus)
 \cite{ATLAS}. Two other detectors 
are ALICE detector \cite{ALICE}, to be used for the 
study of heavy ions, and LHC-B \cite{LHCB}, the  detector for the study of 
B-physics.

The LHC will start to work in 2007  year. There are a lot of 
lines for the research at the LHC \cite{KRMPA}:

a. the search for Higgs boson,

b. the search for supersymmetry,

c. the search for new physics beyond the MSSM (Minimal Supersymmetric Model) 
and the SM,

d. B-physics,

e. heavy ion physics, 
 
f. top quark physics,
 
g. standard physics (QCD, electroweak interactions).

In this paper we briefly review the search for new physics to be performed 
at the LHC. Namely,  we review the expectations for the Higgs boson, the 
supersymmetry and exotica (new physics beyond the SM and 
the MSSM) detection.   We also describe the main parameters of the CMS 
\cite{CMS} and ATLAS \cite{ATLAS} 
detectors. The organisation of the paper is the following. In section 2 we 
describe the main parameters of the CMS and ATLAS detectors. In section 3 we 
review the search for standard Higgs boson  to be done at LHC. 
In section 4 the expectations 
for supersymmetry detection are  discussed. 
Section 5 is devoted to  the review of the search for new physics 
beyond the SM and the MSSM to be done at the LHC. Section 
6 contains concluding remarks.      

\section{CMS and ATLAS  detectors}       

One of the most important tasks for the LHC is the quest for
the origin of the spontaneous symmetry breaking mechanism
in the electroweak sector of the SM. The Higgs boson \cite{HIGGS}  
search is therefore used as a first benchmark
for the detector optimisation for both the CMS and ATLAS. For the SM
Higgs boson, the detector has to be sensitive to the
following processes in order to cover the full mass range
above the  LEP  limit  $M_H \geq 114.4~GeV$ \cite{LEP} on the 
SM Higgs boson mass:

1. $H \rightarrow \gamma \gamma$ for the mass range  $114~ GeV \le m_{H} \leq 
150~GeV$,

2. $H \rightarrow b\bar{b}$ from $WH, ZH, t\bar{t}H$ using
$l^{\pm}(l^{\pm} = e^{\pm}$ or $\mu^{\pm})$- tag and
b-tagging,

3. $H \rightarrow ZZ^{*} \rightarrow 4l^{\pm}$ for the mass range
$130 ~GeV \le m_{H} \le 2m_Z $,

4. $H \rightarrow ZZ \rightarrow 4l^{\pm}, 2l^{\pm}2\nu$ for the
mass range $m_{H} \ge 2m_Z $.

The second supergoal of the LHC project is the 
supersymmetry discovery \cite{SUSY}, i.e. the detection of superparticles. 
Here the main signature are the missing transverse energy events which are
the consequence of undetected lightest stable supersymmetric
particle (LSP) predicted in supersymmetric models with R-parity
conservation. Therefore it is necessary to set stringent requirements
for the hermeticity and $E^{miss}_{T}$ capability of the detector.
Also the search for new physics different from supersymmetry
(new gauge bosons $W^{'}$ and $Z^{'}$, additional dimensions etc.)  
requires high resolution lepton
measurements and charge identification  even in
the $p_{T}$ range of a few TeV. Other possible signature of new
physics - compositeness can be provided by very high $p_{T}$ jet
measurements. Also an important task of the LHC is the study of b- and
t-physics. 

Therefore the basic design considerations for both the ATLAS
and CMS are the following:

1. very good electromagnetic calorimetry for electron and photon
identifications and measurements,

2. good hermetic jet and missing $E_{T}$-calorimetry,

3. efficient tracking at high luminosity for lepton momentum
measurements, for b-quark tagging, and for enhanced electron and photon
identification, as well as tau and heavy-flavour vertexing and
reconstruction capability of some B decay final states at low
luminosity,

4. stand-alone, precision, muon-momentum measurement up to highest
luminosity, and very low-$p_{T}$ trigger capability at lower luminosity,

5. large acceptance in $\eta$ $(\eta \equiv 
  -\ln(\tan(\frac{\theta}{2}))$ coverage.

\subsection{CMS detector}

The CMS detector \cite{CMS} consists of inner detector(tracker), 
electromagnetic calorimeter, hadron calorimeter and  
muon spectrometer. A schematic view of the CMS 
detector is shown in Fig.1.

\begin{figure}[hbt]

\vspace*{0.5cm}
\hspace*{0.0cm}
\epsfxsize=15cm \epsfbox{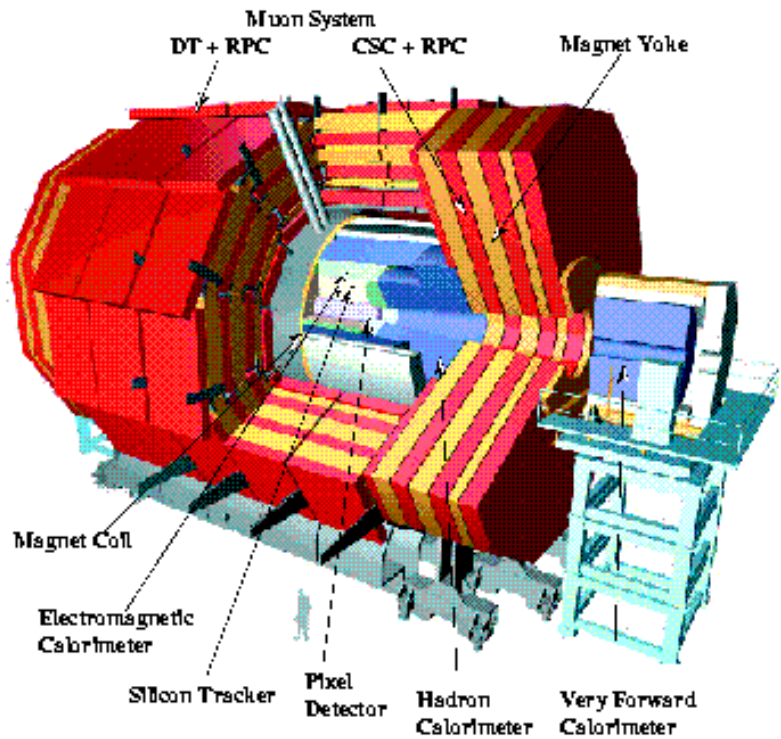}
\vspace*{0.0cm}
\caption[]{\label{fg:CMS} \it Perspective view of the CMS detector}

\end{figure}

The CMS tracking system is contained in a 4 Tesla superconducting 
coil which provides the magnetic field for charged particle tracking. 
The tracking system uses silikon pixels and silikon strip detectors. 
The expected momentum resolution  in the
central rapidity region is $\frac{\delta p_{T}}{p_{T}} =  0.01$
for $p_{T} = 100~GeV$ and it is 5 times worse for $p_{T} = 1~TeV$.

The CMS has a precision electromagnetic calorimeter based on lead 
tungstate ($PbWO_{4})$ crystals, covering $|\eta| < 3$
(with trigger coverage $|\eta| <2.6$).
The energy
resolution at low luminosity is assumed to be
\begin{equation}
\frac{\Delta E}{E} =
\frac{0.03}{\sqrt{E}} \oplus 0.005\,. 
\end{equation}
Estimates \cite{CMS} give the following di-photon mass
resolution for $H \rightarrow \gamma \gamma$ channel $(m_{H} = 100
 ~GeV$):

$\delta m_{\gamma\gamma} = 475~MeV$ (low luminosity
$L = 10^{33}~cm^{-2}s^{-1}$),

$\delta m_{\gamma\gamma} = 775~MeV$
(high luminosity $L =
10^{34}~cm^{-2}s^{-1}$).

The hadron calorimeter surrounds the electromagnetic calorimeter and acts in
conjunction with it to measure the energies and directions of particle jets,
and to provide hermetic coverage for measurement the transverse energy. The
pseudorapidity range $( |\eta| \le 3)$ is covered by the barrel and
endcap hadron calorimeters which sit inside the $4~Tesla$ magnetic 
field of CMS
solenoid. The assumed energy resolution for jets 
is $\Delta E/E = 1.1/\sqrt{E} \oplus 0.05$ . 
The pseudorapidity range 
$( 3.0 \le \eta \le 5.0)$ is covered by a separate very forward
calorimeter. 
The expected energy resolution for jets in the very forward
calorimeter is :
\begin{equation}
\frac{\sigma_{E}}{E} = \frac{1.8 }{\sqrt{E}}
\oplus 0.1\,.
\end{equation}

At the LHC the effective detection of muons from Higgs bosons, $W$, $Z$ and
$t\bar{t}$ decays requires coverage over a large rapidity interval. Muons
from pp collisions are expected to provide clean signatures for a wide
range of new physics processes. Many of these processes are expected
to be rare and will require the highest luminosity. The goal of the muon
detector is to identify these muons and to provide a precision measurement
of their momenta from a few $GeV$ to a few $TeV$. The barrel detector covers
the region $|\eta| \le 1.3$. The endcap detector covers the region
$1.3 \le |\eta| \le 2.4$. 
For $ 0 \le |\eta| \le 2$ the transverse momentum resolution after 
matching with tracker  is  $0.015 - 0.05$ for $p_{T} = 100 ~GeV$
and $0.05 - 0.2$ for $p_{T} = 1~TeV$.

\subsection{ATLAS detector}

The design of the ATLAS detector \cite{ATLAS} is similar 
to the CMS detector. It also 
consists
of inner detector (tracker), electromagnetic calorimeter, hadron calorimeter
and the muon spectrometer.   A schematic view 
of the ATLAS detector is shown in Fig.2.

\begin{figure}[hbt]

\vspace*{0.5cm}
\hspace*{0.0cm}
\begin{turn}{-90}%
\epsfxsize=10cm \epsfbox{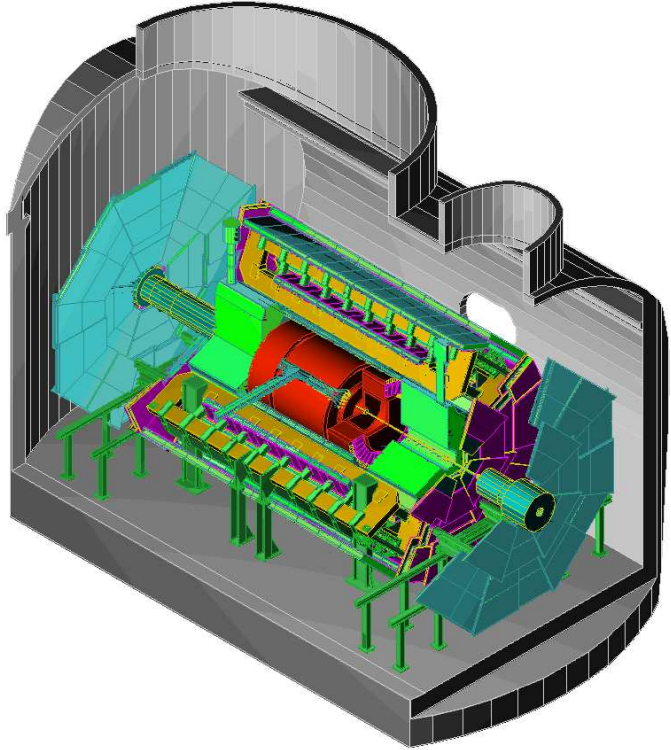}
\end{turn}
\vspace*{0.0cm}

\caption[]{\label{fg:ATLAS} \it Three-dimensional view of the ATLAS detector}
\end{figure}

The ATLAS inner detector consists of silikon pixels, silicon strip detectors  
and a transition radiation detector. The charged track resolution is 
assumed to be 
$\Delta p_{T}/p_{T} = 0.2$ at $p_{T} = 500~GeV$.  

The lead-liquid argon electromagmetic calorimeter covers $|\eta| <~ 3$ 
with trigger coverage $|\eta| <~2.5$. 
The expected energy resolution is of $ \frac{\Delta E}{E} =
\frac{0.1}{\sqrt{E}} \oplus 0.007$ for $|\eta| \le 2.5$. Diphoton
mass resolution is estimated to be $1.4 ~GeV$ for the Higgs boson mass $m_H
= 100~GeV$ for $L = 10^{34}~cm^{-2}s^{-1}$. 

The hadronic calorimeter uses scintillator tiles in the barrel
($|\eta| <1.5$) and liquid 
argon in the endcaps ($1.5  < |\eta| < 3$).
Jet energy resolution is of $\frac{\Delta E}{E} = \frac{0.5}{\sqrt{E}}
\oplus 0.03$.  Forward
calorimeter covers the region $3 \le |\eta| \le 5$ with a resolution 
better than  $\Delta E/E = \frac{1}{\sqrt{E}} \oplus 0.1$. The muon 
system  measures muon trajectories and  the resulting muon 
momentum resolution  is estimated to be  $\frac{\Delta p_{T}}{p_T} = 0.02
(p_T = 100 ~GeV$), 
$\frac{\Delta p_{T}}{p_T} = 0.08(p_T = 1~TeV$) for $|\eta| \le 2.2$.

\section{ Search for standard Higgs boson at the LHC}

\subsection{The Lagrangian of the Standard Model}

The standard  model is the renormalizable model of strong
and electroweak interactions. It has the gauge group
$SU_c(3) \otimes SU_L(2) \otimes U(1)$ and the minimal Higgs
structure consisting of one complex doublet of scalar particles. The 
spontaneous electroweak symmetry breaking $SU_c(3) \otimes SU_L(2)
\otimes U(1) \rightarrow SU_c(3) \otimes U(1)$ due to nonzero vacuum
expectation value of the Higgs doublet provides the simplest
realization of the Higgs mechanism \cite{HIGGS} which generates masses for
$W^{\pm}$, $Z$ gauge bosons and masses to quarks and leptons.  In
this approach, the Goldstone bosons are generated by dynamics of
elementary scalar fields and precisely one neutral Higgs scalar (the
Higgs boson) remains in the physical spectrum. The Lagrangian of
the SM model consists of several pieces \cite{OKUN}:
\begin{equation}
L_{WS} = L_{YM} + L_{HYM} + L_{SH} + L_{f} + L_{Yuk}\,.
\end{equation}
Here $L_{YM}$ is the Yang-Mills Lagrangian without matter fields
\begin{equation}
L_{YM} =
-\frac{1}{4}F^i_{\mu\nu}(W)F^{\mu\nu}_i(W) - \frac{1}{4}
F^{\mu\nu}(W^0)F_{\mu\nu}(W^0) -
\frac{1}{4}F^a_{\mu\nu}(G)F_a^{\mu\nu}(G)\,,
\end{equation}
where $F^i_{\mu\nu}(W)$, $F^a_{\mu\nu}(G)$, $F_{\mu\nu}(W^0)$ are
given by
\begin{equation}
F^i_{\mu\nu}(W) = \partial_{\mu} W^i_{\nu} - \partial_{\nu}W^i_{\mu}
+g_2\epsilon^{ijk}W^j_{\mu}W^k_{\nu}\,,
\end{equation}
\begin{equation}
F_{\mu\nu}(W^0) = \partial_{\mu}W^0_{\nu} -
\partial_{\nu}W^{0}_{\mu}\,,
\end{equation}
\begin{equation}
F^{a}_{\mu\nu}(G) = \partial_{\mu}G^a_{\nu} - \partial_{\nu}G^a_{\mu}
+g_sf^{abc}G^b_{\mu}G^c_{\nu}\,,
\end{equation}
and $W^i_{\mu}$, $W^0_{\mu}$ are the $SU_L(2) \otimes U(1)$ gauge
fields, $G^a_{\mu}$ are the gluon fields,  $\epsilon^{ijk}$,
$f^{abc}$ are the structure constants of the $SU(2)$ and $SU(3)$
gauge groups. The Lagrangian $L_{HYM}$ describes the Higgs doublet
interaction with $SU_L(2)\otimes U(1)$ gauge fields
\begin{equation}
L_{HYM} = (D_{L\mu}H)^{+}(D^{\mu}_LH)\,,
\end{equation}
where covariant derivatives are given by
\begin{equation}
D_{L\mu} = \partial_{\mu} -ig_1\frac{Y}{2}W^0_{\mu} - ig_2
\frac{\sigma^{i}}{2}W^i_{\mu}\,,
\end{equation}
\begin{equation}
D_{R\mu} = \partial_{\mu} -ig_1\frac{Y}{2}W^0_{\mu}\,,
\end{equation}
\begin{equation}
D^q_{L\mu} = \partial_{\mu} - ig_1\frac{Y}{2}W^0_{\mu} - ig_2
\frac{\sigma^{i}}{2}W^{i}_{\mu} - ig_st^aG^a_{\mu}\,,
\end{equation}
\begin{equation}
D^q_{R\mu} = \partial_{\mu} - ig_1\frac{Y}{2}W^0_{\mu} -
ig_st^aG^{a}_{\mu}\,.
\end{equation}
Here $g_1$ is the $U(1)$ gauge coupling constant, $Y$ is the hypercharge
determined by the relation $Q = \frac{\sigma_{3}}{2} + \frac{Y}{2}$,
$\sigma^{i}$ are Pauli matrices, $t^{a}$ are $SU(3)$ matrices in the
fundamental representation,
$H = \left( \begin{array}{cc}
H_1\\
H_2
\end{array}\right)$
is the Higgs $SU(2)$ doublet with $Y = 1$. The
Lagrangian $L_{SH}$ describing Higgs doublet self-interaction has the
form
\begin{equation}
L_{SH} = -V_0(H) = M^2H^{+}H - \frac{\lambda}{2}(H^{+}H)^2\,,
\end{equation}
where $H^{+}H = \sum_{i} H^{*}_iH_i$ and $\lambda$ is the Higgs
self-coupling constant. Lagrangian $L_{f}$ describes the interaction
of fermions with gauge fields. Fermions constitute only doublets and
singlets in $SU_L(2) \otimes U(1)$
\begin{equation}
R_1 = e_R,\;R_2 =\mu_{R},\;R_{3} = \tau_{R} \,,
\end{equation}
\begin{equation}
L_1 = {\nu \choose e}_L \; L_2 ={\nu^{'} \choose \mu}_L \;
L_3 = {\nu^{''} \choose \tau}_L\,
\end{equation}
\begin{equation}
R_{qIu} = (q_{Iu})_R, \;\; (q_{1u} = u, \; q_{2u} = c, \; q_{3u} = t)\,,
\end{equation}
\begin{equation}
R_{qid} =(q_{id})_R, \;\; (q_{1d} = d, \; q_{2d} = s, \; q_{3d} =b)\,,
\end{equation}
\begin{equation}
L_{qI} = {q_{Iu} \choose V^{-1}_{Ii}q_{id}}_L \,,
\end{equation}
where L and R denote left- and right-handed components of the
spinors respectively,
\begin{equation}
\psi_{R,L} = \frac{1 \pm \gamma_{5}}{2} \psi
\end{equation}
and $V_{iI}$ is the Kobayashi-Maskawa matrix. The neutrinos are
assumed to be left-handed and massless. The Lagrangian $L_{f}$ has
the form
\begin{equation}
L_{f} =\sum_{k = 1}^{3}[ i\bar{L}_{k}\hat{D}_{L}L_{k} +
i\bar{R}_{k}\hat{D}_{R}R_{k} + i\bar{L}_{qk}\hat{D}^q_LL_{qk}
+ i\bar{R}_{qku}\hat{D}^q_RR_{qku} +
i\bar{R}_{qkd}\hat{D}^q_RR_{qkd}] \,,
\end{equation}
where $\hat{D}_L = \gamma^{\mu}D_{L\mu}$, $\hat{D}_{R} =
\gamma^{\mu}D_{R\mu}$, $\hat{D}^q_L = \gamma^{\mu}D^q_{L\mu}$,
$\hat{D}^q_R = \gamma^{\mu}D^q_{R\mu}$.
The Lagrangian $L_{Yuk}$ generates fermion mass terms. Supposing the
neutrinos to be massless,  the Yukawa interaction of the
fermions with Higgs doublet has  the form
\begin{equation}
L_{Yuk} = -\sum_{k=1}^{3}[h_{lk}\bar{L}_kHR_k + h_{dk}\bar{L}_{dk}H
R_{dk} + h_{uk}\bar{L}_{uk}(i\sigma^{2}H^{*})R_{uk}] \,+ h.c.\,.
\end{equation}
The potential term $V_0(H)= -M^2H^{+}H  +
\frac{\lambda}{2}(H^{+}H)^2$
for $M^2 > 0$  gives rise to the
spontaneous  symmetry breaking. The doublet $H$ acquires the nonzero
vacuum expectation value
\begin{equation}
<H> = \left( \begin{array}{cc}
0\\
\frac{v}{\sqrt{2}}
\end{array} \right)\,,
\end{equation}
where $v = 246 $ GeV. In the unitare gauge unphysical Goldstone
massless fields are absent and the Higgs doublet scalar field depends
on the single physical scalar field $H(x)$ (Higgs field):
\begin{equation}
H(x) = \left( \begin{array}{cc}
0\\
\frac{v}{\sqrt{2}} + \frac{H(x)}{\sqrt{2}}
\end{array} \right)\,.
\end{equation}
Due to spontaneous gauge symmetry breaking gauge fields except photon
field acquire masses. Diagonalization of mass matrix gives
\begin{equation}
W_{\mu}^{\pm} = \frac{1}{\sqrt{2}}(W^1_{\mu} \mp W^2_{\mu}), \; M_W =
\frac{1}{2} g_2 v \,,
\end{equation}
\begin{equation}
Z_{\mu} = \frac{1}{\sqrt{g^2_2 + g^2_1}}(g_2W^3_{\mu} -g_1W^0_{\mu}),
\; M_Z = \frac{1}{2}\sqrt{g^2_2 +g^2_1}v\,,
\end{equation}
\begin{equation}
A_{\mu} = \frac{1}{\sqrt{q^2_2 + g^2_1}}(g_1W^3_{\mu} + g_2W^0_{\mu}),
\; M_A = 0 \,,
\end{equation}
where $W^{\pm}_{\mu}$, $Z_{\mu}$ are charged and
neutral electroweak bosons,
$A_{\mu}$ is photon. It is convenient to introduce rotation angle
$\theta_{W}$ between $(W^3,W^0)$ and $(Z,A)$ which is called Weinberg
angle
\begin{equation}
\sin{\theta_{W}} \equiv \frac{g_1}{\sqrt{g^2_1 + g^2_2}} \,.
\end{equation}
Experimentally $\sin^{2}{\theta_{W}} \approx 0.23$ \cite{PARTICLE}.
The formula for the electric charge $e$ has the form
\begin{equation}
e = \frac{g_2g_1}{\sqrt{g^2_2 + g^2_1}}\,.
\end{equation}
At the tree level the Higgs boson mass is determined by the
formula
\begin{equation}
m_H = \sqrt{2} M = \sqrt{\lambda} v \,.
\end{equation}
The Lagrangian $L_{Yuk}$ is responsible for the fermion masses
generation. In the unitare gauge the Lagrangian $L_{HYM}$ takes the
form
\begin{equation}
L_{HYM} = \frac{1}{2}\partial^{\mu}H \partial_{\mu}H +
M^2_W(1 + \frac{H}{v})^2 W^{+}_{\mu}W^{\mu} + \frac{1}{2}
M^2_Z(1 + \frac{H}{v})^2Z^{\mu}Z_{\mu} \,.
\end{equation}
The Yukawa Lagrangian in the unitary gauge can be written
in the form
\begin{equation}
L_{Yuk} = -\sum_{i} m_{\psi_{i}}(1 + \frac{H}{v})
 \bar{\psi}_{i} \psi_{i} \,.
\end{equation}

\subsection{Higgs boson mass bounds}

The current lower limit on the SM Higgs boson mass from LEP experiments is 
$m_H \geq 114.4~GeV$  at $95 \%$ C.L. \cite{LEP}. 
Analysis of high-precision measurements of 
electroweak observables leads to indirect upper bound \cite{GRINEW} 
 $m_H \leq 193~GeV$ 
at $95 \%$   C.L. on the Higgs boson mass, so within the SM the Higgs boson 
should be relatively light. 

It is possible to derive bounds on the Higgs 
boson mass from the requirement of the absence of the Landau pole 
singularity for the effective Higgs self-coupling constant \cite{CABIBBO} 
and from the vacuum stability requirement \cite{VACST}. 
The idea of 
the derivation of the bound resulting from the requirement of the absence 
of Landau pole singularity is the following \cite{CABIBBO}. Suppose the SM is 
valid up to scale $\Lambda$. From the requirement of 
the absence of Landau pole singularity for  the effective Higgs 
self-coupling constant $\bar{\lambda}(E)$ for $E \leq \Lambda$ one 
can obtain an upper bound on the Higgs boson mass. For 
$\Lambda = (10^3;10^4;10^6;10^8;10^{12};10^{14})~GeV$ 
and $m_t^{pole} = 175~GeV$  one can find \cite{KRMPA} 
an upper bound on the Higgs boson mass $m_{H} \leq 
(400;300;240;200;180;170;160)~GeV$ respectively. The vacuum stability 
bound  \cite{VACST} comes from the requirement that the electroweak minimum of 
the effective potential is the deepest one for $|H| \leq \Lambda$. For 
$|H| \gg v$ the mass terms in the effective potential are negligible 
compared to the self-interaction term and the vacuum stability 
requirement means that the effective Higgs self-interaction coupling 
constant  $\bar{\lambda}(\mu)$ is nonnegative, $\bar{\lambda}(\mu) \geq 0 $, 
for $\mu \leq \Lambda$. For 
$\Lambda = (10^3;10^4;10^6;10^8;10^{12};10^{14})~GeV$ and 
$m_t^{pole} = 175~GeV$ 
one can find \cite{KRMPA} lower bound on the Higgs boson mass 
$m_{H} \geq (78;101;121;129;136;137)~GeV$ respectively. 
In the MSSM the radiative corrections can increase the
mass of lightest Higgs boson \cite{MHIGGS} up to 
$135~GeV$ ($m_H \leq 135~GeV$) \cite{DEGRAS}.  
As it has been mentioned in refs.\cite{KRPOK} by the measurement of the Higgs 
boson mass it is possible to distinguish between the SM and the 
MSSM or at least to estimate the scale $\Lambda$ where we can expect 
deviations from the SM predictions.

\subsection{Higgs boson decays}

The tree-level Higgs boson couplings to gauge bosons and fermions
can be deduced from the Lagrangians (30, 31). Of these, the
$HW^+W^-$, $HZZ$ and $H\bar{\psi} \psi$ are the most important for
the phenomenology. The partial decay width into fermion-antifermion
pair is \cite{OKUN}
\begin{equation}
\Gamma(H \rightarrow \psi \bar{\psi}) = \frac{G_Fm^2_{\psi}m_HN_c}
{4\pi \sqrt{2}}(1 -\frac{4m^2_{\psi}}{m^2_H})^{\frac{3}{2}}\,,
\end{equation}
where $N_c$ is the number of fermion colours. For $m_{H} \leq 2m_W$
Higgs boson decays mainly with ($\approx$ 90 percent) probability
into b quark-antiquark pair and with $\approx$ 7 percent probability
into $\tau$ lepton-antilepton pair. An account of higher order QCD
corrections can be effectively taken into account in the formula
(32) for the Higgs boson decay width into b quark-antiquark pair by the
replacement of pole b-quark mass in formula (32) by the effective
b-quark mass $\bar{m}_b(m_H)$.
Remember that the relation between the perturbative quark pole mass 
$m_Q$ and the $\bar{MS}$ running quark mass $\bar{m}_Q(m_Q)$
has the form \cite{CRAY}
\begin{equation}
  m_Q = (1 +\frac{4}{3}\frac{\alpha_s(m_Q)}{\pi} + K_Q
(\frac{\alpha_s(m_Q)}{\pi})^2)\bar{m}_Q(m_Q)   ,
\end{equation}
where numerically $K_t \approx 10.9$, $K_b \approx 12.4$ and 
$K_c \approx 13.4$. 

Higgs boson with $m_H \geq 2M_{W}$ will decay into pairs of
gauge bosons with the partial widths
\begin{equation}
\Gamma(H \rightarrow W^{+}W^{-}) = \frac{G_Fm^3_H}{32\pi\sqrt{2}}
(4 -4a_w +3a^2_w)(1-a_w)^{\frac{1}{2}} \,,
\end{equation}
\begin{equation}
\Gamma(H \rightarrow Z^0 Z^0) = \frac{G_Fm^3_H}{64\pi\sqrt{2}}
(4 - 4a_Z + 3a^2_Z)(1 -a_Z) ^{\frac{1}{2}} \,,
\end{equation}
where $a_W = \frac{4M^2_W}{m^2_H}$ and $a_Z=\frac{4M^2_Z}{m^2_H}$.
In the heavy Higgs mass regime $(2m_Z \leq m_H \leq 800$ GeV), the
Higgs boson decays dominantly into gauge bosons. For example,
for $m_H \gg 2m_Z$ one can find that
\begin{equation}
\Gamma(H \rightarrow W^{+}W^{-}) = 2\Gamma(H \rightarrow ZZ)
\simeq \frac{G_Fm^3_H}{8\pi \sqrt{2}}\,.
\end{equation}
The $m^3_H$ behaviour is a consequence of the longitudinal
polarisation states of the $W$ and $Z$. As $m_H$ gets large, so does
the coupling of $H$ to the Goldstone bosons which have been eaten by
the $W$ and $Z$. However, the Higgs boson decay width to a pair of
heavy quarks growth only linearly in the Higgs boson mass. Thus, for
the Higgs masses sufficiently above $2m_Z$, the total Higgs boson
width is well approximated by ignoring the Higgs boson decay to $t\bar{t}$
and including only the two gauge boson modes. For heavy Higgs
boson mass one can find that
\begin{equation}
\Gamma_{total}(H) \simeq 0.48\,TeV(\frac{m_H}{1\,TeV})^3 \,.
\end{equation}

Below threshold the decays into off-shell gauge particles are important. 
The decay width  into single off-shell gauge boson has the form \cite{SPIRA}
\begin{equation}
\Gamma(H \rightarrow V V^*) = \delta_V \frac{3G^2_FM^4_Vm_H}{16\pi^3}
R(\frac{M^2_V}{m^2_H}),
\end{equation}
where $\delta_W =1 $, $\delta_Z = \frac{7}{12} -\frac{10}{9}\sin^2\theta_W 
\frac{40}{27}\sin^4\theta_W$ and 
\begin{equation}
R(x) = 3 \frac{1-8x+20x^2}{\sqrt{4x-1}\arccos(\frac{3x-1}{2x^{3/2}})} -
\frac{1-x}{2x}(2 -13x +47x^2) -\frac{3}{2}(1 -6x + 4x^2)\log(x),
\end{equation}
$x = \frac{M^2_V}{m^2_H}$. For Higgs boson mass slightly larger than the 
corresponding gauge boson mass the decay widths into pairs of off-shell 
gauge bosons play important role. The corresponding formulae can be found in 
ref. \cite{SPIRA}. 

It should be noted that there are a number of important Higgs couplings
which are absent at tree level but appear at one-loop level.
Among them the couplings of the Higgs boson  to two
gluons and two photons are extremely important for the Higgs
boson searches at supercolliders. One-loop induced Higgs
coupling to two gluons is due to t-quark exchange in the loop
\cite{EL76} and it leads to an effective Lagrangian
\begin{equation}
L^{eff}_{Hgg} = \frac{8 g_2\alpha_s}{24\pi m_W}HG^a_{\mu\nu}
G^{a\mu\nu} \,.
\end{equation}
for the interaction of the Higgs boson with gluons.
At lowest order the partial decay width is given by \cite{SPIRA}
\begin{equation}
\Gamma_{LO}(H \rightarrow gg) = \frac{G^2_F\alpha^2_sm^3_H}{36\sqrt{2}\pi^3}
|\sum_{Q}A^H_Q(\tau_Q)|^2,
\end{equation}
\begin{equation}
A^H_Q(\tau) = \frac{3}{2}\tau [1 + (1 -\tau)f(\tau)],
\end{equation}
\begin{equation}
f(\tau) = argsin^2(\frac{1}{\sqrt{\tau}}), \\ \tau \geq 1,
\end{equation}
\begin{equation}
f(\tau) = -\frac{1}{4}[\log(\frac{1 +\sqrt{1 -\tau}}{1 -\sqrt{1 - \tau}} 
- i\pi]^2, \\ \tau < 1
\end{equation}
The parameter $\tau_Q = \frac{4m^2_Q}{m^2_H}$ is defined by the pole mass 
$M_Q$ of the heavy quark  in the loop. For large quark mass 
$ A^H_Q(\tau_Q) \rightarrow 1$. 
It appears that QCD radiative corrections are very large \cite{ZERVA} : the 
decay width is shifted by about (60 -70) percent upwards in the most 
interesting  
mass region  100 GeV $ \leq m_H \leq 500$ GeV. Three loop QCD corrections 
have been calculated in the limit of a heavy top quark. They are positive 
and increase the full next leading order expression by 10 percent 
\cite{CHET}.

Also very important is  the one-loop induced Higgs boson coupling
to two photons due to W- and t-quark exchanges in the loop 
(see Fig.3). The partial decay 
width can be written in the form \cite{SPIRA}
\begin{equation}
\Gamma(H \rightarrow \gamma\gamma) = \frac{G_F\alpha^2m^3_H}{128\sqrt{2}\pi^3}
|\sum_{f}N_{cf}e^2_fA^H_f(\tau_f) +A^H_W(\tau_W)|^2,
\end{equation}
where
\begin{equation}
A^H_f(\tau) = 2\tau[1 +(1-\tau)f(\tau)],
\end{equation}
\begin{equation}
A^H_W(\tau) = -[2 +3\tau +3\tau(2 - \tau)f(\tau)],
\end{equation}
$\tau_i = \frac{4 M^2_i}{m^2_H}$, $i = f, W$ 
and the function $f(\tau)$ is determined by the formulae (43,44). 
The W-loop gives the dominant contribution in the 
intermediate Higgs boson mass range.  
The branching ratios and decay width of the SM Higgs boson 
are presented in Fig.4.

\subsection{Higgs boson production at the LHC}

Typical processes that can be exploited to produce Higgs bosons at the 
LHC are  \cite{KRAS99},  \cite{SPIRA}:

gluon fusion:  $gg \rightarrow H$ (Fig.5),

$WW$, $ZZ$ fusion:  $W^+W^-, ZZ \rightarrow H$ (Fig.6), 

Higgs-strahlung off $W$, $Z$:  $q\bar{q} W,Z \rightarrow W,Z + H$ (Fig7),

Higgs bremsstrahlung off top:   $q\bar{q}, gg \rightarrow t \bar{t} + H$ 
(Fig.8) .

Gluon fusion plays a dominant role throughout the entire Higgs boson 
mass range 
of the SM whereas the $WW/ZZ$ fusion process becomes increasingly 
important with Higgs boson mass rising. The last two reactions are 
important only for light Higgs boson masses. 
                                                                             
The gluon-fusion mechanism \cite{EL76}
\begin{equation}
pp \rightarrow gg \rightarrow H
\end{equation}
is the dominant production mechanism of the Higgs boson at the LHC for 
Higgs boson mass up to 1 TeV. The gluon coupling to the Higgs boson in the 
SM is mediated by triangular loop of top  quark. The 
corresponding form factor approaches a non-zero value for large 
loop-quark masses. At lowest order the partonic cross section can be 
expressed by the gluonic width of the Higgs boson
\begin{equation}
\hat{\sigma}_{LO}(gg \rightarrow H) = \sigma_0 m^2_H \delta(\hat{s} -m^2_H),
\end{equation}
\begin{equation}
\sigma_0 = \frac{\pi^2}{8m^2_H}\Gamma_{LO}(H \rightarrow gg)
|\sum _{Q}A^H_Q(\tau_Q)|^2 ,
\end{equation}     
where $\tau_Q = \frac{4M^2_Q}{m^2_H}$, $\hat{s}$ denotes the partonic 
system of mass energy squared and the form factor $A^H_Q(\tau_{Q})$ 
is determined by the formulae (42-44). In the narrow-width 
approximation hadronic 
cross section can be written in the form 
\begin{equation}
\sigma_{LO}(pp \rightarrow H) = \sigma_0 \tau_H \frac{dL^{gg}}{d\tau_H},
\end{equation}
where $\frac{dL^{gg}}{d\tau_H}$ denotes $gg$ luminosity of the pp collider 
with $\tau_H = \frac{m^2_H}{s}$.  The QCD corrections to the gluon fusion 
process are  essential               \cite{SPIRA},  \cite{ZERVA}, \cite{CHET}.
They stabilise the theoretical predictions 
for the cross section when the renormalization and factorisation scales are 
varied. Moreover, they are large and positive, thus increasing  
the production cross section for Higgs bosons. 

The theoretical prediction for the Higgs boson production cross section 
is presented in Fig.9. 
The cross section decreases with increasing of the Higgs boson mass  
mainly due to the decrease 
of $gg$ partonic luminosity for large invariant masses.

The second important process for the Higgs boson production at the LHC 
is vector-boson fusion,  $ ZZ, ~W^+W^- \rightarrow H $. For large 
Higgs boson mass this mechanism becomes competitive to 
gluon fusion; for intermediate masses the cross section is smaller 
by about an order of magnitude. The corresponding formulae for the 
cross section can be found in refs.\cite{CAHN2}. 

Higgs-strahlung $q\bar{q} \rightarrow V^* \rightarrow VH$ $(V = W,Z)$ is a 
very important process for the search of light Higgs boson at the  
LHC. Though the cross section is smaller than for gluon fusion, 
leptonic decays of electroweak vector bosons are extremely useful to filter 
Higgs boson signal from huge background. The corresponding formulae 
for the cross section are contained in ref.\cite{GLASHOW}.

The process $gg, q\bar{q} \rightarrow t \bar{t} H$ is relevant for small 
Higgs boson masses. The analytical expression for the parton cross section is 
quite involved \cite{KUNSZT}. Note that Higgs boson bremsstrahlung off top 
quarks is an interesting process for measurements of the fundamental 
  $Ht\bar{t}$ Yukawa coupling. The cross section $\sigma(pp 
\rightarrow t\bar{t}H + ...) $ is directly proportional to the square of this 
coupling constant. 

One can say that three classes of processes can be distinguished. The 
gluon fusion of Higgs boson is an universal process, dominant over the 
entire Higgs boson mass range. Higgs-strahlung of electroweak $W,Z$ bosons 
or top quarks is important for light Higgs boson. The $WW/ZZ$ fusion channel, 
by contrast, becomes rather important in the upper part of the Higgs boson 
mass. An overview of the production cross section for the Higgs boson at the 
LHC is presented in Fig.9.

\begin{figure}[hbt]

\vspace*{0.5cm}
\hspace*{0.0cm}
\epsfxsize=15cm \epsfbox{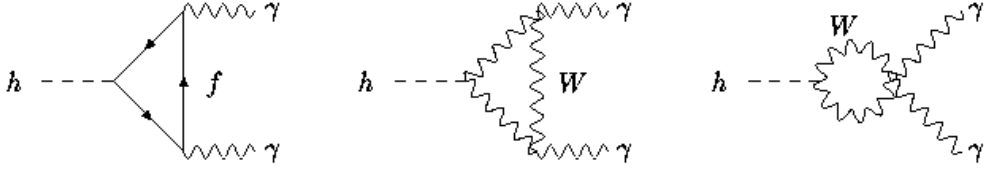}
\vspace*{0.0cm}
\caption[]{\label{fg:CMS} \it One loop diagrams contributing to 
$h \rightarrow \gamma\gamma$ }

\end{figure}

\begin{figure}[hbt]

\vspace*{0.5cm}
\hspace*{0.0cm}
\epsfxsize=15cm \epsfbox{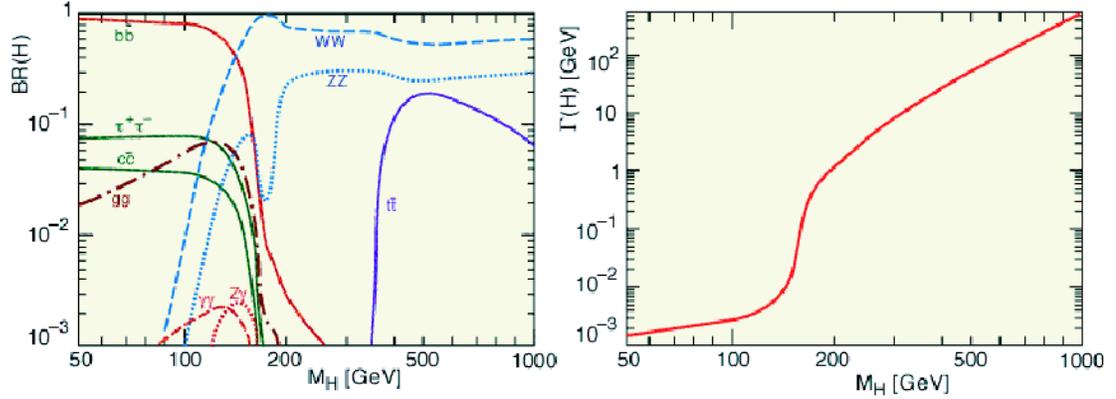}
\vspace*{0.0cm}
\caption[]{\label{fg:CMS} \it Branching ratios and decay width of 
the SM Higgs boson}

\end{figure}

\begin{figure}[hbt]

\vspace*{0.5cm}
\hspace*{0.0cm}
\epsfxsize=15cm \epsfbox{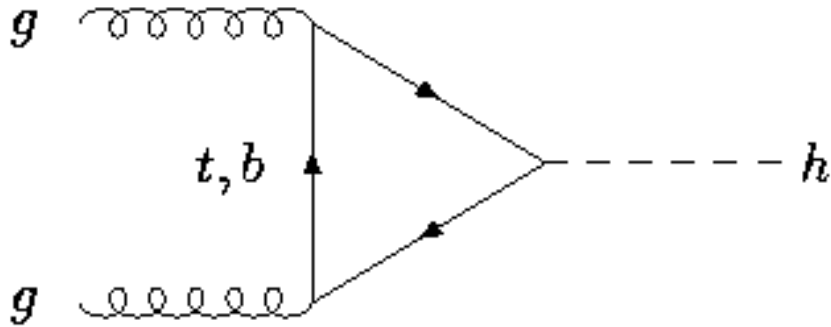}
\vspace*{0.0cm}
\caption[]{\label{fg:CMS} \it Diagram contributing to the production 
of the Higgs boson in gluon-gluon collisions}

\end{figure}

\begin{figure}[hbt]

\vspace*{0.5cm}
\hspace*{0.0cm}
\epsfxsize=15cm \epsfbox{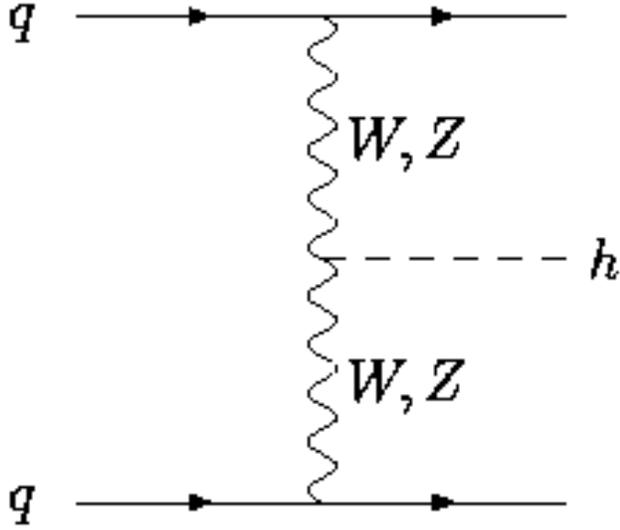}
\vspace*{0.0cm}
\caption[]{\label{fg:CMS} \it Diagram contributing to $ qq 
\rightarrow qqV^{*}V^{*} \rightarrow qqh$ }

\end{figure}

\begin{figure}[hbt]

\vspace*{0.5cm}
\hspace*{0.0cm}
\epsfxsize=15cm \epsfbox{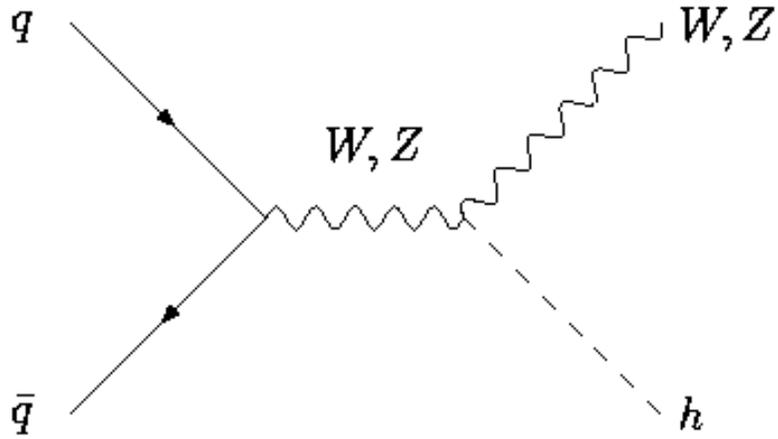}
\vspace*{0.0cm}
\caption[]{\label{fg:CMS} \it Diagram contributing to 
$q\bar{q} \rightarrow V^{*} \rightarrow Vh$}

\end{figure}

\begin{figure}[hbt]

\vspace*{0.5cm}
\hspace*{0.0cm}
\epsfxsize=15cm \epsfbox{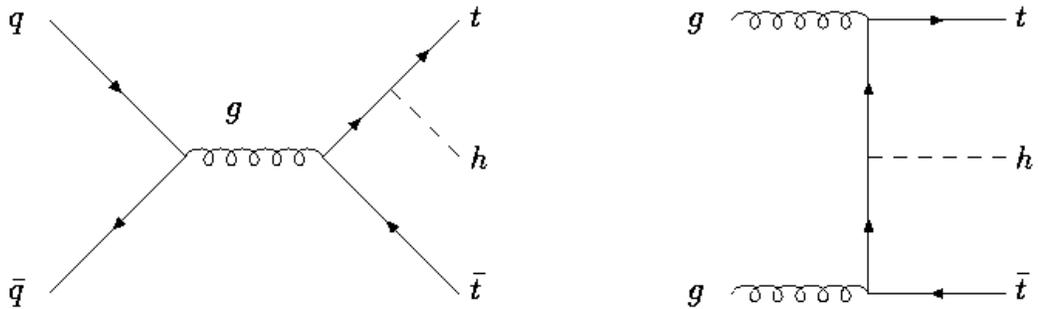}
\vspace*{0.0cm}
\caption[]{\label{fg:CMS} \it Diagrams contributing to $q\bar{q}/gg 
\rightarrow h t\bar{t}$ }

\end{figure}

\begin{figure}[hbt]

\vspace*{0.5cm}
\hspace*{0.0cm}
\epsfxsize=15cm \epsfbox{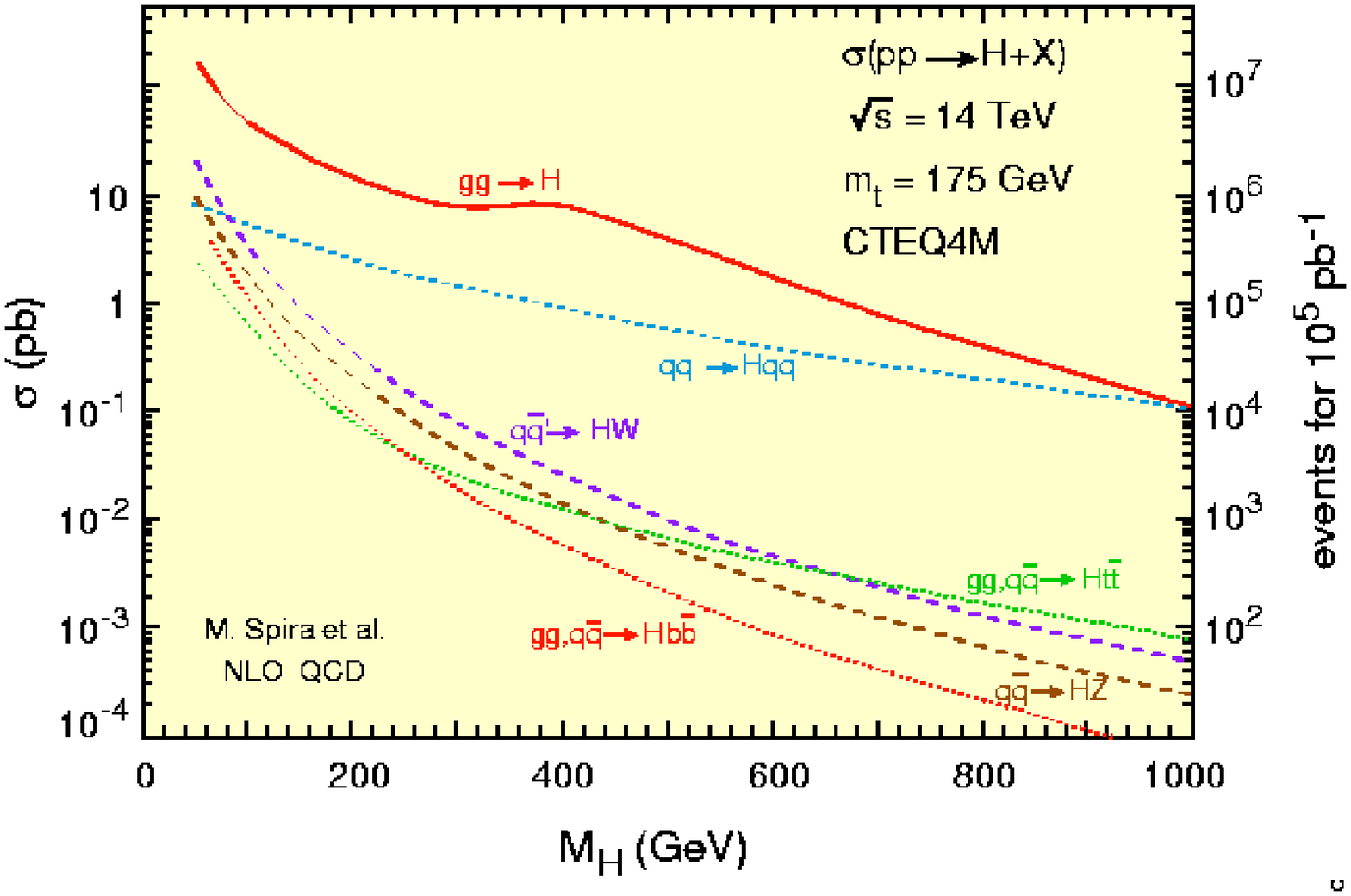}
\vspace*{0.0cm}
\caption[]{\label{fg:CMS} \it Higgs boson production 
cross sections at the LHC for the various production mechanisms 
as a function of the Higgs boson mass}

\end{figure}

\subsection{$H \rightarrow \gamma \gamma$}

One of the most important reactions for the search for Higgs boson at
LHC is
\begin{equation}
pp \rightarrow (H \rightarrow \gamma\gamma) +...\,,
\end{equation}
which is the most promising one \cite{KIN}
for the search for Higgs boson in the
most interesting region $100~GeV \le m_{H} \le 150~GeV$.
The signal  significance $S =
\frac{N_S}{\sqrt{N_B}}$ is estimated to be $6.6(9)$ for
$m_H = 110(130)~GeV$  for low luminosity $L_{low,t} = 30~fb^{-1}$ 
and $10(13)$ for $m_H = 110(130) ~GeV$ and
for high luminosity $L_{high,t} = 100~fb^{-1}$ \cite{CMS}. The general
conclusion is that at $5\sigma$ level CMS detector 
will be able to discover
Higgs boson \footnote{It should be noted that  more appropriate 
characteristic for future experiments \cite{BIT} 
is the probability of the discovery, 
i.e. the probability that future experiment will measure the number 
of events $N_{ev}$ such that the probability that standard physics 
reproduces $N_{ev}$ is less than $5.7 \cdot10^{-7}$ ($5\sigma$). 
For instance, for the standard Higgs boson search with 
$m_H = 110~GeV$ and for 
$L = 30~fb^{-1}(20~fb^{-1})$ the standard 
significance is $6.6(5.4)$. At the language of the probabilities it 
means \cite{BIT} that the CMS will discover at $\geq 5\sigma$ the 
Higgs boson with the probability $96(73)$ percent.}       
for $95 ~GeV \le m_{H} \le 145 ~GeV$ at low 
luminosity and at high
luminosity the corresponding Higgs boson mass discovery interval is
$85 ~GeV \le m_{H} \le 150~GeV$ (see Fig.10).  
The similar estimates have been made for the
ATLAS detector \cite{ATCOL} and they coincide with the CMS 
estimates (in terms of signal significances) up to $30 \%$ \cite{ATCOL}.

\begin{figure}[hbt]

\vspace*{0.5cm}
\hspace*{0.0cm}
\epsfxsize=15cm \epsfbox{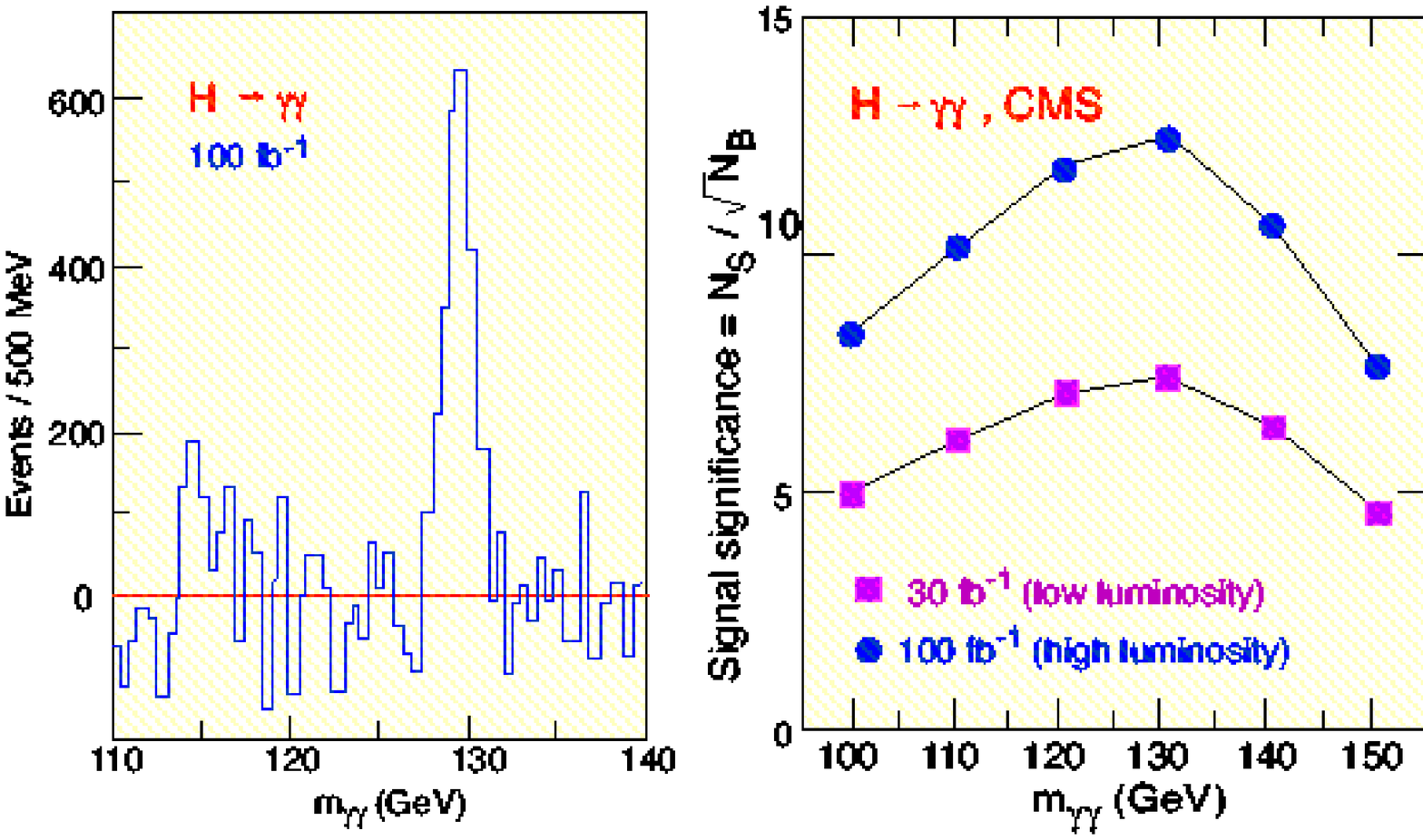}
\vspace*{0.0cm}
\caption[]{\label{fg:CMS} \it Mass peak and signal significance for 
$H \rightarrow \gamma \gamma$ }

\end{figure}

\subsection{$H \rightarrow \gamma\gamma$ in association
with high $E_{T}$ jets.}

The idea to look for Higgs boson signal associated with a high $p_t$ 
jet in the final state was considered in ref.\cite{DUB}, where the 
matrix elements of signal subprocesses $gg \rightarrow g + H$, 
$gq \rightarrow q + H$ and $q\bar{q} \rightarrow g + H$ have been calculated 
in the leading order $\alpha^3_s$. 
For the Higgs boson mass 
$ 100 ~GeV \leq M_H \leq 150 ~GeV$ and for an integrated luminosity 
$10 ~fb^{-1}$ this channel has dozens  of signal events with a number of 
background events only by factor 2-3 higher \cite{DUB}. The significance 
$N_S/\sqrt{N_B} \sim 4, 5$ and 4 for $M_H = 100, 120 $ and $140~GeV$ 
respectively indicating good prospects for discovery of the light 
Higgs boson at low  luminosity. These results also imply that 
at high luminosity phase with year luminosity $100 ~fb^{-1}$ LHC will 
give hundred of events with high $p_t$ associated with hard jet with the 
signal significance $\sim 15$.

\subsection{$H \rightarrow WW^{*} \rightarrow l^+\nu l^{'-} \nu$}

The signature 
$pp \rightarrow H \rightarrow WW^{*} 
\rightarrow l^+\nu {l^{'}}^-\bar{\nu^{'}}$ \cite{DIT}
provides the Higgs boson discovery for the Higgs boson mass 
region between $155 ~GeV$ and $180 ~GeV$ at the LHC. 
Especially important is that the signature
$H \rightarrow WW^{*} \rightarrow l^+\nu l^{'-} \nu$ allows to discover 
Higgs boson in the mass region around $170~GeV$ where the branching ratio 
for $H \rightarrow 4l$ is small and the use of four lepton signature 
for the Higgs boson discovery does not help at least for low luminosity. 
This signature 
does not require extraordinary detector performance and only requires a 
relatively low integrated luminosity of about $5 ~fb^{-1}$. 
The results of the analysis \cite{DIT} demonstrate that this 
signature provides not only the Higgs boson discovery channel 
for a mass range between $(155 - 180)~GeV$ with $S/B \geq 0.35$ 
but also helps to establish a LHC Higgs boson signal for masses between 
$(120 -500)~GeV$. Simulations  \cite{DGRE} based on PYTHIA to 
generate events and CMSCIM calorimeter simulation for the jet veto 
confirm qualitatively the results of ref.\cite{DIT}.

\subsection{$pp \rightarrow H  +2~forward~jets$} 

The weak boson fusion channels $qq \rightarrow qqH$ lead to energetic 
jets in the forward and backward directions, and the absence of colour 
exchange in the hard process  \cite{RAIN},\cite{RAIND},\cite{RAINH}  that 
allows to obtain a large reduction of backgrounds from $t\bar{t}$, 
QCD jets, W- and Z-production and compensate the smallness of the 
Higgs weak boson fusion cross section compared to inclusive 
$gg \rightarrow H$. Note that the process of Higgs boson 
production in the weak boson fusion with forward jet tagging has been 
considered first for the channels 
$H \rightarrow ZZ \rightarrow 4l, 2l2\nu$ in ref.\cite{CAHN}.
 The reaction  $pp \rightarrow (H \rightarrow \gamma \gamma) 
+~2~forward~jets$ has been investigated at parton level in 
ref.\cite{RAIN} and at fast  CMS detector
simulation level  in ref.\cite{DUB1}. The main conclusion of 
ref.\cite{DUB1} is that the significance $S = \frac{N_S}{\sqrt{N_B}} = 5$ 
is reached at the luminosities $(25 -35)~fb^{-1}$ for 
$m_H = 115 - 145~GeV$.    
Additional advantage of this signature is that the ratio of signal to 
background $S/B \sim 1$ in comparison with $S/B \sim 1/15$ for inclusive 
$pp \rightarrow (H\rightarrow \gamma\gamma)+ ...$ reaction.

The signature $H \rightarrow 
W^{(*)}W \rightarrow e^{\pm}\mu^{\mp}p^{mis}_T$ in weak boson 
fusion mechanism with forward jet tagging has been investigated in 
ref.\cite{RAIND}. 
The spin correlations, leading to small opening angles 
between  two charged leptons, are used to suppress the backgrounds.
This mode provides the Higgs boson discovery for 
$m_H \geq 120~GeV$.

\subsection{$H \rightarrow ZZ^{*}(ZZ) \rightarrow 4$ leptons}

The channel $H \rightarrow ZZ^{*} \rightarrow 4\,l$ is the most  promising 
one to observe Higgs boson in the mass range $130~GeV - 180~GeV$.  
Below  $2M_Z$    the event rate is small and the background
reduction more difficult, as one of the $Z$s is off mass shell. In
this mass region the width of the Higgs boson is small $\Gamma_{H}
< 1~GeV$, and the observed width is  entirely determined by the 
instrumental mass resolution.    
 The significance of the signal is proportional to the
square root of the four-lepton mass resolution ($S = N_S/\sqrt{N_B}$ and 
$N_{S,B} \sim \sigma_{4l}$), so the lepton 
energy/momentum resolution is of decisive importance 
\footnote{ Typical Higgs boson mass resolutions in this mass range are: 
$\sigma_{4\mu} \approx 1~GeV,\, \sigma_{4e} \approx 1.5~GeV$ (CMS) \cite{CMS} 
and $\sigma_{4\mu} \approx 1.6 ~GeV$, $\sigma_{4e} \approx 1.6~GeV$ (ATLAS) 
\cite{ATLAS}.}. 
The comparison of the CMS and ATLAS discovery potentials 
with $H \rightarrow ZZ^{*} \rightarrow 4~leptons$ 
based on the analyses presented  in the two Technical Proposals 
has been performed in ref.\cite{POG}. The main conclusion of the 
ref.\cite{POG} 
is that in terms of significances ATLAS and CMS discovery potentials coincide 
up to $20 \% $.    For the region $130 ~GeV \le m_{H} \le 180 ~GeV$ 
and  for $L_{t} = 100~fb^{-1}$ CMS detector 
\cite{KIN} 
will  discover the Higgs boson
with $\ge 5\sigma$ signal significance  except 
narrow mass region around $170~GeV$ where $\sigma \times Br$ 
has a minimum due to the opening 
of the $H \rightarrow WW$ channel and drop of the 
$H \rightarrow ZZ^{*}$ branching ratio just 
below the $ZZ$ threshold.  

For $180~GeV \leq m_H \leq 600~GeV$, the four-lepton 
signature is considered to be 
the most reliable one for the Higgs boson discovery at LHC, since the 
expected signal rates are large and the background is small.
The main background to the $H \rightarrow ZZ \rightarrow  4l^{\pm}$ 
process is the irreducible $ZZ$ production from $q\bar{q} \rightarrow ZZ$ 
and $gg \rightarrow ZZ$. The $t\bar{t}$ and $Zb\bar{b}$ backgrounds are  
small and reducible by a $Z$-mass cut. 
The use of this signature  allows to detect the Higgs boson 
at $\geq 5\sigma$ level up to $m_H \approx 400~GeV$ 
at $10~fb^{-1}$ and up to $m_H \approx 650~GeV$ at $100~fb^{-1}$ 
\cite{KIN}.

\subsection{$WH(t\bar{t}H) \rightarrow \gamma \gamma + 
lepton$ + ...}

The $WH \rightarrow
l\gamma\gamma + X $ and $t\bar{t}H \rightarrow l\gamma\gamma + X$
final states are other promising signature for the Higgs boson
search. The production cross section is smaller than the inclusive
$H \rightarrow \gamma \gamma $ by a factor $\approx 30$. However
the isolated hard lepton from the W and t decays allows to obtain a
strong background reduction and to indicate the primary vertex at
any luminosity.
The main conclusion \cite{DUB} is that  for an integrated luminosity 
$165 fb^{-1}$ in both channels $pp \rightarrow WH$ and 
$pp \rightarrow t\bar{t}H$ in the two-photon invariant mass interval 
$M_H - 1 ~GeV \leq M_{\gamma \gamma} \leq M_H + 1~GeV$ there 
are $\sim 100$ signal for $M_H = 120~GeV$ 
and $\sim 20$ irreducible background events if the photon transverse 
momentum cuts are $20~GeV$. 
Only in the high luminosity phase the use of this signature 
allows to make an 
important cross-checking if the Higgs boson signal has shown up 
before in $pp \rightarrow H + ...
\rightarrow \gamma \gamma +...$ classical signature.
      
\subsection{$t\bar{t}H \rightarrow t\bar{t}b\bar{b}$}

The large $H \rightarrow b\bar{b}$  branching ratio for $m_H \leq 130~GeV$ 
can be used in the associated production channel $t\bar{t}H$. To extract 
the Higgs signal in $t\bar{t}H \rightarrow l^{\pm}\nu 
q\bar{q} b\bar{b} b\bar{b}$ 
channel requires tagging of up to 4 jets, reconstruction of the Higgs boson 
mass from two b-jets and the reconstruction of the associated leptonic and 
hadronic top. The main conclusion \cite{KIN} is that this channel 
allows to discover light Higgs boson ($m_H \leq 120~GeV$).

\subsection{$H \rightarrow WW \rightarrow ll\nu\nu$,
$H \rightarrow WW \rightarrow l \nu jj$ and $H \rightarrow ZZ
\rightarrow ll jj$.}

The channel $H \rightarrow l l \nu \nu$ has a six times larger
branching than $H \rightarrow 4l^{\pm}$. The  main background comes
from $ZZ$, $ZW$, $t\bar{t}$ and $Z + jets$.  
The conclusion \cite{CMS}, \cite{ATCOL} is that using this mode it 
would be possible to
discover Higgs boson in the interval $400 ~GeV
\le m_{H} \le (800 - 900) ~GeV$ for $L_{t} = 100~fb^{-1}$.

The channels $H \rightarrow WW \rightarrow l \nu jj$ and $H \rightarrow ZZ
\rightarrow ll jj$ are important for $m_H \approx 1~TeV$ region, where 
the large $W, Z \rightarrow q\bar{q}$ branching ratios are used. Two hard 
jets from hadronic decays of $W/Z$ plus one or two high $p_T$ leptons from 
$W/Z$ decays are required. The backgrounds are: $Z +~jets$, 
$W +~jets$, $ZW$, $WW$, $t\bar{t}$. For $m_H \approx 1~TeV$ the Higgs boson 
is very broad ($\Gamma_H \approx 0.5~TeV$) and $WW/ZZ$ fusion mechanism 
gives  about 50 percent of the total cross section, therefore the forward 
region signature is essential. The main conclusion \cite{CMS}, \cite{ATCOL} 
is that the use of the decays      
$H \rightarrow WW \rightarrow l \nu jj$ and $H \rightarrow ZZ
\rightarrow ll jj$ allows to discover the heavy Higgs boson with a mass 
up to $1~TeV$ for $L_t = 100~fb^{-1}$.

\subsection{Investigation of the Higgs boson properties}

For the most interesting Higgs boson mass region 
$114.4~GeV \leq m_H \leq 193~GeV$ the $H \rightarrow \gamma \gamma$ and 
$H \rightarrow ZZ/ZZ^{*} \rightarrow  4l^{\pm}$ 
channels provide a precision in mass 
determination better than $10^{-3}$ \cite{KIN}, \cite{BRAN}, \cite{ATCOL}.
Direct measurement of the SM Higgs boson width is possible only for $m_H \geq 
200~GeV$ where the natural width exceeds the experimental mass resolution 
$\sim 1~GeV$. Precision at the $O(10^{-2})$ level is expected from 
$H \rightarrow ZZ^{*} \rightarrow 4l^{\pm}$.

The use of weak boson channels and 
$H \rightarrow WW^{*}$, $H \rightarrow \gamma \gamma$ decays 
allows to extract information on the $HWW$ coupling. The ratio of the 
Higgs boson decay widths $\Gamma_{W}/\Gamma_{Z}$ can be measured in the 
direct Higgs boson production using $\sigma_H \times BR(H \rightarrow 
WW^{*}/\sigma_H \times BR(H \rightarrow ZZ^{*}) = \Gamma_W/\Gamma_Z$.   
The simultaneous use of the  channels $H \rightarrow \gamma\gamma$ 
and $H \rightarrow ZZ^*$ allows to determine $\sigma_H \times 
BR(H \rightarrow \gamma \gamma)/ \sigma_H \times BR(H \rightarrow 
ZZ^{*}) = \Gamma(H \rightarrow \gamma \gamma)/ 
\Gamma(H \rightarrow ZZ^{*})$.
Precision of better than 20 percent is expected for these measurements 
with $300~fb^{-1}$ \cite{KIN}, \cite{BRAN}.       

\subsection{Summary}

LHC will be able to discover the SM Higgs boson from the lower LEP2 limit 
$m_H \geq 114.4~GeV$ up to $m_H = 1~TeV$ value 
(see Figs.11-13) where the Higgs boson is 
very broad $\Gamma_H \approx 0.5~TeV$ and it is no longer sensible to 
consider it as an elementary particle. 
The most reliable signatures for the LHC Higgs boson search are:

1. $H \rightarrow \gamma\gamma$,

2. $H \rightarrow ZZ^{*},ZZ \rightarrow 4l^{\pm}$,
 
3. $H \rightarrow WW^{*} \rightarrow l^{+}\nu l^{'-} \nu$,

4.  $H \rightarrow ZZ,WW, \rightarrow ll \nu \nu, lljj, l\nu jj$.

The simultaneous use of different channels  allows to extract the 
ratio of the Higgs boson decay widths.

\begin{figure}[hbt]

\vspace*{0.5cm}
\hspace*{0.0cm}
\epsfxsize=15cm \epsfbox{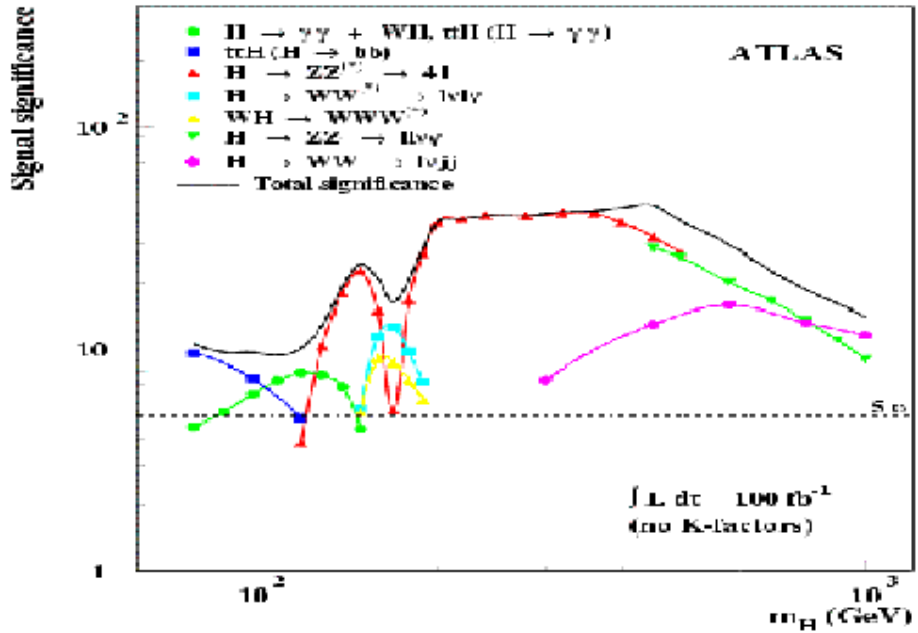}
\vspace*{0.0cm}

\caption[] {\label{fg:kun1} \it The discovery reach of the SM Higgs boson 
for $100~fb^{-1}$ in ATLAS }

\end{figure}


\begin{figure}[hbt]

\vspace*{0.5cm}
\hspace*{0.0cm}
\epsfxsize=15cm \epsfbox{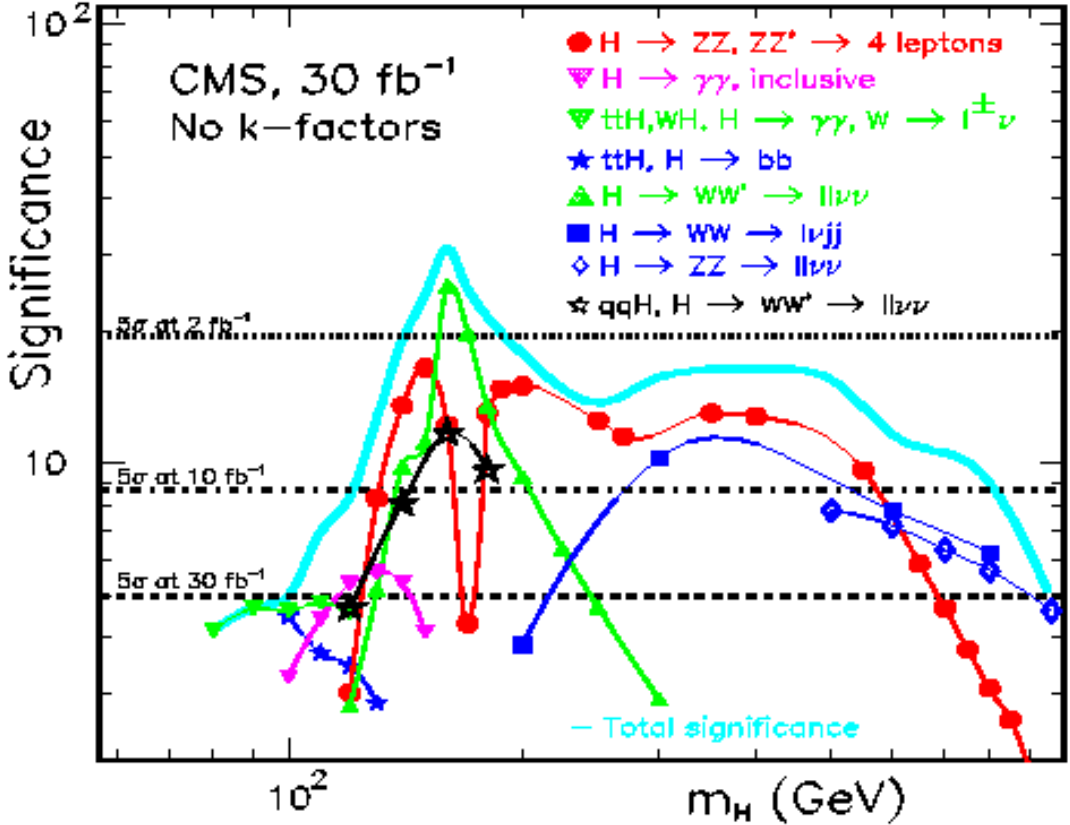}
\vspace*{0.0cm}
\caption[]{\label{fg:CMS} \it The discovery reach of the SM Higgs boson in 
the CMS for $30~fb^{-1}$ } 

\end{figure}

\begin{figure}[hbt]

\vspace*{0.5cm}
\hspace*{0.0cm}
\epsfxsize=15cm \epsfbox{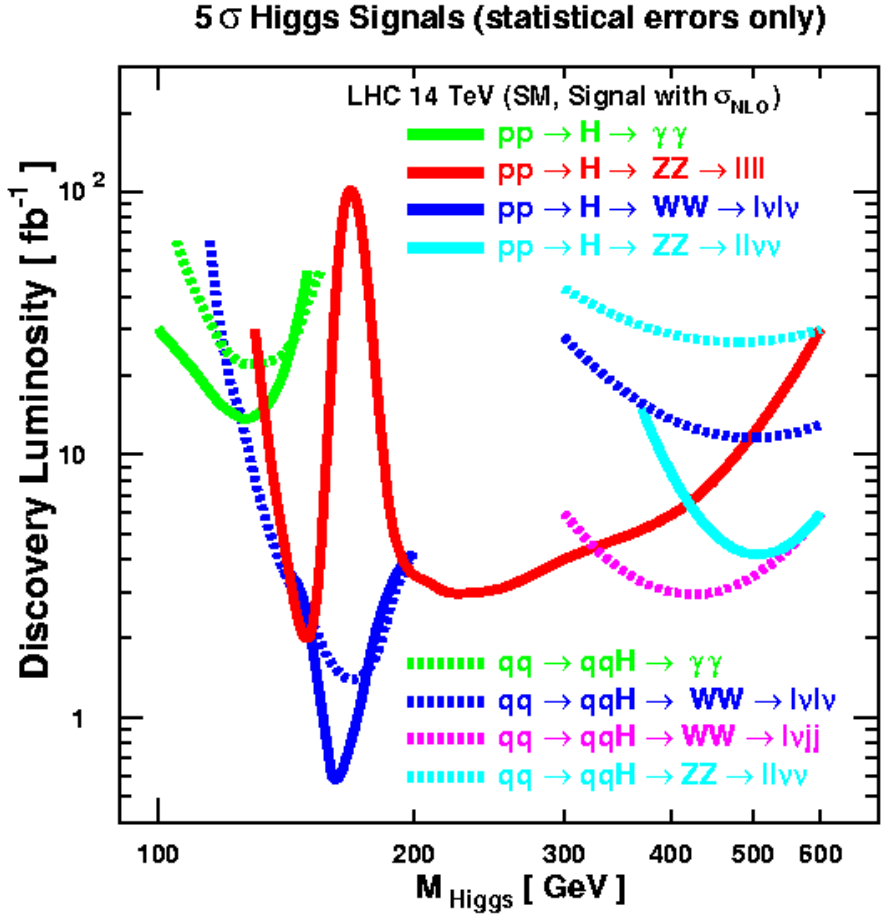}
\vspace*{0.0cm}
\caption[]{\label{fg:CMS} \it The minimum luminosity to reach 
$5\sigma$ discovery in CMS } 

\end{figure}

\begin{figure}[hbt]

\vspace*{0.5cm}
\hspace*{0.0cm}
\epsfxsize=15cm \epsfbox{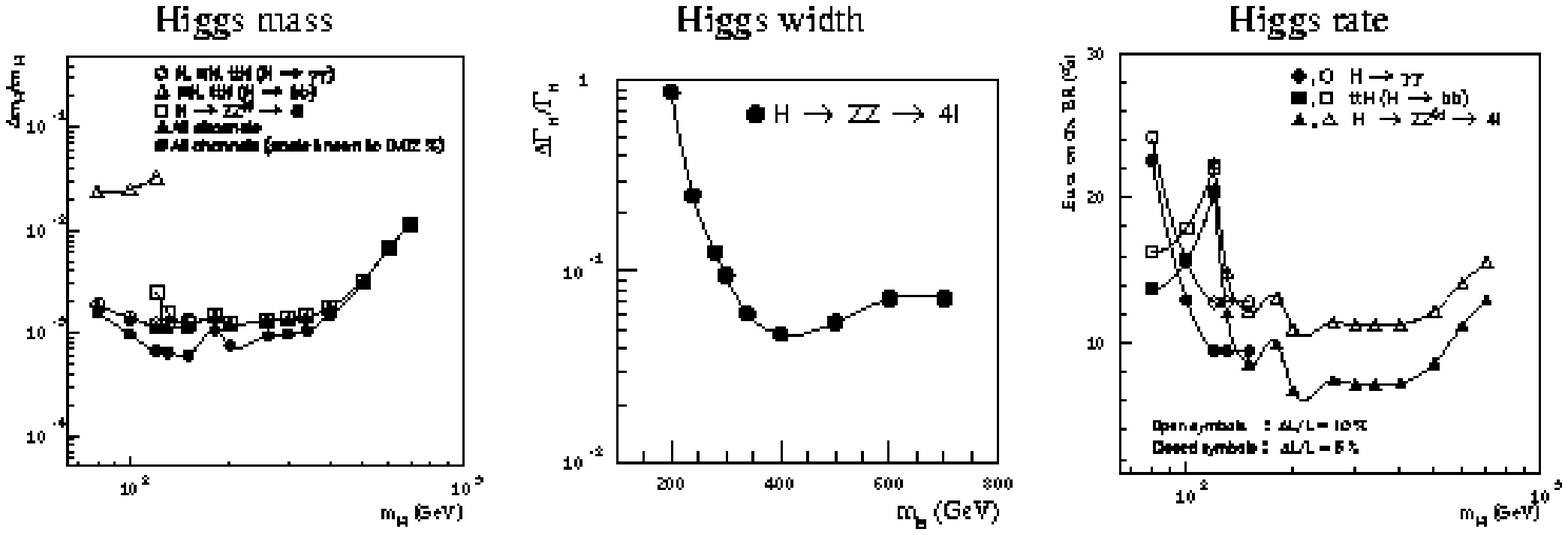}
\vspace*{0.0cm}

\caption[] {\label{fg:kun2} \it The accuracy in measuring the SM Higgs boson 
mass, width and production rates in ATLAS  }

\end{figure}


\section{Supersymmetry search within MSSM}

\subsection{The MSSM model}

Supersymmetry (SUSY) is a new type of symmetry that 
relates bosons and fermions 
\cite{GOLDFAND}, \cite{SUSY}. Locally supersymmetric theories 
necessary incorporate 
gravity \cite{WESS}. SUSY is also an essential ingredient of superstring 
theories \cite{STRING}. The  interest in SUSY is due to the observation 
that measurements of the gauge coupling constants at LEP1 are in favour of 
the Grand Unification in a supersymmetric theory with superpartners of 
ordinary particles which are lighter than $O(1)~TeV$ \cite{SUSY}. 
Besides supersymmetric electroweak models offer the simplest solution of the 
gauge hierarchy problem \cite{SUSY}. In real life supersymmetry has to be 
broken and to solve the gauge hierarchy problem the masses of superparticles 
have to be lighter than $O(1)~TeV$.  
Supergravity provides 
natural explanation of the supersymmetry breaking \cite{NIL}, 
namely, an account 
of the supergravity breaking in hidden sector leads to soft supersymmetry 
breaking in observable sector. 

An elegant formulation of supersymmetry can be achieved in the framework of 
superspace \cite{WESS}.
Two new Grassmanian (anticommuting) coordinates $\theta_{\alpha}$,  
$\bar{\theta}_{\dot{\alpha}}$ are introduced. 
Thus we extend our four dimensional 
space $x_{\mu}$ to superspace ($x_{\mu}, \theta_{\alpha}, 
\bar{\theta}_{\dot{\alpha}}$). There are two often used types of fields on 
a superfield: chiral superfield and vector superfield \cite{WESS}. 
For the chiral superfield Grassmannian Taylor expansion has the form
\begin{equation}
\Phi(y, \theta) = A(y) +\sqrt{2}\theta\psi(y) + \theta\theta F(y) ,
\end{equation}
where $y = x +i\theta\sigma\bar{\theta}$. Chiral superfield 
$\Phi(y, \theta)$ has 2 bosonic (complex scalar field $A$) 
and 2 fermionic (Weyl spinor $\psi$) degrees of freedom. The component fields 
$A$ and $\psi$ are called superpartners. The field $F$ is an auxiliary field 
and it has no physical meaning. One can get rid of the auxiliary field using 
the equations of motion. For any arbitrary function of chiral superfields one 
has
\begin{equation}
W(\Phi_i) = W(A_i + \sqrt{2}\theta \psi_i +\theta\theta F_i) = \\
W(A_i) +\frac{\partial W}{\partial A_i} \sqrt{2}\theta \psi_i + \theta\theta 
( \frac{\partial W}{\partial A_i} F_i - \frac{1}{2} \frac{\partial^2 W}
{\partial A_i \partial A_j} \psi_i \psi_j)
\end{equation}
The  $W$ is usually referred to as a superpotential which replaces the usual 
potential for the scalar potentials.
To construct the gauge invariant interactions, one needs a real vector 
superfield $V = V^{+}$. Under the abelian (super)gauge transformation 
the superfield $V$ is transformed as
\begin{equation}
V \rightarrow V + \Phi + \Phi^{+},
\end{equation}
where  $\Phi$ is chiral superfield.  One can choose a gauge 
(the Wess-Zumino gauge ) where 
\begin{equation}
V = -\theta\sigma^{\mu}\bar{\theta}v_{\mu}(x) +
i\theta\theta\bar{\theta}\bar{\lambda}(x) -  
i\bar{ \theta}\bar{\theta} \theta\lambda (x) + 
\frac{1}{2}\theta\theta \bar{\theta}\bar{\theta} D(x) \,.
\end{equation}
The physical degrees of freedom corresponding to a real vector superfield 
$V$  are the vector gauge field $v_{\mu}(x)$ and the Majorana spinor 
field $\lambda(x)$. The field $D(x)$ is an auxiliary field and 
can be eliminated 
with the help of equations of motion.
One can also define a field strength tensor  (an analog of $F_{\mu\nu}$ 
in gauge theories) as:
\begin{equation}
W_{\alpha} = - \frac{1}{4}\bar{D}^2 e^{gV}D_{\alpha}e^{-gV} \,,
\end{equation}

\begin{equation}
\bar{W}_{\dot{\alpha}} = - \frac{1}{4}D^2 e^{gV} 
\bar{D}_{\dot{\alpha}}e^{-gV} \,.
\end{equation}
The strength tensor is a chiral superfield and in the Wess-Zumino gauge it is 
a polynomial over component fields:
\begin{equation}
W_{\alpha} = T^{a}(-i\lambda^a_{\alpha} + \theta^{\alpha}D^a - \frac{i}{2}
(\sigma^{\mu}\bar{\sigma}^{\nu}\theta)_{\alpha}F^a_{\mu\nu} + \theta^2 
\sigma^{\mu}D_{\mu}\bar{\lambda}^a),
\end{equation}
where
\begin{equation}
F^a_{\mu\nu} = \partial_{\mu}v^a_{\nu} - \partial_{\nu}v^a_{\mu} + 
gf^{abc}v^b_{\mu}v^c_{\nu} \,, 
\end{equation}
\begin{equation}
D_{\mu}\bar{\lambda}^a = \partial \bar{\lambda}^a + 
gf^{abc}v^b_{\mu}\bar{\lambda}^c \,.
\end{equation}
Here $T^a$ and $f^{abc}$ are generators and structure constants of some 
group $G$.
 Supersymmetry invariant Lagrangians can be written in an elegant 
way in superspace if we use integration over the superspace according to 
the rules of Grassmannian integration \cite{WESS}
\begin{equation}
\int d\theta_{\alpha} = 0, \\ \int \theta_{\alpha}d\theta_{\beta} = 
\delta_{\alpha\beta}\,.
\end{equation}
Let us start with the case of chiral superfields without local gauge 
invariance. The space-time Lagrangian density of the renormalizable Lagrangian 
is represented in the form \cite{WESS}
\begin{equation}
L = \int d^2 \theta d^2 \bar{\theta} \Phi_i^{+} \Phi_i +
(\int d^2 \theta W_3 +h.c.)\,,
\end{equation}
where
\begin{equation}
W_3 = \lambda_i \Phi_i + \frac{1}{2} m_{ij}\Phi_i \Phi_j + \frac{1}{3}
h_{ijk}\Phi_i\Phi_j\Phi_k\,,
\end{equation} 
Performing explicit integration over the Grassmannian parameters one 
can find that

\begin{equation}
L = i \partial_{\mu} \bar{\psi}_i \bar{\sigma}^{\mu}\psi_i + \\
\partial^{\mu} A^{*}_i \partial_{\mu} A_i + F^{*}_i F_i + 
[\lambda_i F_i + m_{ij}(A_iF_j -\frac{1}{2} \psi_i\psi_j) +   \\ 
h_{ijk}(A_i A_j F_k - \psi_i \psi_j A_k) +h.c.]\,.
\end{equation}

Eliminating the auxiliary fields $F_i$ and $F^{*}_i$ using the equations of 
motion, one finally gets
\begin{equation}
L = i \partial_{\mu} \bar{\psi}_i \bar{\sigma}^{\mu}\psi_i + \\
\partial^{\mu} A^{*}_i \partial_{\mu} A_i - 
(\frac{1}{2} m_{ij} \psi_i\psi_j  +  \\ 
h_{ijk}\psi_i \psi_j A_k +h.c.) - \\
|\lambda_k + m_{ik}A_i + h_{ijk}A_iA_j|^2 \,.
\end{equation}
 Let us consider the case of gauge fields. The supersymmetric generalisation 
of the Yang-Mills Lagrangian has the form 
\begin{equation}
L_{SYM} =    \frac{1}{4}\int d^2\theta Tr(W^{\alpha}W_{\alpha}) + h.c.
\end{equation}
In terms of component fields the Lagrangian (67) has the form
\begin{equation}
L_{SYM} = -\frac{1}{4}F^a_{{\mu\nu}}F^{a \mu \nu} - i \lambda^a\sigma^{\mu} 
D_{\mu}\bar{\lambda}^a +\frac{1}{2}D^a D^a \,.
\end{equation}
A complete SUSY and gauge invariant renormalizable Lagrangian looks 
like
\begin{equation}
L_{SUSY YM} = \frac{1}{4}(\int d^2 \theta Tr(W^{\alpha}W_{\alpha}) +h.c.) \\
+\int d^2 \theta d^2 \bar{\theta} \Phi^{+}_{ia}(e^{gV})^a_b\Phi^b_i + \\
(\int d^2 \theta W_3(\Phi_i) +h.c.) \,,
\end{equation}
where $W_3(\Phi_i)$ is a gauge invariant superpotential. In terms of the 
component fields the above Lagrangian takes the form
\begin{eqnarray}
&&L_{SYM} = -\frac{1}{4}F^a_{{\mu\nu}}F^{a \mu \nu} - i \lambda^a\sigma^{\mu} 
D_{\mu}\bar{\lambda}^a +\frac{1}{2}D^a D^a \\ \nonumber
&&+(\partial_{\mu}A_i -igv^a_{\mu}T^aA_i)^{+}
(\partial^{\mu} A_i - igv^a_{\mu}T^a
A_i) \\ \nonumber 
&& -i\bar{\psi}_i\bar{\sigma}^{\mu}(\partial_{\mu}\psi_i - igv^a_{\mu}T^a
\bar{\psi}_i) -  
gD^aA^{+}_iT^aA_i -(i\sqrt{2}gA^{+}_iT^a\lambda^a \psi_i 
+ h.c.) \\ \nonumber 
&&+F^+_iF_i 
+(\frac{\partial W}{\partial A_i}F_i  -  
\frac{1}{2}\frac{\partial^2 W}{\partial A_i \partial A_j} \psi_i \psi_j 
+h.c.)\,.
\end{eqnarray}
Integrating out the auxiliary fields $D^a$ and $F_i$ one reproduces the 
usual Lagrangian.
  
The simplest supersymmetric generalisation of 
the SM is the Minimal Supersymmetric Standard Model (MSSM) \cite{SUSY},
\cite{KAZ}. It is 
supersymmetric model based on standard $SU_c(3) \otimes SU_L(2) \otimes U(1)$ 
gauge group with electroweak symmetry spontaneously broken via vacuum 
expectation values of two different Higgs doublets. The MSSM consists of 
taking the SM and adding the corresponding supersymmetric partners. It should 
be stressed that the MSSM contains two hypercharges $Y = \pm 1$ 
Higgs doublets, which is the minimal structure for the Higgs sector of an 
anomaly-free supersymmetric extension of the SM. The supersymmetric 
electroweak models also require at least two Higgs doublets to generate 
masses for both ``up''-type and ``down''-type fermions. 

The SUSY generalisation of the SM model Lagrangian can be presented in the form
\begin{equation}
L_{SUSY}  = L_{Gauge,M} + L_{Yukawa},
\end{equation}
where
\begin{equation}
L_{Gauge,M} = \sum_{SU(3),SU(2),U(1)} \frac{1}{4}(\int d^2\theta Tr W^{\alpha}
W_{\alpha} + h.c.) +\sum_{matter}\int d^2\theta d^2\bar{\theta}
\Phi^{+}_ie^{gV}\Phi_i \,,
\end{equation}
\begin{equation}
L_{Yukawa} = \int d^2 \theta (W_R + W_{NR}) +h.c.\,.
\end{equation}
The renormalizable 
superpotential $W_R$ of the MSSM determines the Yukawa 
interactions of quarks and leptons 
and  preserves global $B - L$. Here $B$ is the baryon number and $L$ is the 
lepton number. 
It has the form 
\begin{equation}
W_R = \epsilon_{ij}(h^U_{ab}Q^j_aU^c_bH^i_2 + h^D_{ab}Q^J_aD^c_bH^i_1  \\ 
+ h^L_{ab}L^j_aE^c_bH^i_1 + \mu H^i_1H^j_2), 
\end{equation}
where $i, j = 1,2,3$ are the $SU(2)$ and $a, b = 1,2,3$ are the 
generation indices; colour indices are omitted. The last term in the 
superpotential (74) describes the Higgs boson mixing. The 
most general expression for the superpotential 
$W_{NR}$ has the form
\begin{equation}
W_{NR} = \epsilon_{ij}(\lambda^L_{abd}L^i_aL^j_bE^c_d + \lambda^L_{1abd}
L^i_aQ^j_bD^c_d + \mu^{'}_aL^i_aH^j_2) + \lambda^B_{abd}U^c_aD^c_bD^c_d
\end{equation}
Note that the most general expression for the effective 
superpotential ( 75) contains renormalizable terms violating $B - L$ that can 
lead to the problems with proton decay. 
To  get rid of such dangerous terms 
in the superpotential, R-parity conservation \cite{R-parity} 
is postulated. Here 
$ R = (-1)^{3(B-L)+2S}$ for a particle with spin $S$. For ordinary particles 
$R = 1$, whereas for the corresponding supersymmetric partners $R = -1$. 
If we postulate $R$-parity conservation then the $W_{NR} = 0$.
Experimental bounds on  the $R$-parity violating coupling constants are 
rather strong \cite{PARTICLE}, \cite{R-bound}
\begin{equation}
\lambda^L_{abc} < O(10^{-4})\,,
\end{equation}
\begin{equation}
\lambda^L_{1abc} < O(10^{-4})\,,
\end{equation}
\begin{equation}
\lambda^B_{abc} < O(10^{-9})\,.
\end{equation}
The $R$-parity conservation has a crucial 
impact on supersymmetric phenomenology. 
An important consequence of $R$-parity conservation is that the lightest 
supersymmetric particle (LSP) is stable. The cosmological constraints  imply 
that the LSP is weakly-interacting electrically neutral and colourless 
particle. Other important consequence of the $R$-parity conservation is that 
at supercolliders superparticles are produced in pairs, therefore at least 
two LSP have eventually to be produced at the end of the decays of heavy 
unstable supersymmetric particles. Being weakly interacting particle LSP  
escapes detector registration, therefore the classic signature for 
$R$-parity conserving supersymmetric models is the transverse missing 
energy due to the LSP-escape. 

In real life supersymmetry has to be broken. At present the most popular 
scenario for producing low-energy supersymmetry breaking is called the 
hidden sector scenario  \cite{SUSY},   \cite{NIL}, \cite{DINE}. 
According to this scenario, there exists two 
sectors: the usual matter belongs to the visible sector. The second hidden 
sector of the theory contains fields which lead to the supersymmetry breaking. 
These two sectors interact with each other by exchange of some fields 
which mediate SUSY breaking from the hidden to visible sector. At present 
there are two most elaborated scenario of SUSY breaking:

1. Gravity mediation (SUGRA) ,

2. Gauge mediation.

In SUGRA scenario  \cite{NIL}, \cite{SUSY} visible and hidden 
sectors interact with each other via 
gravity. Some scalar fields in hidden sector develop nonzero 
vacuum expectation values 
for their $F$-components which lead to spontaneous SUSY breaking. 
Since in SUGRA theory supersymmetry is local, spontaneous SUSY breaking 
leads to Goldstone particle which is a Goldstone fermion. With the help 
of a super-Higgs effect this particle is absorbed into an additional 
component of a spin $3/2$ particle -gravitino which becomes massive in 
close analogy with the standard Higgs mechanism. SUSY breaking is 
mediated to a visible sector via gravitational interactions leading to 
SUSY breaking scale $M_{SUSY} \sim m_{3/2}$, where $m_{3/2}$ is the gravitino 
mass.  The effective low-energy Lagrangian contains explicit soft 
supersymmetry breaking terms 
\begin{equation}
L_{soft} = -\sum_{i,j}m^2_{ij}A_iA_j^{*} - 
\sum_i M_i(\lambda_i \lambda_i + 
\bar{\lambda}_i\bar{\lambda}_i) -(BW^{(2)}(A) + BW^{(3)}(A) +h.c.)\,.
\end{equation}
Here $W^{(2)}$ and $W^{(3)}$ are the quadratic and cubic terms of the visible 
superpotential, respectively. 
The mass parameters in soft SUSY breaking Lagrangian (79) are proportional to 
the gravitino mass $m_{3/2}$.

In gauge mediation mechanism \cite{DINE} the SUSY breaking is mediated to the 
observable world via gauge interactions. The messengers are the gauge 
bosons or matter fields of the SM or of its GUT generalisation. In such 
scenario it is possible to have a renormalizable model with dynamic 
SUSY breaking where (in principle) all the parameters are calculated numbers.
In gauge mediated scenario all soft SUSY breaking masses are correlated to 
the gauge couplings. Besides there is no problem with flavour violating 
couplings as well.  In this scenario the lightest stable superparticle (LSP) 
is the gravitino. In general soft SUSY breaking mechanisms lead to 
the generation of soft SUSY breaking operators of the dimension $\leq 3$.
 
In the MSSM supersymmetry is softly broken at some high scale $M$ by generic
soft terms
\begin{eqnarray}
&&-L_{soft} =
m_{0}(A^u_{ij}U^c_iQ_jH_2 +A^d_{ij}D^c_i G_j H_1 + \\ \nonumber
&&A^l_{ij}E^c_iL_jH_1 + h.c.) +
(m^2_q)_{ij}Q^+_iQ_j  +(m^2_u)_{ij}(U^c_i)^+U^c_j \\ \nonumber
&&+ (m^2_d)_{ij}(D^c_i)^+ D^c_j + (m^2_l)_{ij}(L^c_i)^+ L^c_j +
(m^2_e)_{ij}(E^c_i)^+ \\ \nonumber
&&E^c_j + m^2_1 H_1H^+_1
+ m^2_2H_2H^+_2
+ (B{m_{0}}^2 H_1H_2 + h.c.) +  (\frac{1}{2}m_a (\lambda_{a}
\lambda_{a})+h.c.)  \,.
\end{eqnarray}

In general all soft SUSY breaking terms are arbitrary that complicates the 
analysis and spoils the predictive power of the theory. 
In MSUGRA model \cite{SUSY}, \cite{KAZ}               
the universality  of different soft parameters at GUT scale is 
postulated. Namely, all the spin 0 particle masses 
(squarks, sleptons, higgses) are postulated to be equal to the 
universal value $m_0$ at GUT scale. All gaugino particles masses 
are  postulated to be equal 
to $m_{1/2}$ at GUT scale and all 
the cubic and quadratic terms proportional to $A$ and $B$ are postulated 
to repeat the structure  of the Yukawa superpotential (74). It should be 
stressed that the MSUGRA model is not the general model and it could be 
considered as a ``toy model''  for concrete applications. 

So, in MSUGRA model  soft SUSY breaking masses and coupling constants 
 are supposed to be equal  at $M_{GUT}$ scale, namely:
\begin{equation}
A^u_{ij}(M_{GUT}) = Ah^u_{ij}(M_{GUT}), \;
A^d_{ij}(M_{GUT}) = Ah^d_{ij}(M_{GUT}), \; A^l_{ij}(M_{GUT}) = A
h^l_{ij}(M_{GUT}),
\end{equation}
\begin{eqnarray}
&&(m^2_q)_{ij}(M_{GUT}) = (m^2_u)_{ij}(M_{GUT}) = (m^2_d)_{ij}
(M_{GUT}) =  \\ \nonumber
&&(m^2_l)_{ij}(M_{GUT}) = (m^2_e)_{ij}(M_{GUT}) =
\delta_{ij} m^2_1(M_{GUT}) = \delta_{ij} m^2_2(M_{GUT}) =
\delta_{ij} m_0^2 \,,
\end{eqnarray}
\begin{equation}
m_1(M_{GUT}) = m_2(M_{GUT}) = m_3(M_{GUT}) = m_{1/2} \,.
\end{equation}
Note that it is more appropriate to impose boundary conditions
not at GUT scale but at Planck scale $M_{PL} = 2.4\cdot 10^{18}~GeV$.
An account of the renormalization effects between Planck  and
GUT scales can drastically change the features of the spectrum.
For instance, if we assume that the physics between Planck scale and
GUT scale is described by SUSY $SU(5)$ model then an account of
the evolution between Planck  and GUT scales \cite{POLONSKY}, 
\cite{POPOV} can change
qualitatively the spectrum of sleptons for $m_{0} \ll m_{1/2}$ \cite{POPOV}.
The renormalization group equations for soft SUSY
breaking parameters in neglection of all Yukawa coupling constants
except top-quark Yukawa in one loop 
approximation read \cite{LOPEZ}, \cite{KAZ}
\begin{equation}
\frac{d\tilde{m}^2_L}{dt} = (3\tilde{\alpha}_2M^2_2 +\frac{3}{5}
\tilde{\alpha}_1 M^2_1) \,,
\end{equation}
\begin{equation}
\frac{d\tilde{m}^2_E}{dt} = (\frac{12}{5}\tilde{\alpha}_1M^2_1) \,,
\end{equation}
\begin{equation}
\frac{d\tilde{m}^2_Q}{dt} = (\frac{16}{3}\tilde{\alpha}_3M^2_3 +
3\tilde{\alpha}_2M^2_2 + \frac{1}{15} \tilde{\alpha}_1 M^2_1) -
\delta_{i3}Y_t(\tilde{m}^2_Q +\tilde{m}^2_U + m^2_2 + A^2_tm^2_0 -
\mu^{2}) \,,
\end{equation}
\begin{equation}
\frac{d\tilde{m}^2_U}{dt} = (\frac{16}{3}\tilde{\alpha}_3M^2_3 +
\frac{16}{15}\tilde{\alpha}_1M^2_1) - \delta_{i3}2Y_t(\tilde{m}^2_Q
+\tilde{m}^2_U + m^2_2 +A^2_tm^2_0 - \mu^{2}) \,,
\end{equation}
\begin{equation}
\frac{d\tilde{m}^2_D}{dt} = (\frac{16}{3}\tilde{\alpha}_3M^2_3 +
\frac{4}{15}\tilde{\alpha}_{1}M^2_1) \,,
\end{equation}
\begin{equation}
\frac{d\mu^{2}}{dt} = 3(\tilde{\alpha}_2 +\frac{1}{5}\tilde{\alpha}_1
-Y_t)\mu^{2} \,,
\end{equation}
\begin{equation}
\frac{dm^2_1}{dt} = 3(\tilde{\alpha}_2M^2_2 +
\frac{1}{5}\tilde{\alpha}_1M^2_1)
  + 3(\tilde{\alpha}_2 + \frac{1}{5}
\tilde{\alpha}_1 - Y_t)\mu^{2} \,,
\end{equation}
\begin{equation}
\frac{dm^2_2}{dt} =  3(\tilde{\alpha}_2M^2_2
+\frac{1}{5}\tilde{\alpha}_1M^2_1) + 3(\tilde{\alpha}_2 + \frac{1}{5}
\tilde{\alpha}_1)\mu^{2}
-3Y_t(\tilde{m}^2_Q + \tilde{m}^2_U + m^2_2 + A^2_t m^2_0) \,,
\end{equation}
\begin{equation}
\frac{dA_t}{dt} = -(\frac{16}{3}\tilde{\alpha}_3\frac{M_3}{m_0} +
3\tilde{\alpha}_2\frac{M_2}{m_0} + \frac{13}{15} \tilde{\alpha}_1\frac
{M_1}{m_0}) - 6Y_tA_t \,,
\end{equation}
\begin{equation}
\frac{dB}{dt} = -3(\tilde{\alpha}_2\frac{M_2}{m_0} +
\frac{1}{5} \tilde{\alpha}_1 \frac{M_1}{m_0}) - 3Y_tA_t \,,
\end{equation}
\begin{equation}
\frac{dM_i}{dt} = -b_i\tilde{\alpha}_iM_i \,,
\end{equation}
\begin{equation}
b_1 =\frac{33}{5}, \; b_2 =1,\;b_3=-3 \,.
\end{equation}
Here $\tilde{m}_U$, $\tilde{m}_D$, $\tilde{m}_E$ refer to the masses
of the superpartner of the quark and lepton singlets, while
$\tilde{m}_Q$ and $\tilde{m}_L$ refer to the masses of the isodoublet
partners; $m_1$, $m_2$, $m_3$ and $\mu$ are the mass parameters in 
the Higgs potential, $A$ and $B$ are the coupling constants 
 of the $L_{soft}$ as defined before; 
$M_i$ are  gaugino  masses before mixing. The
renormalization group equation for the top
Yukawa coupling constant has the form
\begin{equation}
\frac{dY_t}{dt} =  Y_t(\frac{16}{3}\tilde{\alpha}_3 +3
\tilde{\alpha}_{2} +\frac{13}{15} \tilde{\alpha}_{1}) - 6Y^2_t \,,
\end{equation}
while the RG equations for the gauge couplings are
\begin{equation}
\frac{d\tilde{\alpha}_i}{dt} = -b_i\tilde{\alpha}^2_i \,.
\end{equation}
Here
\begin{equation}
\tilde{\alpha}_i = \frac{\alpha_{i}}{4\pi},\;
Y_t =\frac{h^2_t}{16\pi^{2}},\; t = \ln{(\frac{M^2_{GUT}}{Q^2})} \,,
\end{equation}
and the top Yukawa coupling $h_t$ is related to the running top
mass by the relation
\begin{equation}
m_t = h_t(m_t)\frac{v}{\sqrt{2}}\sin{\beta} \,.
\end{equation}
The boundary conditions at $Q^2 = M^2_{GUT}$ are
\begin{equation}
\tilde{m}^2_Q = \tilde{m}^2_U = \tilde{m}^2_D = \tilde{m}^2_E =
\tilde{m}^2_L = m^2_0 \,,
\end{equation}
\begin{equation}
\mu = \mu_{0}; \; m^2_1 = m^2_2 = \mu^{2}_{0} +m^2_0; \;
m^2_3 = B\mu_{0} m_0 \,,
\end{equation}
\begin{equation}
M_i = m_{1/2}, \; \tilde{\alpha}_{i}(0) = \tilde{\alpha}_{GUT}; \;
i =1,2,3 \,.
\end{equation}
For the gauginos of the $ SU_{L}(2) \otimes U(1)$ gauge group one has to
consider the mixings with the Higgsinos. The mass terms in the full
Lagrangian are 
\begin{equation}
L_{Gaugino-Higgsino} = -\frac{1}{2}
   M_3 \bar{\lambda}^a_3 \lambda^{a}_{3}
   -\frac{1}{2} \bar{\chi} M^{(0)} \chi -
   (\bar{\psi}M^{(c)}\psi +h.c.) \,,
\end{equation}
where $\lambda^{a}_{3}$ are the 8 Majorana gluino fields, and
\begin{equation}
\chi = \left( \begin{array}{cccc}
\tilde{B}^0\\
\tilde{W}^3\\
\tilde{H}^0_1\\
\tilde{H}^0_2
\end{array} \right) \,,
\end{equation}
\begin{equation}
\psi = \left( \begin{array}{cc}
\tilde{W}^+\\
\tilde{H}^+
\end{array} \right) \,,
\end{equation}
are the Majorano neutralino and Dirac chargino fields. The mass
matrices are:
\begin{equation}
M^{(0)} = \left( \begin{array}{cccc}
M_1 & 0 & -A & B\\
0 & M_2 & C & -D \\
-A & C & 0 & -\mu\\
B & -D & -\mu & 0
\end{array} \right) \,,
\end{equation}
\begin{equation}
M^{(c)} = \left( \begin{array}{cc}
M_2 & \sqrt{2}M_W \sin{\beta} \\
\sqrt{2}M_W\cos{\beta} & \mu
\end{array} \right) \,,
\end{equation}
where:
\begin{equation}
A = M_Z\cos{\beta}\sin{\theta_{W}}, \; B = M_Z
\sin{\beta}\sin{\theta_{W}} \,,
\end{equation}
\begin{equation}
C =M_Z\cos{\beta}\cos{\theta_{W}}, \; D =
M_Z\sin{\beta}\cos{\theta_{W}} \,.
\end{equation}
After the solution of the corresponding renormalization group
equations for $\alpha_{GUT} = \frac{1}{24.3}$, $M_{GUT} =
2.0\cdot10^{16}$ Gev, $\sin^{2}{\theta_{W}} =0.2324$ , 
$\tan \beta = 1.65$ and
$A_t(0) = 0$ one finds the numerical formulae for effective  squark and
slepton square masses at electroweak scale \cite{KAZ}
\begin{equation}
\tilde{m}^2_{E_{L}}(M_Z) = m^2_0 + 0.52m^2_{1/2} -0.27\cos{2\beta}
M^2_Z \,,
\end{equation}
\begin{equation}
\tilde{m}^2_{\nu_{L}}(M_Z) =m^2_0 + 0.52m^2_{1/2}
+0.5\cos{2\beta}M^2_Z \,,
\end{equation}
\begin{equation}
\tilde{m}^2_{E_{R}}(M_Z) = m^2_0 +0.15m^2_{1/2}
-0.23\cos{2\beta}M^2_Z  \,,
\end{equation}
\begin{equation}
\tilde{m}^2_{U_{L}}(M_Z) = m^2_0 + 6.5m^2_{1/2}
+0.35\cos{2\beta}M^2_Z  \,,
\end{equation}
\begin{equation}
\tilde{m}^2_{D_{L}}(M_Z) = m^2_0 + 6.5m^2_{1/2} -
0.42\cos{2\beta}M^2_Z \,,
\end{equation}
\begin{equation}
\tilde{m}^2_{U_{R}}(M_Z) = m^2_0 +6.1m^2_{1/2}
+0.15\cos{2\beta}M^2_Z  \,,
\end{equation}
\begin{equation}
\tilde{m}^2_{D_{R}}(M_Z) = m^2_0 +6.0m^2_{1/2}
-0.07\cos{2\beta}M^2_Z  \,,
\end{equation}
\begin{equation}
\tilde{m}^2_{b_{R}}(M_Z) = \tilde{m}^2_{D_{R}}  \,,
\end{equation}
\begin{equation}
\tilde{m}^2_{b_{L}}(M_Z) = \tilde{m}^2_{D_{L}} - 0.49m^2_0
-1.21m^2_{1/2}  \,,
\end{equation}
\begin{equation}
\tilde{m}^2_{t_{R}}(M_Z) = 
\tilde{m}^2_{U_{R}}(M_Z) +m^2_t -0.99m^2_0 -2.42m^2_{1/2}  \,,
\end{equation}
\begin{equation}
\tilde{m}^2_{t_{L}}(M_Z) = \tilde{m}^2_{U_{L}}(M_Z) + m^2_t - 0.49m^2_0
-1.21 m^2_{1/2} \,.
\end{equation}
After mixing the mass eigenstates of the stop matrix are:
\begin{eqnarray}
&&\tilde{m}^2_{t_{1,2}}(M_Z) \approx
\frac{1}{2}[0.5m^2_0 +9.1m^2_{1/2} +2m^2_t +
0.5\cos{2\beta}M^2_Z] \\ \nonumber
&&\mp \frac{1}{2}[(1.5m^2_{1/2} + 0.5 m^2_0 + 0.2\cos{(2\beta)}M^2_Z)^2
+4m^2_t(A_tm_o -\mu/\tan{\beta})^2]^{1/2} \,.
\end{eqnarray}
The gauginos and Higgsinos have similar quantum numbers which causes
a mixing between the weak interaction eigenstates and the mass
eigenstates. The two chargino eigenstates $\chi^{\pm}_{1,2}$ are:
\begin{equation}
M^2_{1,2} = \frac{1}{2}[M^2_2 + \mu^{2} + 2M^2_W] \mp
\frac{1}{2}[(M^2_2 -\mu^{2})^2 +
4M^4_W\cos^2{2\beta} + 4M^2_W(M^2_2 +\mu^{2} +
2M_2\mu\sin{2\beta})]^{1/2} \,,
\end{equation}
where at GUT scale the masses of  
gaugino fields of the $SU(3)$,
$SU_L(2)$ and $U(1)$ groups are equal to $m_{1/2}$. The eigenvalues of
the $4\times 4$ neutralino mass matrix can be solved by a numerical
diagonalization. If the parameter $\mu$ is much larger than $M_1$
and $M_2$, the mass eigenstates become
\begin{equation}
\chi^{0}_{i} = [\tilde{B}, \tilde{W}_{3},
\frac{1}{\sqrt{2}}(\tilde{H}_{1} - \tilde{H}_{2}),
\frac{1}{\sqrt{2}}(\tilde{H}_{1} + \tilde{H}_{2})] \,
\end{equation}
with eigenvalues $|M_1|$, $|M_2|$, $|\mu| $ and $|\mu|$ respectively
( the bino and neutral wino do not mix with each other and  with the
Higgsino eigenstates).

The tree level Higgs potential in MSSM has the form
\begin{equation}
V_{0}(H_1,H_2) = m^2_1|H_1|^2 + m^2_2|H_2|^2 - m^2_3(H_1H_2 + h.c)
+\frac{g^2_2 + {g}^2_1}{8}(|H_1|^2 - |H_2|^2)^2 + \frac{g^2_2}{2}
|H^{+}_1H_2|^2 \,.
\end{equation}

The minimisation of the effective potential $V_0(H_1,H_2)$ leads to
the equations:
\begin{equation}
v^2 = \frac{8(m^2_1-m^2_2\tan^2{\beta})}{(g^2_2 + {g}^2_1)
(\tan^2{\beta} -1)} \,,
\end{equation}
\begin{equation}
\sin{2\beta} = \frac{2m^2_3}{m^2_1 + m^2_2} \,.
\end{equation}
 After the diagonalization of the corresponding mass matrices
CP-odd neutral Higgs boson $A(x)$ acquires a mass $m^2_A = m^2_1 +
m^2_2$, charged Higgs boson $H^{+}(x)$ acquires a mass $m^2_{H^{+}} =
m^2_{A} + M^2_W$ and CP-even Higgs bosons $H(x)$ and $h(x)$ have
masses
\begin{equation}
m^2_{H,h} = \frac{1}{2}[m^2_A +M^2_Z \pm
\sqrt{(m^2_A + M^2_Z)^2 - 4m^2_AM^2_Z\cos^2{2\beta}} ~] \,,
\end{equation}
where $<H_1> = v_1 = \frac{v\cos{\beta}}{\sqrt{2}}$,
$<H_2> = v_2 = \frac{v\sin{\beta}}{\sqrt{2}}$ ,
$\tan{\beta} = \frac{v_2}{v_1}$.
At tree level we have the following mass relations:
\begin{equation}
m^2_h + m^2_H = m^2_A + M^2_Z \,,
\end{equation}
\begin{equation}
m_h \leq m_A \leq m_H \,,
\end{equation}
\begin{equation}
m_h \leq M_Z|\cos{2\beta}| \leq M_Z \,.
\end{equation}
Therefore at tree level the lightest Higgs boson is lighter than the
Z-boson. However the radiative corrections due to big top quark
Yukawa coupling constant increase the mass of the lightest Higgs
boson in MSSM \cite{MHIGGS}. The upper limit on the Higgs
boson mass in MSSM depends
on the value of the top quark mass and on the value of stop
quark masses. 
For $m_{t,pole} = 175 $ Gev and stop quark
masses lighter than 1 Tev the Higgs boson mass is  
lighter than $135~GeV$ \cite{DEGRAS}.

After the solution of the corresponding equations for
the determination of nontrivial electroweak potential the number
of unknown parameters is decreased by 2. At present
more or less standard choice of free parameters in MSSM includes
$m_0$, $m_{1/2}$, $\tan{\beta}$, $A$ and sign($\mu$).

\subsection{Superparticle cross sections.}

At LHC sparticles can be produced via the following tree level
reactions \cite{EICHT}:

 a. $gg,qq, qg \rightarrow \tilde{g} \tilde{g},\, \tilde{g}
\tilde{q}, \, \tilde{q} \tilde{q}$ ,

b. $qq, gq \rightarrow \tilde{g} \tilde{\chi}^0_{i},\, \tilde{g}
\tilde{\chi}^{\pm}_i,\, \tilde{q}\tilde{\chi}^0_i,\,
\tilde{q} \tilde{\chi}^{\pm}_i$ ,

c. $qq \rightarrow \tilde{\chi}^{\pm}_i \tilde{\chi}^{\mp}_j,\, 
\tilde{\chi}^{\pm}_i
\tilde{\chi}^0_j,\, \tilde{\chi}^0_i \tilde{\chi}^0_j $ ,

d. $qq \rightarrow \tilde{l} \tilde{\nu}, \, \tilde{l} \tilde{l},
\, \tilde{\nu} \tilde{\nu}$ ,

It is straightforward to calculate the elementary (tree level)
cross sections for the production of superparticles in collisions of
quarks and gluons. Here following ref.\cite{EICHT}
we collect the main formulae for elementary cross sections.

The differential cross section of the production of two gauge
fermions in quark-antiquark collisions is

\begin{eqnarray}
&&\frac{d\sigma}{dt}(q \bar{q}^{'} \rightarrow gaugino1 +
gaugino2) = \\ \nonumber
&&\frac{{\pi}}{s^2}
[ A_s \frac{(t - m^2_2)(t-m^2_1) + (u -m^2_1)(u -m^2_2) +2s m_1m_2}
{(s -M^2_s)^2} +
A_t \frac{(t-m^2_1)(t-m^2_2)}{(t-M^2_t)^2} + \\ \nonumber
&&A_u \frac{(u-m^2_1)(u-m^2_2)}{(u-M^2_u)^2} +
A_{st} \frac{(t-m^2_1)(t-m^2_2) +
m_1m_2s}{(s - M^2_s)(t-M^2_t)} + \\ \nonumber
&&A_{tu} \frac{m_1m_2s}{(t-M^2_t)(u-M^2_u)}
+ A_{su} \frac{(u-m^2_1)(u-m^2_2) +m_1m_2s}{(s-M^2_s)(u-M^2_u)}] \,,
\end{eqnarray}
where $m_1$ and $m_2$ are the masses of the produced gauginos,
$M_s$, $M_t$ and $M_u$ are the masses of the particles
exchanged in the s,t, and u channels respectively.
The coefficients $A_x$ are given in ref. \cite{EICHT}. For instance,
for the case of the gluino pair
production in quark-antiquark collisions the coefficients $A_x$ are
\cite{EICHT}:

$A_t = \frac{4}{9} A_s$, $A_u = A_t$, $A_{st} = A_s$, $ A_{su} =
A_{st}$, $A_{tu} = \frac{1}{9} A_s$, $A_s = \frac{8\alpha^2_s}{3}
 \delta_{qq^{'}}$

The differential cross section for the production of gluino pairs
in gluon-gluon collisions is
\begin{eqnarray}
&&\frac{d\sigma}{dt}(gg \rightarrow \tilde{g} \tilde{g}) =  \\ \nonumber
&&\frac{9\pi\alpha^2_s}{4s^2}[
\frac{2(t - m^2_{\tilde{g}})(u - m^2_{\tilde{g}})}{s^2} +
[[\frac{(t-m^2_{\tilde{g}})(u-m^2_{\tilde{g}}) - 2m^2_{\tilde{g}}
(t + m^2_{\tilde{g}})}{(t-m^2_{\tilde{g}})^2} \\ \nonumber
&&+ \frac{(t-m^2_{\tilde{g}})(u-m^2_{\tilde{g}}) + m^2_{\tilde{g}}
(u - t)}{s(t-m^2_{\tilde{g}})}] + (t \leftrightarrow u) ] \\ \nonumber
&&+\frac{m^2_{\tilde{g}}(s-4m^2_{\tilde{g}})}{(t-m^2_{\tilde{g}})
(u-m^2_{\tilde{g}}}] \,.
\end{eqnarray}
The total cross section has the form
\begin{equation}
 \sigma(gg \rightarrow \tilde{g}\tilde{g}) =
\frac{3\pi\alpha^2_s}{4s}[3[1 + \frac{4m^2_{\tilde{g}}}{s} -
\frac{4m^4_{\tilde{g}}}{s^2}]\ln{[\frac{s+L}{s-L}]}
- [4 + \frac{17m^2_{\tilde{g}}}{s}]\frac{L}{s}] \,,
\end{equation}
where $ L = [s^2 - 4m_{\tilde{g}}^2s]^{1/2}$.

The differential cross section  for the reaction
$q_iq_j \rightarrow \tilde{q}_i \tilde{q}_j$ for the case of
equal masses of right-handed and left-handed squarks is
\begin{eqnarray}
&&\frac{d\sigma}{dt}(q_iq_j \rightarrow \tilde{q}_i\tilde{q}_j)
= \\ \nonumber
&&\frac{4\pi \alpha^2_s}{9s^2}[- \frac{(t-m^2_i)(t-m^2_j) +st}
{(t- m^2_{\tilde{g}})^2} - \delta_{ij}\frac{(u-m^2_i)(u-m^2_j) +su}
{(u-m^2_{\tilde{g}})^2} +  \\ \nonumber
&&\frac{sm^2_{\tilde{g}}}{(t-m^2_{\tilde{g}})^2} +
\frac{sm^2_{\tilde{g}}}{(u-m^2_{\tilde{g}})^2} \delta_{ij}
- \frac{2sm^2_{\tilde{g}}}{3(t-m^2_{\tilde{g}})(u-m^2_{\tilde{g}})}
\delta_{ij}] \,,
\end{eqnarray}
where $m_i$ and $m_j$ are the masses of produced squarks and
$m_{\tilde{g}}$ is the gluino mass.

For the reaction $q_i\bar{q}_j \rightarrow \tilde{q}_i
\tilde{q}^{*}_j$ the differential cross section has the form
\begin{eqnarray}
&&\frac{d\sigma}{dt}(q_i\bar{q}_j \rightarrow \tilde{q}_i
\tilde{q}^{*}_j) = \\ \nonumber
&&\frac{4\pi\alpha^2_s}{9s^2}[[\frac{ut -m^2_im^2_j}
{s^2}][\delta_{ij}[2 -\frac{2}{3}\frac{s}{(t-m^2_{\tilde{g}})}]
+\frac{s^2}{(t-m^2_{\tilde{g}})^2}] +
\frac{sm^2_{\tilde{g}}}{(t-m^2_{\tilde{g}})^2}] \,.
\end{eqnarray}

For the reaction $gg \rightarrow \tilde{q}_i \tilde{q}_i^{*}$ the
differential cross section is
\begin{eqnarray}
&&\frac{d\sigma}{dt}(gg \rightarrow \tilde{q}_i \tilde{q}_i^{*})
= \\ \nonumber
&&\frac{\pi \alpha^2_s}{s^2}[\frac{7}{48} + \frac{3(u-t)^2}{16s^2}]
[1 + \frac{2m^2t}{(t-m^2)^2} + \frac{2m^2u}{(u-m^2)^2} +
\frac{4m^4}{(t-m^2)(u-m^2)}] \,.
\end{eqnarray}
Here m is the mass of the corresponding squark (we assume the left-
and right-handed squarks are degenerated in mass).

The differential cross section for the reaction
$gq_i \rightarrow gaugino + \tilde{q}_i$ has the form
\begin{eqnarray}
&&\frac{d\sigma}{dt}(gq_i \rightarrow gaugino  + \tilde{q}_i)
= \\ \nonumber
&&\frac{\pi}{s^2}[B_s \frac{(\mu^2 - t)}{s} +
B_t\frac{[(\mu^2 -t)s +2\mu^2(m^2_i-t)]}{(t -\mu^2)^2} + \\ \nonumber
&&B_u\frac{(u-\mu^2)(u+m^2_i)}{(u-m^2_i)^2} +
B_{st}\frac{[(s-m^2_i + \mu^2)(t-m^2_i) -\mu^2s]}{s(t-\mu^2)}
+ \\ \nonumber
&&B_{su}\frac{[s(u  + \mu^2) + 2(m^2_i -\mu^2)(\mu^2 -u)]}
{s(u-m^2_i)} +  \\ \nonumber
&&B_{tu}\frac{[(m^2_i -t)(t +2u +\mu^2) + (t - \mu^2)(s + 2t - 2m^2_i)
+(u - \mu^2)(t + \mu^2 + 2m^2_i)]}{2(t - \mu^2)(u - m^2_i)}] \,,
\end{eqnarray}
where $\mu$ is the mass of the gauge fermion and $m_i$ is the mass of
the scalar quark. The coefficients $B_x$ are contained in
ref.\cite{EICHT}. For instance, for the case when
$gaugino \equiv gluino$
the coefficients $B_x$ are: $B_s = \frac{4\alpha^2_s}{9}
\delta_{ij}$, $B_t = \frac{9}{4}B_s$, $B_u = B_s$, $B_{st} = -B_t$,
$B_{su} = \frac{1}{8} B_s$, $B_{tu} = \frac{9}{8}B_s$.

 Consider finally the production of sleptons. The differential cross
section for the production of charged slepton-sneutrino pairs is
\begin{equation}
\frac{d\sigma}{dt}(d\bar{u}  \rightarrow W^{*} \rightarrow \tilde{l}_L
\bar{\tilde{\nu}}_L) = \frac{g^4_2|D_W(s)|^2}{192\pi s^2}
(tu - m^2_{\tilde{l}_L}m^2_{\tilde{\nu}_L}) \,.
\end{equation}
For $\tilde{l}_L$ pair production the differential cross section is
\begin{eqnarray}
&&\frac{d\sigma}{dt}(q\bar{q} \rightarrow \gamma^{*}, Z^{*} \rightarrow
\tilde{l}_L \bar{\tilde{l}}_L) =
\frac{2\pi \alpha^2}{3s^2}[tu - m^4_{\tilde{l}_L}]
[\frac{q^2_l q^2_q}{s^2} + \\ \nonumber
&&(\alpha_l
- \beta_l)^2(\alpha^2_q + \beta^2_q)|D_Z(s)|^2 + \\ \nonumber
&&\frac{2q_lq_q\alpha_q(\alpha_l - \beta_l)(s  - M^2_Z)}{s}
|D_Z(s)|^2] \,,
\end{eqnarray}
where $D_V(s) = 1/(s - M^2_V + iM_V \Gamma_V)$, $q_l = -1$, $q_{\nu}
= 0$, $q_u = 2/3$, $q_d = -1/3$, $\alpha_l = \frac{1}{4}(3t - c)$,
$\alpha_{\nu} = \frac{1}{4}(c + t)$, $\alpha_u = -\frac{5}{12}
t + \frac{1}{4}c$, $\alpha_{d} = -\frac{1}{4}c +\frac{1}{12}t$,
$\beta_l =\frac{1}{4}(c + t)$, $\beta_{\nu} = -\frac{1}{4}(c + t)$,
$\beta_{u} = -\frac{1}{4}(c + t)$, $\beta_{d} = \frac{1}{4}(c + t)$,
$c = \cot{\theta_W}$, $t = \tan{\theta_W}$.
The differential cross section for sneutrino pair production can
be obtained by the replacement $\alpha_l$, $\beta_l$, $q_l$ and
$m_{\tilde{l}}$ by $\alpha_{\nu}$, $\beta_{\nu}$, $0$ and $m_{\tilde{\nu}}$
respectively, whereas for $\tilde{l}_R$ pair production one has
substitute $\alpha_l - \beta_l \rightarrow \alpha_l + \beta_l$ and
$m_{\tilde{l}_L} \rightarrow m_{\tilde{l}_R}$.

The QCD corrections to the squark and gluino tree level cross sections 
are essential \cite{SPIRA2}

%
%

\begin{figure}[hbt]

\vspace*{0.5cm}
\hspace*{0.0cm}
\epsfxsize=15cm \epsfbox{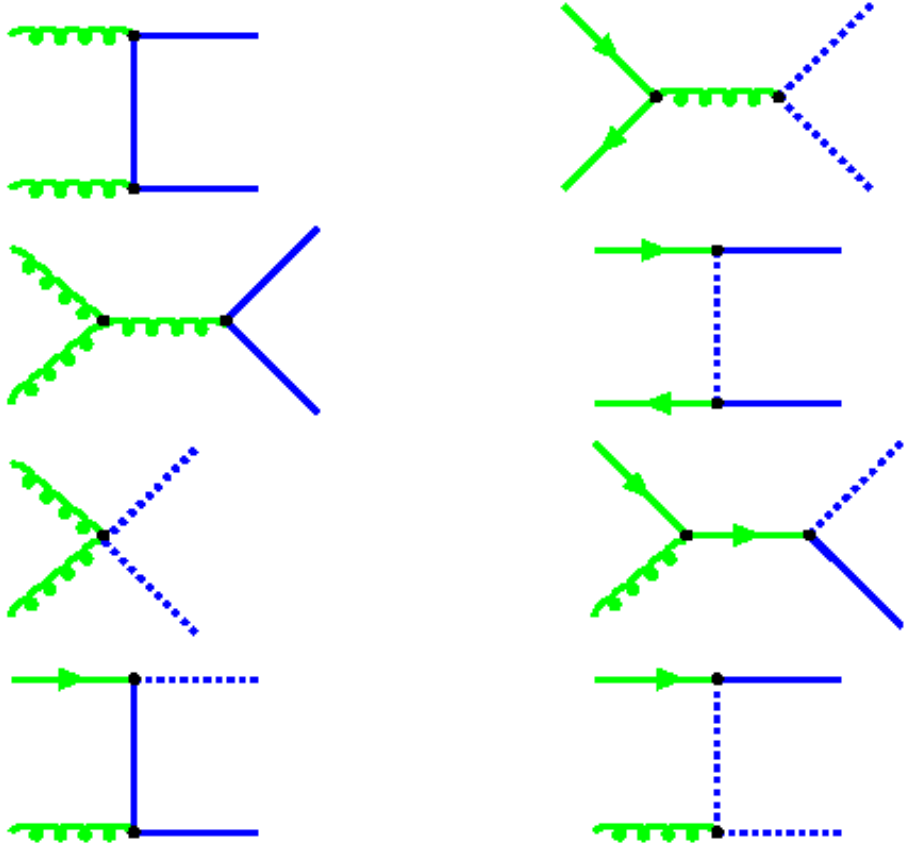}
\vspace*{0.0cm}
\caption[]{\label{fg:CMS} \it  Squark and gluino production diagrams} 

\end{figure}

\subsection{Superparticle decays.}

The decay widths of the superparticles depend rather
strongly on the relations between superparticle masses.
Here we outline the main decay channels only.
The formulae for the decay widths are contained in refs.\cite{SDECAY}.
Consider at first the decays of gluino and squarks. For
$m_{\tilde{g}} > m_{\tilde{q}}$ the main decays
are the following:
\begin{equation}
\tilde{g} \rightarrow \tilde{q}_i \bar{q}_i\\, \bar{\tilde{q}}_k q_k \,,
\end{equation}
\begin{equation}
\tilde{q}_k \rightarrow \tilde{\chi}^{0}_{i} q_k \,,
\end{equation}
\begin{equation}
\tilde{q}_k \rightarrow \tilde{\chi}^{+}_j q_m,\, \tilde{\chi}^{-}_j q_l \,,
\end{equation}
For $m_{\tilde{g}} < m_{\tilde{q}}$ the main decays are:
\begin{equation}
\tilde{q}_i \rightarrow \tilde{g} q_i \,,
\end{equation}
\begin{equation}
\tilde{g} \rightarrow q \bar{q}^{'} \tilde{\chi}^{+}_{k} \,,
\end{equation}
\begin{equation}
\tilde{g} \rightarrow q^{'} \bar{q} \tilde{\chi}^{-}_{k} \,,
\end{equation}
\begin{equation}
\tilde{g} \rightarrow q \bar{q} \tilde{\chi}^{0}_{k} \,.
\end{equation}

Charginos and neutralinos have a lot of decay modes. Especially interesting 
for the LHC SUSY discovery are their leptonic modes, for instance:
\begin{equation}
\tilde{\chi}^{\pm}_1 \rightarrow \tilde{\chi}^0_1 l^{\pm} \nu \,,
\end{equation}
\begin{equation}
\tilde{\chi}^{\pm}_1 \rightarrow (\tilde{l}^{\pm}_L  
\rightarrow \tilde{\chi}^0_1 l^{\pm})\nu \,,
\end{equation}
\begin{equation}
\tilde{\chi}^{\pm}_1 \rightarrow (\tilde{\nu}  
\rightarrow \tilde{\chi}^0_1 \nu) l^{\pm} \,,
\end{equation}
\begin{equation}
\tilde{\chi}^{\pm}_1 \rightarrow \tilde{\chi}^0_1(W^{\pm} 
\rightarrow l^{\pm} \nu)\,,
\end{equation}
\begin{equation}
\tilde{\chi}^0_2    \rightarrow   \tilde{\chi}^0_1   l^+l^- \,,
\end{equation}
\begin{equation}
\tilde{\chi}^0_2    \rightarrow (\tilde{\chi}^{\pm}_1 
\rightarrow \tilde{\chi}^0_1 l^{\pm}\nu)l^{\mp}\nu  \,,
\end{equation}
\begin{equation}
\tilde{\chi}^0_2    \rightarrow (\tilde{l}^{\pm}_{L,R} 
\rightarrow \tilde{\chi}^0_1 l^{\pm})l^{\mp} \,.
\end{equation}


Two-body decays of neutralinos and charginos
into Higgs bosons are:
\begin{equation}
\tilde{\chi}^0_i \rightarrow \tilde{\chi}^0_j +h(H) \,,
\end{equation}
\begin{equation}
\tilde{\chi}^0_i \rightarrow \tilde{\chi}^{\pm}_k + H^{\mp} \,,
\end{equation}
\begin{equation}
\tilde{\chi}^{\pm}_i \rightarrow   \tilde{\chi}^0_k + H^{\pm} \,,
\end{equation}
\begin{equation}
\tilde{\chi}^{\pm}_i \rightarrow \tilde{\chi}^{\pm}_j + h(H)   \,.
\end{equation}

The left sleptons dominantly decay via gauge interactions
into charginos or neutralinos via two body decays
\begin{equation}
\tilde{l}_L \rightarrow l + \tilde{\chi}_i^0 \,,
\end{equation}
\begin{equation}
\tilde{l}_L \rightarrow \nu_{L} +\tilde{\chi}^{-}_j \,,
\end{equation}
\begin{equation}
\tilde{\nu}_L \rightarrow \nu_{L} + \tilde{\chi}_{i}^0 \,,
\end{equation}
\begin{equation}
\tilde{\nu}_L \rightarrow l + \tilde{\chi}^{+}_j  \,.
\end{equation}
For relatively light sleptons only the decays into the LSP are
possible, so that light sneutrino decays are invisible. Heavier
sleptons can decay via the chargino or other (non LSP)
channels. These decays are important because they proceed
via the larger $SU(2)$ gauge coupling constant and can dominate
the direct decay to LSP. The $SU(2)$ singlet charged sleptons
$\tilde{l}_R$ only decay via their $U(1)$ gauge interactions
and in the limit of vanishing Yukawa coupling their decays to
charginos are forbidden. Therefore the main decay mode of
right-handed slepton is
\begin{equation}
\tilde{l}_R \rightarrow l + \tilde{\chi}^{0}_{i} \,.
\end{equation}
In many cases the mode into LSP dominates.

\subsection{Search for sparticles at LHC}

\paragraph{ Squarks and gluino.}

The gluino and squark production cross sections at  LHC are 
the biggest ones compared to slepton or gaugino cross sections.
Therefore  gluinos and squarks production at  LHC is the most interesting 
reaction from the SUSY discovery point of view with the cross sections around 
$1~pb$ for squark and gluino masses around $1~TeV$. 
The squark and gluino decays produce 
missing transverse energy from the LSP plus multiple jets and varying 
numbers of leptons from the intermediate gauginos \cite{BAER}. 

It is natural to divide the signatures used for the squark and 
gluino detections into the
following categories \cite{BAER}:

a. multi jets plus $E^{miss}_T$    events,

b. 1l  plus jets plus $E_T^{miss}$ events,

c. 2l plus jets plus $E_T^{miss}$ events,

d. 3l plus jets plus $E_T^{miss}$ events,

e. 4l plus jets plus $E_T^{miss}$ events,

f. $\geq 5 l$  plus jets plus $E_T^{miss}$ events.

Multileptons arise as a result of the cascade decays of neutralinos
and charginos into W- and Z-bosons with subsequent decays of
W- and Z-bosons into leptonic modes. For instance, the same sign
and opposite sign dilepton events  arise as a result of the cascade decay
\begin{equation}
\tilde{g} \rightarrow q^{'} \bar{q} \tilde{\chi}^{\pm}_{i}, \; 
\tilde{\chi}^{\pm}_{i}
\rightarrow W^{\pm}\tilde{\chi}^{0}_1  \rightarrow l^{\pm} 
\nu \tilde{\chi}^0_1 \,,
\end{equation}
where l stands for both $e$ and $\mu$.
Opposite sign dilepton events can arise also as a result of cascade
decay
\begin{equation}
\tilde{g} \rightarrow q \bar{q} \tilde{\chi}^{0}_{i} , \;
\tilde{\chi}^0_i \rightarrow Z \tilde{\chi}^0_1 \rightarrow
l^{+}l^{-} \chi^0_1 \,.
\end{equation}

The main conclusion \cite{ABDUL}, \cite{ATCOL} is 
that for the MSUGRA model
 LHC(CMS) will be able to discover SUSY 
with squark or gluino masses up to $(2 - 2.5)~TeV$ for 
$L_{tot} = 100~fb^{-1}$. 
The most powerful signature for squark and gluino detection in 
MSUGRA model is the signature with multijets and the $E_T^{miss}$
 (signature a).
It should be noted that for the case of 
the MSSM model with arbitrary squark 
and gaugino masses the LHC SUSY discovery potential 
depends rather strongly on the 
relation between  the LSP, squark and gluino masses and it decreases with 
the increase of the LSP mass \cite{BIT1}. 
For the LSP mass close to the squark or 
gluino masses it is possible to discover SUSY with the squark or gluino 
masses up to $(1.2 - 1.5)~TeV$ \cite{BIT1}.

Let us stress that multilepton supersymmetry
signatures $(b~ -~f)$ arise as a result of decays
of squarks or gluino into charginos or neutralinos different from
LSP with subsequent decays of charginos or neutralinos into W-,
Z-bosons plus LSP. Leptonic decays of W-, Z-bosons is the
origin of leptons.  However, for the case of nonuniversal gaugino
masses it is possible to realize the
situation  when all charginos and neutralinos except LSP are heavier
than gluino and squarks. Therefore, gluino and squarks will
decay mainly into quarks or gluons plus LSP, so cascade decays and as a
consequence multilepton events will be negligible.

\paragraph{ Neutralino and chargino search.}

Chargino and neutralino pairs, produced through the Drell-Yan mechanism
$pp \rightarrow \tilde{\chi}^{\pm}_1 \tilde{\chi}_2^0$ 
may be detected through their leptonic decays 
$ \tilde{\chi}^{\pm}_1 \tilde{\chi}_2^0 \rightarrow 
 lll + E^{miss}_T$. So, the signature is three isolated leptons with 
$E^{miss}_T$.
The three-lepton signal is produced through the decays (148-153)
and the undetected neutrino and $\chi^0_1$  in decays 
 (148-153) produce $E^{miss}_{T}$.
The main backgrounds to the three lepton  channel arise from $WZ/ZZ$,
$t\bar{t}$, $Zb\bar{b}$ and $b\bar{b}$  production.
There could be also SUSY background arising as a result of squark
and gluino cascade decays into multileptonic modes.

Typical cuts are the following \cite{ABDUL}:

i. Three isolated leptons with $p^l_t > 15$ GeV.

ii. Veto central jets with $E_t > 25$ GeV in $|\eta| < 3.5$.

iii. $m_{l\bar{l}} < 81$ GeV
 or $m_{l\bar{l}} \neq M_Z \pm
\delta M_Z$.

The main conclusion is that neutralino and chargino could be detected
provided their masses are lighter than 350 GeV \cite{ABDUL}. Moreover, it is
possible to determine the $M(\tilde{\chi}^0_2) - 
M(\tilde{\chi}^0_1)$ mass difference by
the measurement of the distribution on $l^+l^-$ invariant mass arising
as a result of the decay $\tilde{\chi}^0_2 \rightarrow 
\tilde{\chi}^0_1 + l^+l^-$
\cite{ABDUL}.

\paragraph{ Sleptons search.}

Slepton pairs, produced through the Drell-Yan mechanism 
$pp \rightarrow \gamma^{*}/Z^{*} \rightarrow \tilde{l}^{+} 
\tilde{l}^{-} $ can be
detected through their leptonic decays $\tilde{l} \rightarrow l +
\chi^0_1$. So the typical signature used for sleptons 
detection  is the  dilepton pair with
missing energy and no hadronic jets \cite{ABDUL}. 
For $L_{t} = 100~fb^{-1}$ LHC(CMS)  will be able to discover
sleptons with the masses up to 400 Gev \cite{ABDUL}, \cite{KRAS4}.

\paragraph{The search for flavour lepton number violation in
slepton decays.}

In supersymmetric models with explicit flavour lepton number violation
due to soft supersymmetry breaking terms there could be detectable
flavour lepton number violation in slepton decays \cite{KRAS2}.
For instance, for the case of nonzero mixing $\sin{\phi} \neq 0$ between
right-handed selectrons and smuons we have flavour lepton number
violation in slepton decays, namely \cite{KRAS2}:
\begin{equation}
\Gamma(\tilde{\mu}_R \rightarrow \mu + LSP) =
\Gamma \cos^2{\phi}\,,
\end{equation}
\begin{equation}
\Gamma(\tilde{\mu}_R \rightarrow e + LSP) = \Gamma \sin^2{\phi}\,,
\end{equation}
\begin{equation}
\Gamma(\tilde{e}_R \rightarrow e + LSP) = \Gamma \cos^2{\phi}\,,
\end{equation}
\begin{equation}
\Gamma(\tilde{e}_R \rightarrow \mu + LSP) = \Gamma \sin^2{\phi}\,,
\end{equation}
\begin{equation}
\Gamma = \frac{g^2_1}{8\pi}(1 - \frac{M^2_{LSP}}{M^2_{SL}})^2 \,.
\end{equation}

The typical consequence of  nonzero smuon-selectron mixing is
the existence of accoplanar $e^{\pm}\mu^{\mp}$ signal events with missing
energy arising as a result of the production of slepton pairs with their
subsequent decays with flavour lepton number violation. The possibility
to detect flavour lepton number violation in slepton decays at LHC
has been discussed in refs. \cite{KRAS4}. The main conclusion is that for
the most optimistic case of the maximal mixing $\sin{\phi} = \frac{1}
{\sqrt{2}}$ between right-handed sleptons $\tilde{e}_R$ 
and $\tilde{\mu}_R$ it would be possible to discover slepton mixing at LHC
for slepton masses up to $275~GeV$ \cite{KRAS4}. Other possibilities 
to detect the effects of flavour lepton number violation in slepton 
decays at LHC have been considered in refs. \cite{OTHER}. 

\paragraph{The measurement of sparticle masses.}

After the LHC SUSY discovery the main problem will be 
 to separate many different 
channels produced by the SUSY cascade decays and to extract the values 
of SUSY parameters (squark, gluino, neutralino, chargino and 
slepton masses).  In the MSSM model, the decay products of 
SUSY particles always contain an 
invisible LSP $\tilde{\chi}^0_1$, so SUSY particles can not be 
reconstructed directly. The most promising approach to determine  sparticle 
masses is to use kinematical endpoints 
 \cite{HINCH} in different distributions. 
For example, the $l^{+}l^{-}$ distribution from 
$\tilde{\chi}^0_2 \rightarrow \tilde{\chi}^0_1 l^{+}l^{-}$ decay has 
an endpoint that determines $M_{\tilde{\chi}^0_2} - M_{\tilde{\chi}^0_1}$.     
However, the distribution from the two-body decay 
$\tilde{\chi}^0_2 \rightarrow \tilde{l}^{\pm}l^{\mp} \rightarrow 
\tilde{\chi}^0_1 l^{+}l^{-}$ has a sharp edge at the endpoint
$\sqrt{\frac{(M^2_{\tilde{\chi}^0_2} -M^2_{\tilde{l}})(M^2_{\tilde{l}} -
M^2_{\tilde{\chi}^0_1} )}{M^2_{\tilde{l}}}}~  $.  
When a longer decay chain can be identified, more combinations of 
masses can be measured \cite{ATCOL}, \cite{ALESSIA}. Note also that  as 
proposed in ref.\cite{HINCH} the ``hardness'' of an event is 
characterised by the scalar sum of transverse energies of the four hardest 
jets and the missing transverse energy:
\begin{equation}
E^{sum}_T = E^1_T + E^2_T +E^3_T + E^4_T + E^{miss}_T
\end{equation}
The peak value of the $E^{sum}_T$  spectrum for the inclusive 
SUSY signal provides a good estimate of the SUSY signal 
in MSUGRA model with the peak value $M_{peak} \equiv M_{SUSY} 
\approx min(M_{\tilde{g}}, M_{\tilde{q}})$ \cite{HINCH}, \cite{TOVEY}. 
Here  $M_{\tilde{q}}$ is the average mass of squarks 
from the first two generations. By the measurement of $E^{sum}_T$ 
distribution it is possible to estimate the $M_{SUSY}$ scale 
with (10-20) percent accuracy.

\paragraph{Gauge mediated supersymmetry breaking.}

In GMSM (Gauge Mediated Supersymmetry Breaking) models \cite{GRDGOR} 
the gravitino 
$\tilde{G}$ is very light and the phenomenology depends on the type of 
the next lightest SUSY  particle (NLSP), either the $\tilde{\chi}^0_1$ or a 
slepton, and by its lifetime decay into $\tilde{G}$. If the NLSP is the 
$\tilde{\chi}^0_1$ and it decays mainly into $\tilde{\chi}^0_1 \rightarrow 
\tilde{G} \gamma$, then SUSY signature contains in addition two hard, 
isolated photons. If the NLSP is charged slepton and it is long-lived, 
then it penetrates the calorimeter like a muon but with $\beta < 1$. The 
slepton mass in this case can be measured directly using the muon chambers 
as  a time-of-flight system \cite{ATCOL}, \cite{WROCH}.

\subsection{  SUSY Higgs bosons search}

\paragraph{  }

The MSSM has three neutral and one charged Higgs bosons: $h$, $H$, $A$ and 
$H^{\pm}$ \footnote{LEP2 experiments give lower bounds $91.0~GeV$ and 
$91.9~GeV$ for light $h$ and pseudoscalar $A$-bosons. Besides, the 
excluded $\tan{\beta}$ regions are $0.5 \leq \tan{\beta} \leq 2.4$ 
for the maximal mixing scenario and  $0.7 \leq \tan{\beta} \leq 10.5$  
for the no mixing scenario \cite{LEP}.}. 
As it has been mentioned before at tree level
the lightest Higgs boson mass is predicted to be lighter than
$m_{Z}$. However an account of radiative corrections  \cite{MHIGGS}
can increase the Higgs boson mass up to 
$135~GeV$ for stop masses less or equal to $1~TeV$ 
\cite{DEGRAS}.
In  MSUGRA the Higgs sector 
is described mainly by two parameters: 
the mass of $A$ boson and $\tan(\beta)$ -  the ratio of the vacuum 
expectation values of the Higgs  fields that couple to 
up-type and down-type quarks.  
In the limit of large $A$ boson mass, the couplings of $h$ boson 
coincide with the corresponding couplings of the SM Higgs boson. 

At high $\tan \beta$ the $H$, $A$ decay mainly into $b\bar{b}$.
 However this mode is not very useful due to huge 
$b\bar{b}$ background.
The decays of $H$ and $A$ to 
$\tau^{+}\tau^{-}$ and $\mu^{+}\mu^{-}$  
are the most important 
for the $A$ and $H$ bosons detection \cite{ATCOL}, \cite{DENEG}. 
In the MSSM, 
the $H \rightarrow \tau^{+}\tau^{-}$ and 
$A \rightarrow \tau^{+}\tau^{-}$ rates are enhanced over the SM for large 
$\tan(\beta)$. The production of the heavy neutral MSSM Higgs 
bosons is mainly through $gg \rightarrow H_{SUSY}$ and 
$gg \rightarrow b\bar{b}H_{SUSY}$. The Higgs boson coupling to $b$-quarks is 
enhanced at high $\tan \beta$ and the associated    
$gg \rightarrow b\bar{b}H_{SUSY}$ 
production dominates ($\sim 90 \%$ of the total rate) for 
$\tan \beta  \geq 10$ and $M_{H} \geq 300~GeV$. 
The gluon fusion cross section is determined by quark loops and can 
be significantly reduced in the case of large stop mixing and 
light stop mass \cite{DJOUADI}. Due to the dominance of the  
$gg \rightarrow b\bar{b}H_{SUSY}$ production mechanism at high $\tan \beta$ 
production rates for the heavy Higgs bosons $H$ and $A$ are not sensitive 
to the loop effects.

\paragraph{Light Higgs boson.}

For  SUSY masses  bigger than $O(300)$ Gev the decay widths 
and the production rates for the 
lightest Higgs boson $h$ are approximately the same as 
for the SM Higgs boson (decoupling regime) and the most 
promising signature here  is   
$h \rightarrow \gamma\gamma$.
Also the signatures $pp \rightarrow t\bar{t}(h \rightarrow 
b\bar{b})$  
and $ pp \rightarrow qq^{~} (h \rightarrow WW^{*} 
\rightarrow l^{+}l^{'-} \nu\bar{\nu}^{~})$
are important. 
Note that in the case of large stop mixing and for 
light stop $m_{\tilde{t}_1} \leq 200~GeV$ the rate 
$gg \rightarrow h \rightarrow \gamma\gamma$ could be significantly reduced 
due to the stop and top loops destructive interferences in  
$gg \rightarrow h $ which could lead to no discovery for this 
signature. For the most difficult region $m_h \sim m_A 
\sim m_H \sim 100~GeV$ and high $\tan \beta$  the use 
of $gg \rightarrow b\bar{b}h \rightarrow b\bar{b} \mu^{+}\mu^{-}$ 
helps to detect the Higgs boson \cite{KIN}, \cite{DENEG}, 
\cite{BOOS}.

\paragraph{Heavy neutral bosons $A$ and $H$.}

The $\tau \bar{\tau}$ final states can be searched for in a $ 2~lepton$, 
$lepton + \tau~jet$, $2~\tau~jet$ final states \cite{KIN}, \cite{ATCOL}.
For the one lepton plus one hadron final states, intermediate
backgrounds are due to $Z, \gamma^{*} \rightarrow \tau \bar{\tau};
t\bar{t} \rightarrow \tau \bar{\tau} + X, \tau + X$ and
$b\bar{b} \rightarrow \tau \bar{\tau} + X, \tau X$. 
Efficient $\tau$-jet identification has been developed based on low energy 
multiplicity, narrowness and isolation of the $\tau$-jet in 
$H,A \rightarrow \tau\tau$. This identification has been shown to provide 
a rejection factor $\geq 1000 $ per QCD jet. The Higgs boson can be 
reconstructed in the $H \rightarrow \tau \tau$ channel from the 
visible $\tau$ momenta (leptons or $\tau$-jets) and $E^{miss}_t$ using the 
collinearity approximation for the neutrinos from $\tau$ decays. 
Precision of the Higgs boson mass measurement is estimated to 
be $\leq 10 \%$ for $A,H \rightarrow \tau\tau$ at high $\tan \beta$.
The  $A,H$ bosons can be discovered using
these $\tau\tau$ modes with the masses up to $(600- 800)~ GeV$ 
\cite{KIN}, \cite{DENEG}, \cite{ATCOL} 
(see Fig.23).

In the MSSM 
the branching ratio for $A, H \rightarrow \mu\mu$ is 
small, $\sim 3 \cdot 10^{-4}$, however the associated $gg \rightarrow 
b \bar{b} H(A)$ production is enhanced at large $\tan \beta$. 
The Drell-Yan production
$\gamma^{*},Z^{*} \rightarrow \mu^{+}\mu^{-}$
is the dominant background and it is suppressed  with 
$b$-tagging \cite{KIN}. 
Precision of the Higgs boson mass measurement is estimated to 
be $0.1 -1.5 \%$ for this mode.

So the heavy $H,A$ bosons are expected to be found at high $\tan \beta$ 
using $\tau\tau$ and $\mu\mu$ decay modes. LHC discovery range is 
$\tan \beta \geq 10$ for $m_{A}\leq 200~GeV$ \cite{KIN}. For $\tan 
\beta \leq 10$ the $H,A$ decays to sparticles may be used. 
The channel    
$A,H \rightarrow \tilde{\chi}^0_2 \tilde{\chi}^0_2 
\rightarrow 4l^{\pm} +X$ has been found \cite{PHIL} to be the most promising 
one for heavy neutral Higgs bosons discovery, 
provided  neutralinos and sleptons are 
light enough so that the  $\tilde{\chi}^0_2 \rightarrow \tilde{l}l 
\rightarrow \tilde{\chi}^0_1 l^{+}l^{-}$ branching ratio is significant. 
Using this channel it is possible to discover $H, A$ bosons with the 
masses $(200 - 400)~GeV$ \cite{DENEG}, \cite{ATCOL}.

\paragraph{Charged Higgs boson.}

Search for the charged Higgs boson at LHC is important for the 
understanding of the nature of the Higgs sector. Really, discovery 
of a charged Higgs boson at LHC would be a clear proof for 
physics beyond SM. For $m_{H^{\pm}} < m_{top}$ $H^{\pm}$  decays mainly 
into $\tau \nu$. For $m_{H^{\pm}} > 200~GeV$ the $H^{\pm} \rightarrow 
tb$ decay dominates but the $BR(H^{\pm} \rightarrow \nu \tau)$ 
approaches $0.1$ for $m_{H^{\pm}} \geq 400~GeV$. For light charged Higgs 
boson, $m_{H^{\pm}} < m_{top}$, the main $H^{\pm}$ production 
mechanism is through the $t\bar{t}$ events followed by $t \rightarrow 
H^{\pm}b$. The use of the $H^{\pm} \rightarrow \tau \nu$ decay mode 
allows to discover $H^{\pm}$ almost independently of $\tan \beta$ 
for light charged Higgs boson \cite{DENEG} \cite{ATCOL}.  
The heavy charged Higgs boson, $m_{H^{\pm}} > m_{top}$, is mainly 
produced in association with the top quark through the processes 
$gb \rightarrow t H^{\pm}$ and $gg \rightarrow tbH^{\pm}$. Again in this 
case the decay mode $H^{\pm} \rightarrow \nu \tau$ is the most perspective 
for $H^{\pm}$ detection. The use of the $\tau$ polarisation from 
 $H^{\pm} \rightarrow \nu \tau$ decay \cite{ROY} allows to suppress 
the backgrounds from $t\bar{t}, Wtb, W \rightarrow \tau \nu$ since due to 
spin correlations, the single pion from a one-prong 
$\tau$ decay is harder  
when it originates from an $H^{\pm}$ than from a $W$. In purely hadronic 
final states in $gb \rightarrow t(H^{\pm} \rightarrow \nu \tau)$ 
with hadronic top quark decay the transverse mass reconstructed from 
the $\tau$ jet and the $E^{miss}_T$ vector has a Jacobian peak structure 
with an endpoints at $m_W$ for the backgrounds.. 
This allows to extract the $H^{\pm}$ mass with the accuracy better than 
$10 \%$. The discovery reach for this signature is shown in Fig.23.

The $H^{\pm} \rightarrow tb$ decay for the $gb \rightarrow tH^{\pm}$ 
production reaction has been studied requiring one isolated lepton from 
the decay of one of the top quarks \cite{DENEG}, \cite{ATCOL}. 
To extract the Higgs signature 
in these multijet events requires tagging of three $b$-jets, 
reconstruction of the leptonic and hadronic top quark decays and the 
Higgs boson mass reconstruction.        
The discovery reach for this signature is shown in Fig.23.
Note also that the s-channel production of $H^{\pm}$ in 
$q\bar{q}^{'} \rightarrow H^{\pm} \rightarrow 
\tau \nu$ can be used for the $H^{\pm}$ 
detection \cite{SLAB} but the reduction of huge $q\bar{q}^{`} 
\rightarrow W \rightarrow \tau\nu$ background is rather 
difficult. 
A precision of $\sim (1 - 2) \%$ is expected for the charge Higgs 
boson mass measurement \cite{KIN}. Moreover due to the $\sigma \sim 
\tan^2 \beta$ behaviour of the cross section it is possible to 
determine $\tan \beta$ with a precision better than $7 \%$ for 
$\tan \beta > 20$ and $m_{H^{\pm}} = 250~GeV$ \cite{KIN}.

The main conclusion \cite{KIN}, \cite{DENEG}, \cite{ATCOL} concerning 
the situation with the search for
MSSM Higgs bosons at LHC for different $(m_{A},\tan(\beta))$ values is that
almost the full  $(m_{A},\tan(\beta))$ -values can be explored with 
the $h \rightarrow \gamma \gamma $ $h \rightarrow b \bar{b}$ 
decay modes for total luminosity $L_t = 30~fb^{-1}$. The heavy 
$H,A$ bosons will be discovered for $\tan \beta  \geq 10$ 
using the $H,A \rightarrow \tau\tau \,, \mu\mu$ decay modes with 
the $A,H$ boson masses up to $800~GeV$. 
For the search of the charged Higgs boson $H^{\pm}$ , 
the $gb \rightarrow tH^{\pm}, H^{\pm} \rightarrow \tau\nu$ channel is the most 
important one with a discovery reach for $\tan \beta \geq 20$  
up to $m_{H^{\pm}} \approx 400~GeV$.  
The most  difficult region with $110~GeV~ \le m_{A} \le 200~GeV$,  
$3 \le \tan(\beta) \le 10$ could be explored with the SUSY 
particle decay modes provided the neutralinos and sleptons are light enough.

\begin{figure}[hbt]

\vspace*{0.5cm}
\hspace*{0.0cm}
\epsfxsize=15cm \epsfbox{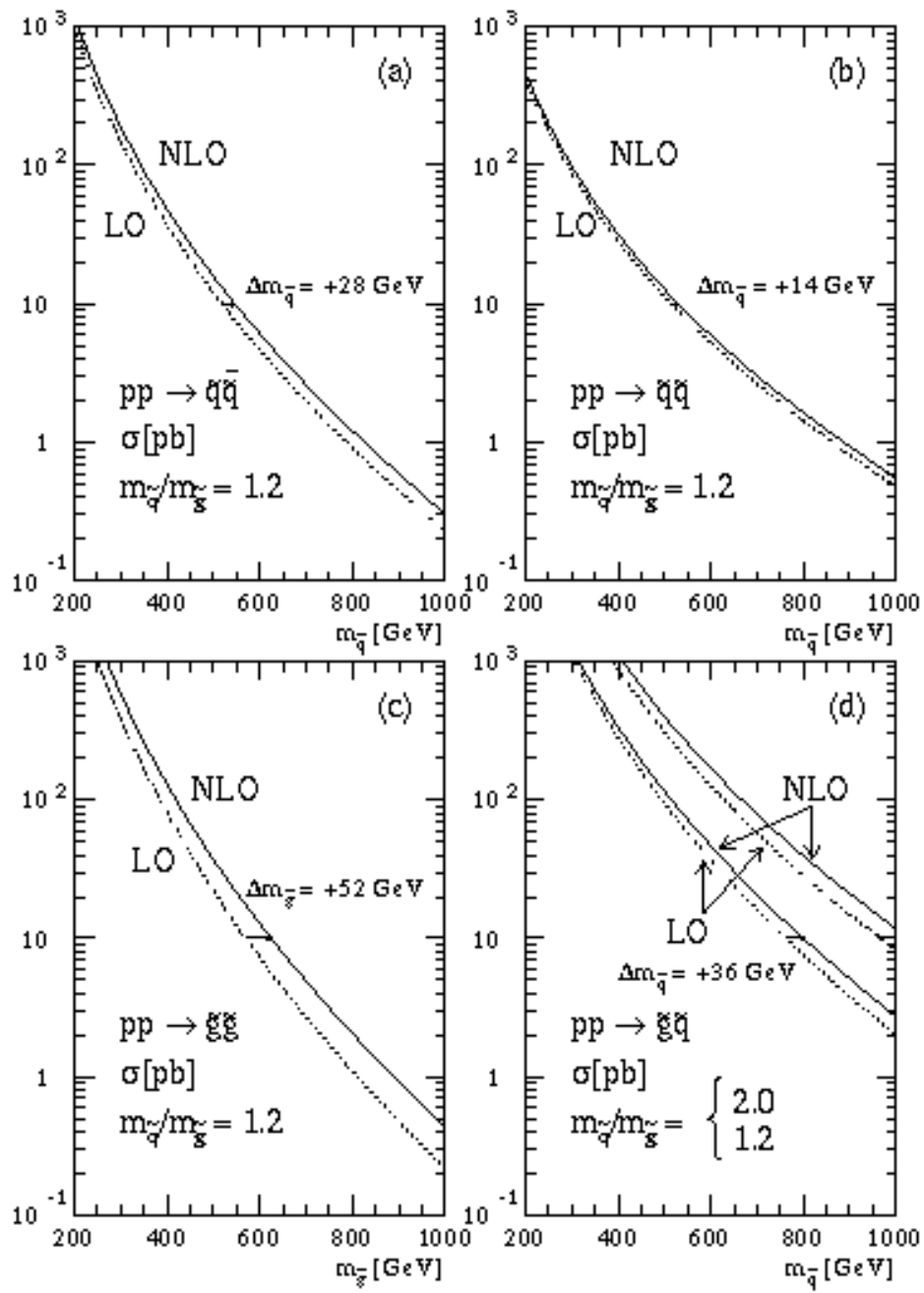}
\vspace*{0.0cm}
\caption[]{\label{fg:CMS} \it The total SUSY cross sections for the 
LHC }

\end{figure}

\begin{figure}[hbt]

\vspace*{0.5cm}
\hspace*{0.0cm}
\epsfxsize=15cm \epsfbox{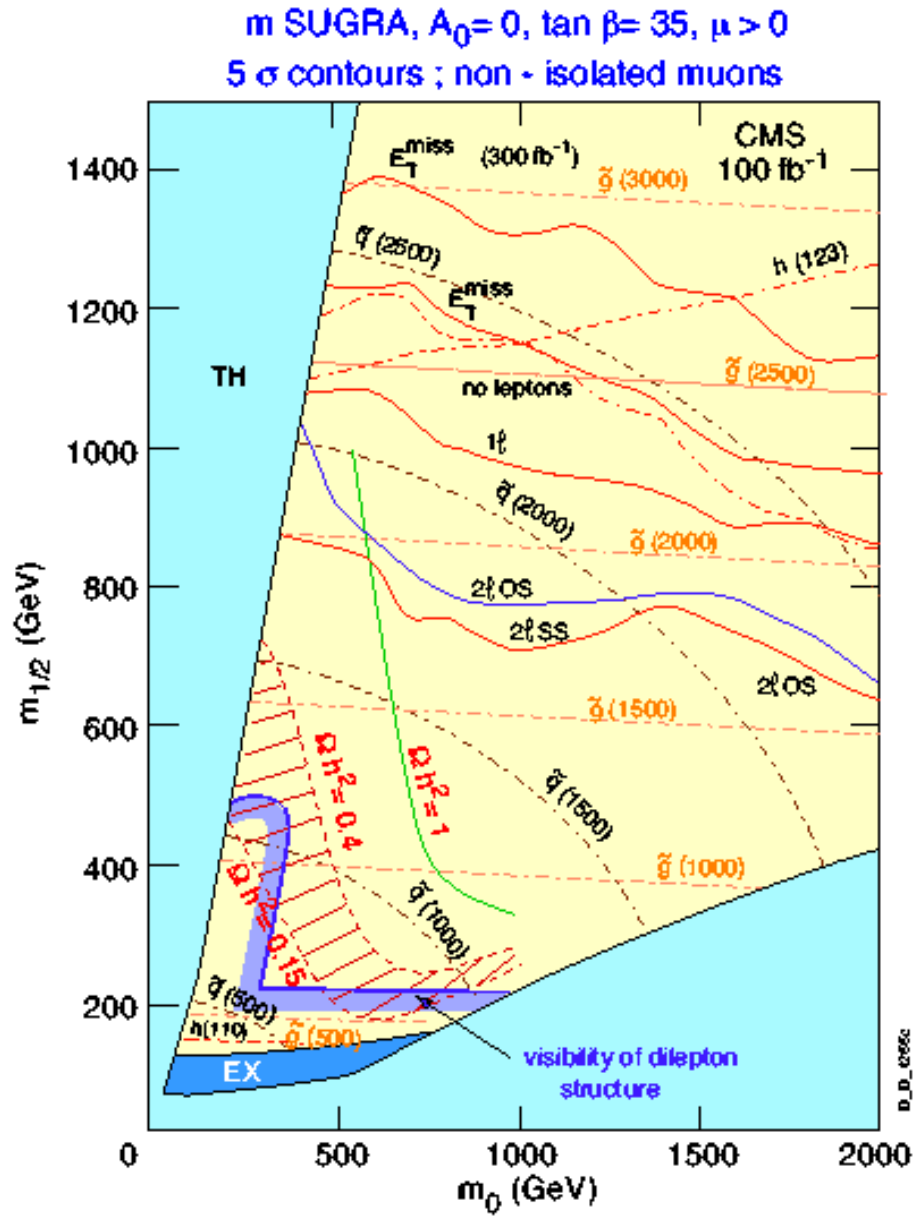}
\vspace*{0.0cm}
\caption[]{\label{fg:CMS} \it Discovery potential in MSUGRA model 
for $\tan{\beta} = 35$ and $\mu = +$ }

\end{figure}

\begin{figure}[hbt]

\vspace*{0.5cm}
\hspace*{0.0cm}
\epsfxsize=15cm \epsfbox{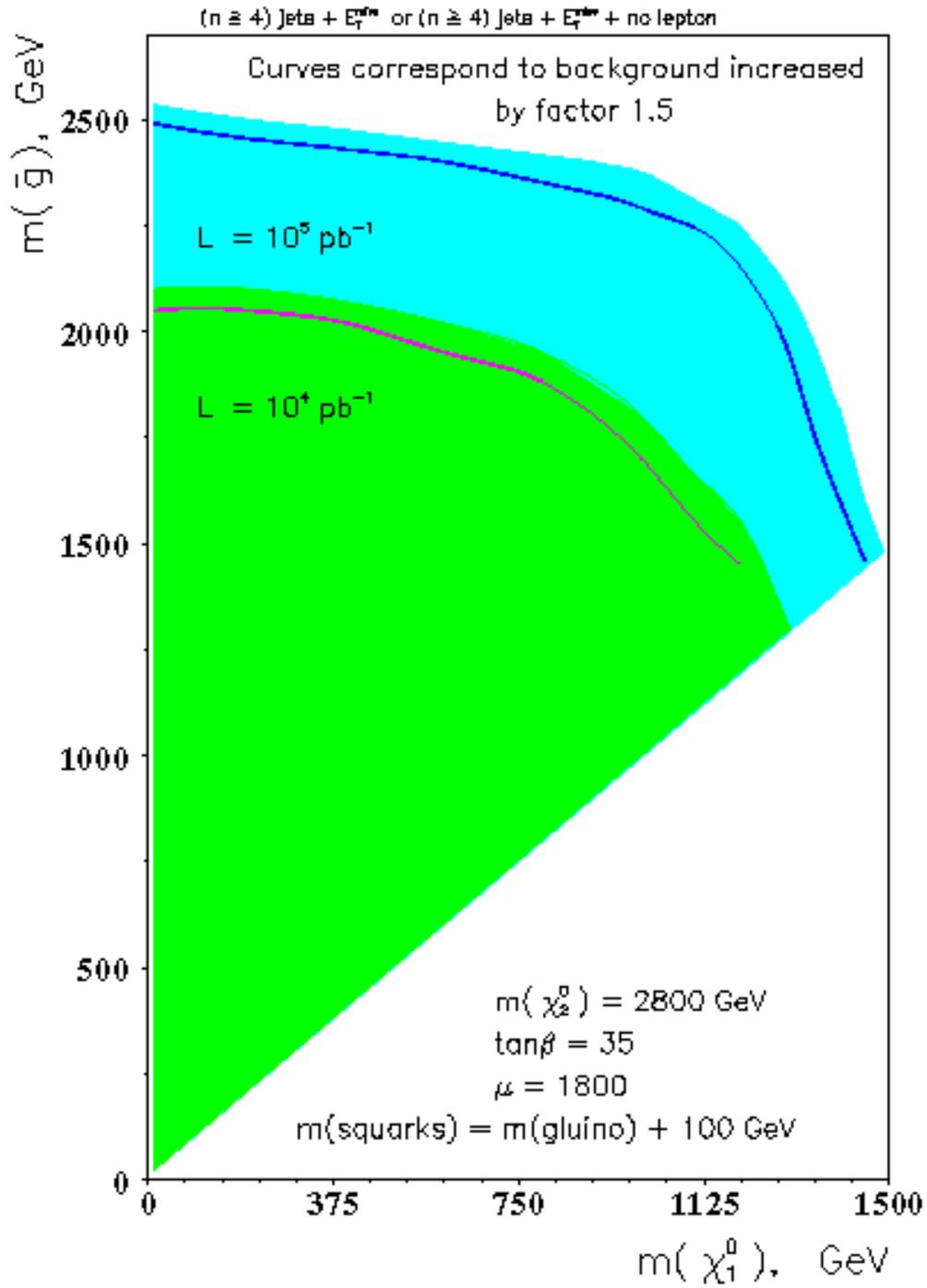}
\vspace*{0.0cm}
\caption[]{\label{fg:CMS} \it The CMS discovery potential at $100^fb^{-1}$ 
for different values of $m_{\tilde{\chi}^0_1}$ and $m_{\tilde{g}}$ 
in the case $m_{\tilde{q}} = m_{\tilde{g}} + 100~GeV$ }

\end{figure}

\begin{figure}[hbt]

\vspace*{0.5cm}
\hspace*{0.0cm}
\epsfxsize=15cm \epsfbox{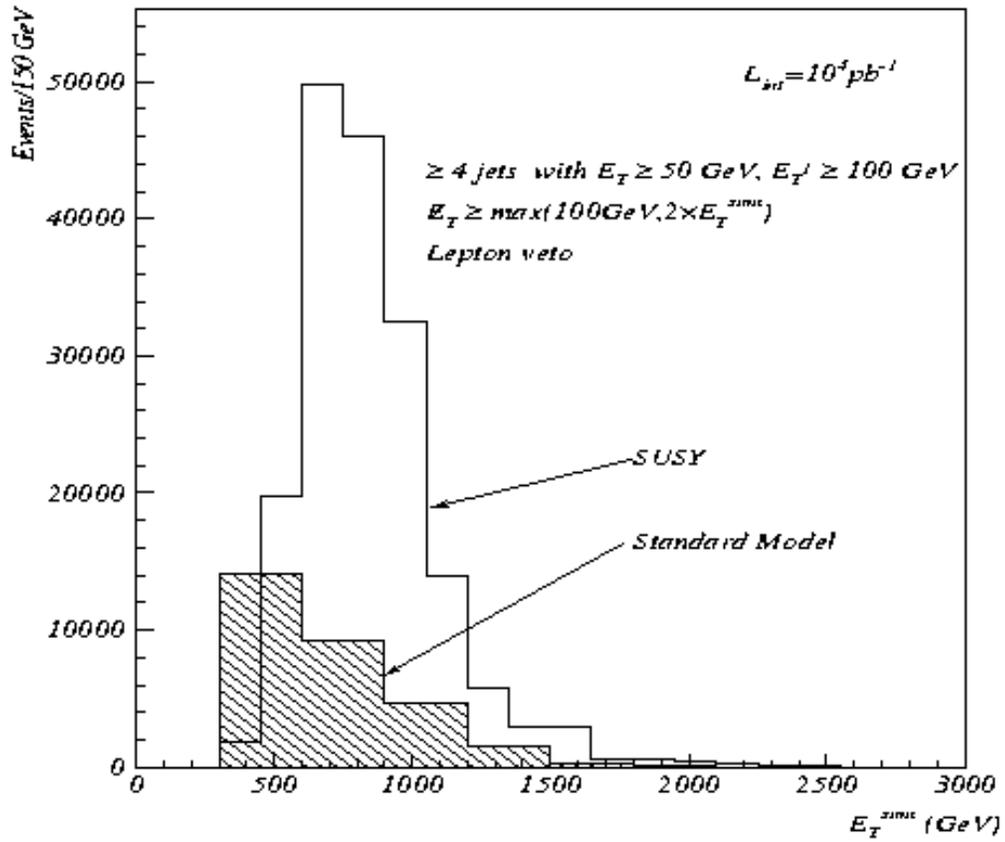}
\vspace*{0.0cm}
\caption[]{\label{fg:CMS} \it $E^{sum}_T$ distribution for both the 
inclusive SUSY and the SM backgrounds after event selection cuts 
have been applied  }

\end{figure}

\begin{figure}[hbt]

\vspace*{0.5cm}
\hspace*{0.0cm}
\epsfxsize=15cm \epsfbox{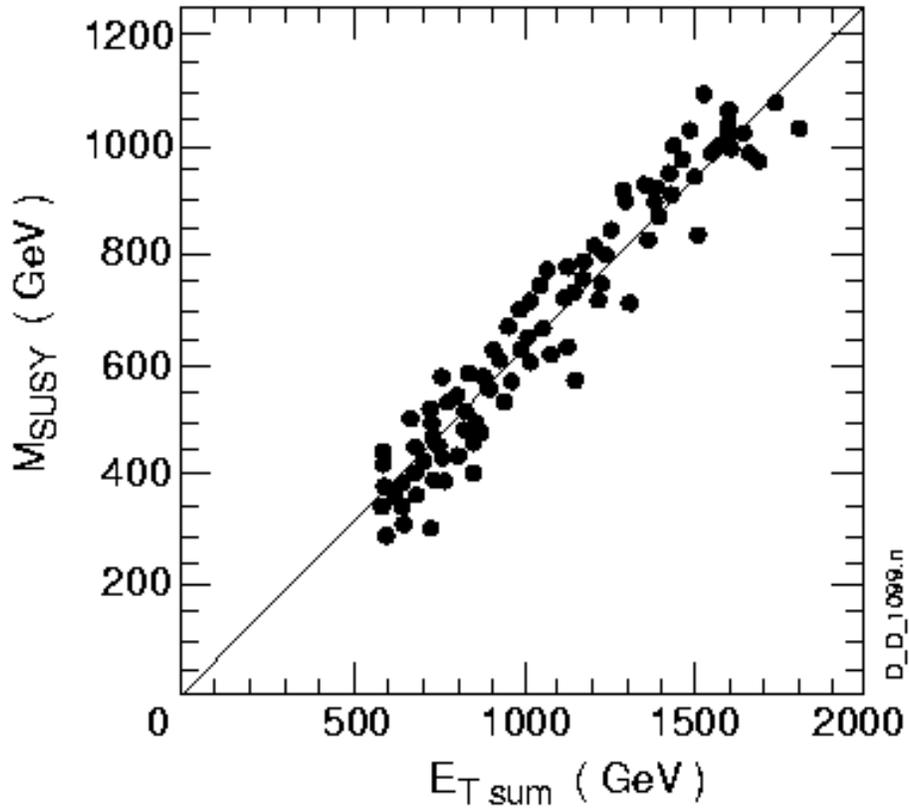}
\vspace*{0.0cm}
\caption[]{\label{fg:CMS} \it The relation between the peak value 
of the $E^{miss}_T$ distribution and the $M_{SUSY}$ value }

\end{figure}

\begin{figure}[hbt]

\vspace*{0.5cm}
\hspace*{0.0cm}
\epsfxsize=15cm \epsfbox{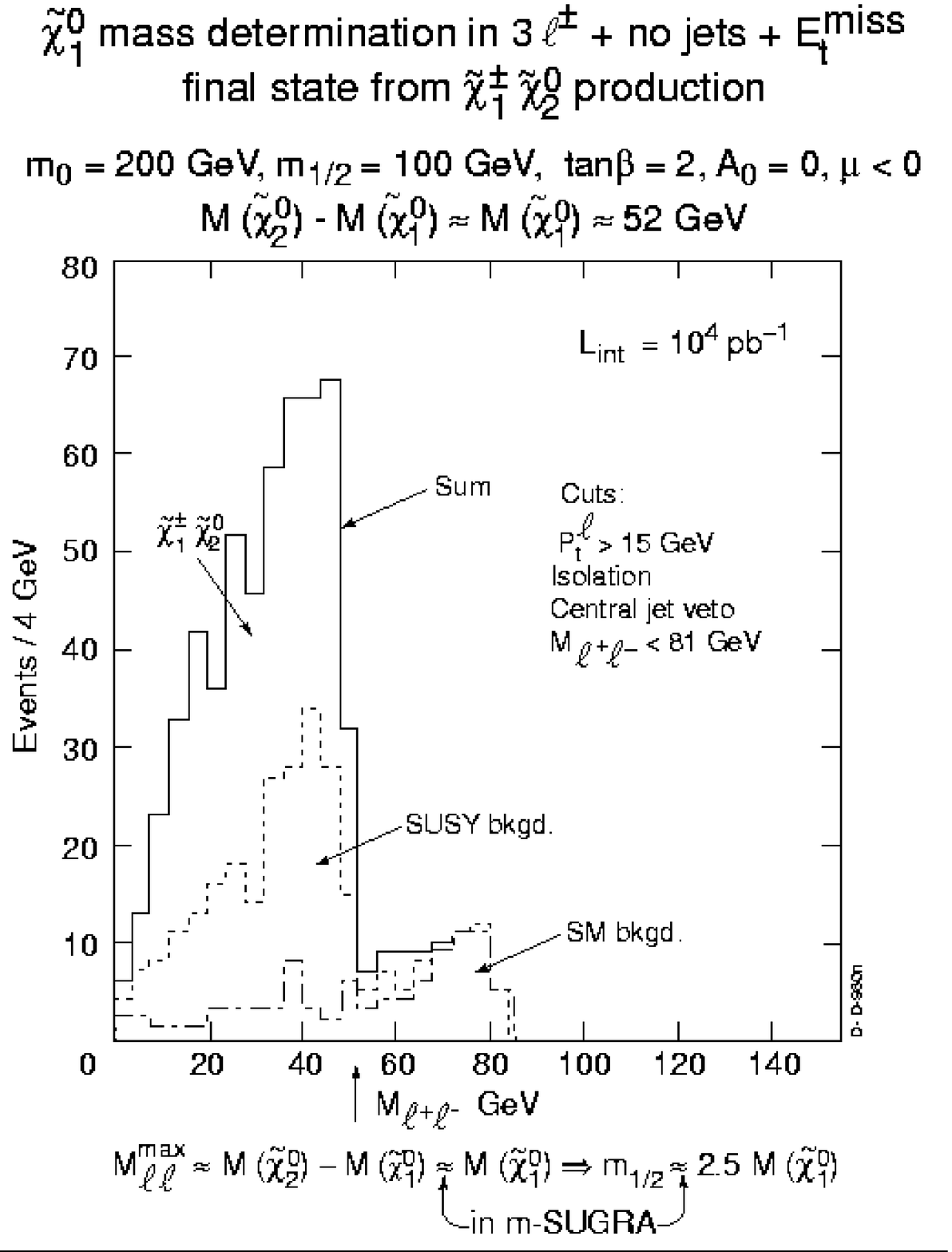}
\vspace*{0.0cm}
\caption[]{\label{fg:CMS} \it Dilepton invariant masses distribution for 
MSUGRA point ($m_0 =200~GeV$, $m_{1/2} = 100~GeV$ in the $3l + no jets + 
E^{miss}_T$ events. Contribution from SM and SUSY backgrounds  are 
also shown }

\end{figure}

\begin{figure}[hbt]

\vspace*{0.5cm}
\hspace*{0.0cm}
\epsfxsize=15cm \epsfbox{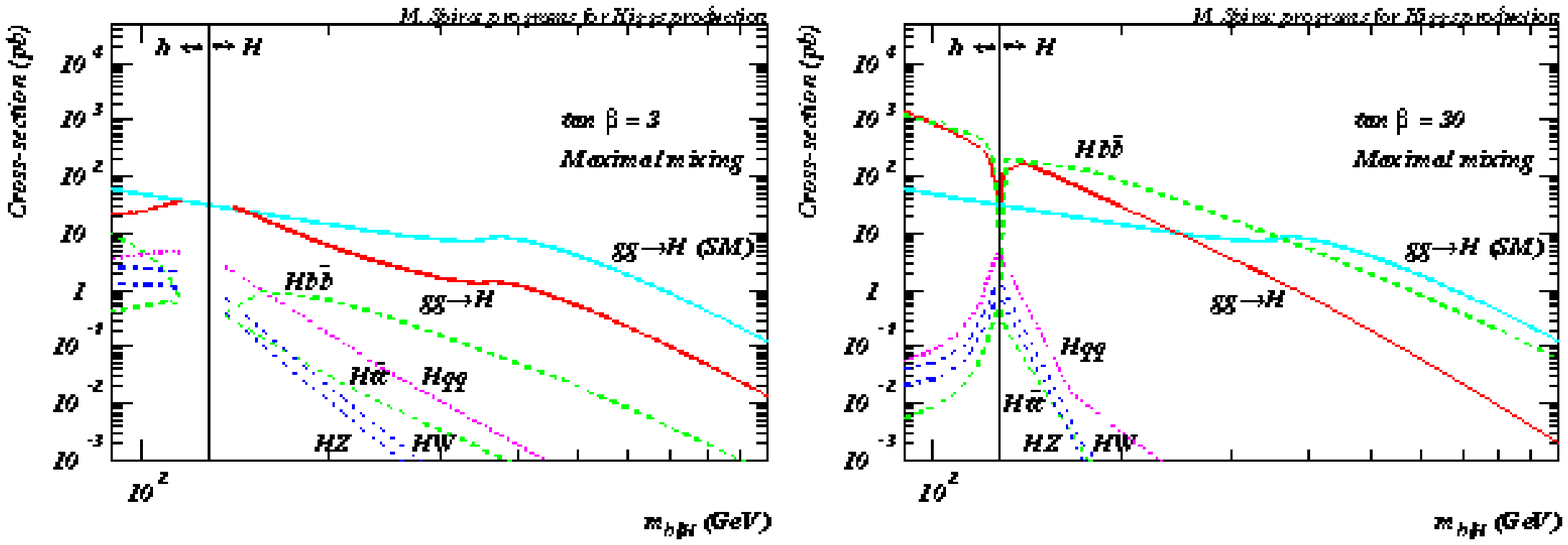}
\vspace*{0.0cm}
\caption[]{\label{fg:CMS} \it The branching ratios of light $h$ and 
heavy $H$ as a function of their masses }

\end{figure}

\begin{figure}[hbt]

\vspace*{0.5cm}
\hspace*{0.0cm}
\epsfxsize=15cm \epsfbox{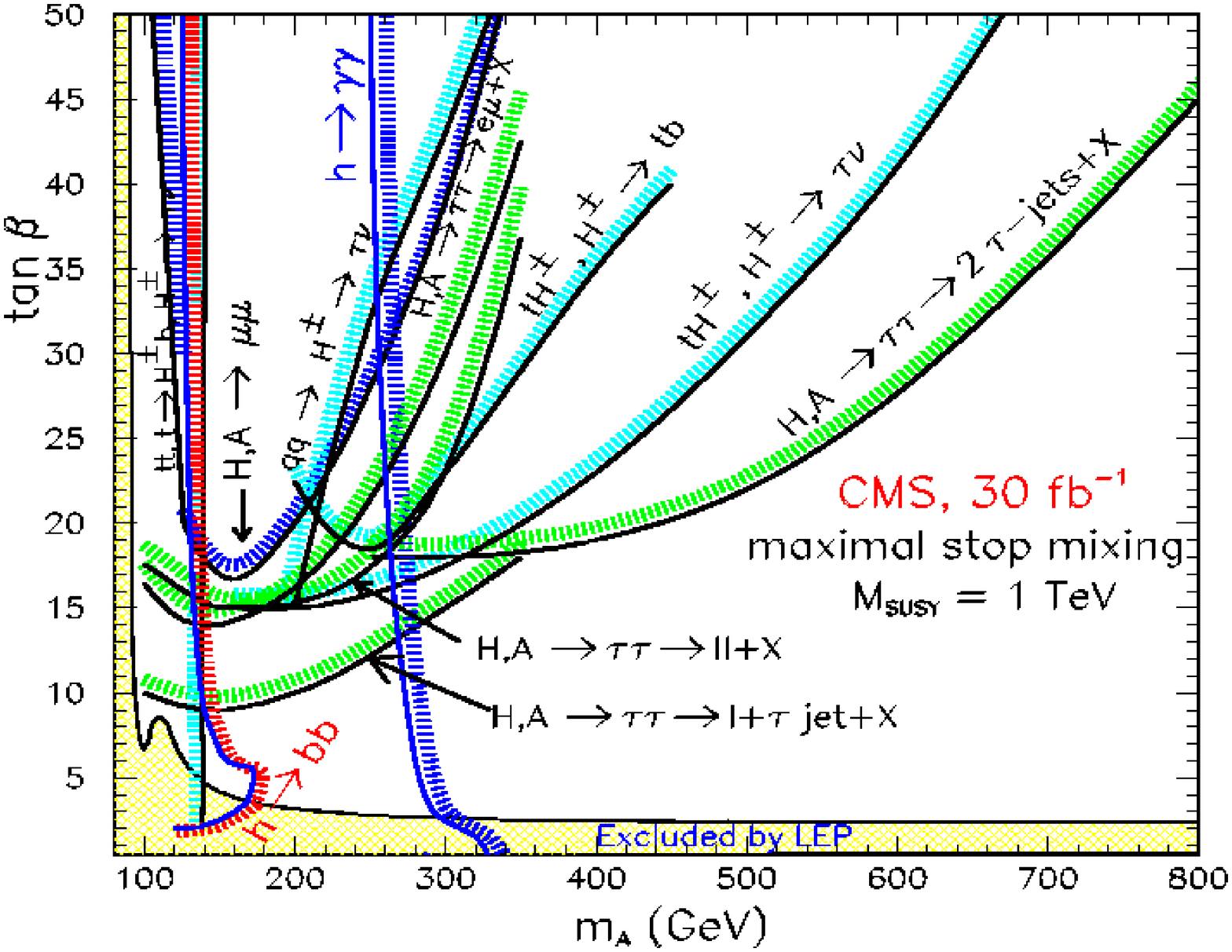}
\vspace*{0.0cm}
\caption[]{\label{fg:CMS} \it Expected $5\sigma$ discovery range for the MSSM 
Higgs bosons with maximal stop mixing in the CMS detector for $30_fb^{-1}$ }

\end{figure}

\section{Search for new physics beyond the SM and the MSSM}

There are a lot of models different from the SM and  the MSSM. Here we 
briefly describe some of them.

\subsection{Extra dimensions}

There is much theoretical interest in models  that have extra space 
dimensions \cite{ADD}, \cite{RS}, \cite{ANT}, \cite{KR1},
\cite{GUID}, \cite{RUBAK}, \cite{KIS}. The main 
motivation is that  models with big
$R_c \geq O(1)~TeV^{-1}$ extra space dimensions can explain the 
hierarchy between the electroweak and Planck scale. 
 In such models new physics can appear at a 1 TeV 
scale and therefore can be  tested at the LHC. 

In the ADD model \cite{ADD}  the metric  looks like
\begin{equation}
ds^2 = g_{\mu\nu}(x)dx^{\mu}dx^{\nu} + \eta_{ab}(x,y)dy^ady^b,
\end{equation}
where $\nu, \nu = 0,1,2,3$ and $a,b = 1,...d$.
All  $d$ additional dimensions 
are compactified with a characteristic size $R_c$. 
The relation between a fundamental mass scale in $D = 4+d$ dimensions, $M_D$, 
and 4-dimensional Planck scale $M_{PL}$ has the form
\begin{equation}
M^2_{PL} = V_dM^{2+d}_D,
\end{equation}
where $V_d$ is a volume of the compactified dimensions ($V_d = (2\pi R_c)^d$ 
for toroidal form of extra dimensions). 
In the ADD model there are 2 free 
parameters, the number $d$ of additional dimensions and the fundamental 
scale $M_D$.  
From the requirement 
that $M_D \sim 1~TeV$ one can find that the compactification 
radius $R^{-1}_c$ ranges from $10^{-3}~eV$ to $10~MeV$  if $d$ 
runs from 2 to 6. 
In the ADD model all SM gauge and matter fields are to 
be confined to a 3-dimensional brane embedded into a $(3+d)$-dimensional 
space and  only  gravity lives in the bulk. It means that the 
energy-momentum tensor of matter has the form 
\begin{equation}
T_{AB}(x,y) = \eta^{\mu}_A\eta^{\nu}_B T_{\nu\nu}(x)\delta(y),
\end{equation}
with $A,B = 0,1...3+d$. The graviton interaction Lagrangian is 
\begin{equation}
L_g = - \frac{1}{\bar{M}_{PL}}G^{(n)}_{\mu\nu}T_{\mu\nu},
\end{equation}
where $n$ labels the KK (Kaluza - Klein) excitation level and 
$\bar{M}_{PL} = M_{PL}/\sqrt{8\pi} = 2.4\cdot10^{18}~GeV$. From the 
Lagrangian (174) we see that that the couplings of graviton 
excitations are universal and very small. The masses of the KK 
graviton excitations are 
\begin{equation}
m_n = \frac{\sqrt{(n_an^a)}}{R_c},
\end{equation}
where $n_a = (n_1,n_2...n_d)$. A mass splitting  $\Delta m \sim R^{-1}_c $ 
is extremely small and we have an almost continuous spectrum of the gravitons.
The production cross section of the KK gravitons with masses $m_n \leq E$ is 
\begin{equation}
\sigma_{KK} \sim \frac{E^d}{M^{d+2}_D} \,.
\end{equation}
The lifetime of the massive graviton is \cite{ADD}
\begin{equation}
\tau_n \sim \frac{1}{M_{PL}}(\frac{M_{PL}}{m_n})^3 \,.
\end{equation}
 
Thus the KK gravitons behave like massive, almost stable non-interacting 
spin-2 particles. Their collider signature is an imbalance in missing mass 
of final states with a continuous mass distribution.
The most promising signature of the graviton production at the LHC 
originates from the reaction $pp \rightarrow jet + E^{miss}_T$. Note 
that at parton level the subprocess $gq \rightarrow qG^{(n)}$ gives 
the largest contribution.
The main background arises from the $Z + jet$, $Z \rightarrow \nu\bar{\nu} $ 
production.  
The use of this reaction allows to  discover extra space dimensions at the 
LHC(ATLAS) with the inverse radius less than $9~TeV$ \cite{ATCOL}.
Very  interesting signature for  the direct production of the massive 
gravitons is the process $pp \rightarrow \gamma + E^{miss}_T$  
which can be used as an independent test, although it has the much 
lower rate.
The SM background comes mainly from $pp \rightarrow \gamma 
(Z \rightarrow \nu\bar{\nu}$).  
Another prediction of the ADD model is that the contribution of virtual 
massive graviton resonances to the matrix elements modifies the SM 
cross sections at large momentum transfer (for instance the Drell-Yan 
cross section). At tree level the contribution of virtual massive 
gravitons to a matrix element is proportional to 
\begin{equation}
M \sim \frac{1}{\bar{M}_{PL}^2}\sum_{n} \frac{1}{s - m^2_n}
\end{equation}
The sum in (178) diverges for $d \geq 2$, the cutoff $M_c$ is to be calculated 
in full theory. The following very rough substitution 
$M \sim \frac{1}{M^4_c}$ is usually made to estimate lower bound on 
$M_c$ which will be extracted from the  LHC data. 
It appears that the diphoton and Drell-Yan 
productions at the LHC lead to sensitivity of $M_c$ up to $7.4~TeV$.

\begin{figure}[hbt]

\vspace*{0.5cm}
\hspace*{0.0cm}
\epsfxsize=15cm \epsfbox{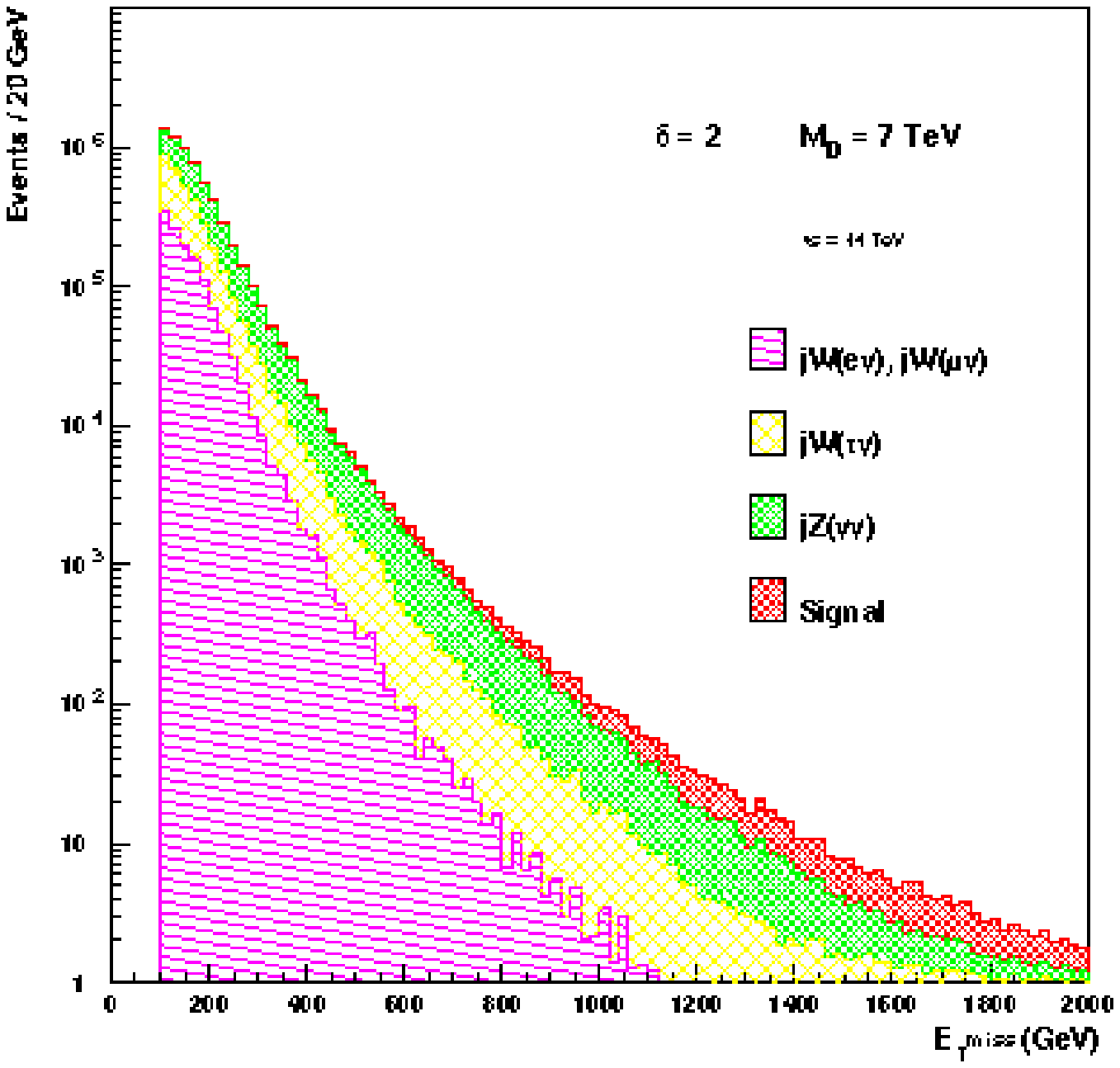}
\vspace*{0.0cm}

\caption[] {\label{fg:kun2} \it Distributions of the missing transverse 
energy in extra dimensions signal and background events after 
the selection and for $100~fb^{-1}$. $d = 2$, $M_d = 7~TeV$ is shown for 
signal  }

\end{figure}


In RS (Randall-Sundrum) model \cite{RS} gravity lives in a 5-dimensional 
Anti-de Sitter space with a single extra dimension compactified to the 
orbifold $S^1/Z_2$. 
The metric has the form 
\begin{equation}
ds^2 = e^{-2k|y|}\eta_{\mu\nu}dx^{\mu}dx^{\nu} +dy^2\\,
\end{equation}
where $y = r_c\theta (0 \leq \theta \leq \pi)$, $r_c$ being a "radius" of the 
extra dimension. The parameter $k$ defines the scalar curvature of 
the space. From the 5-dimensional action one can derive the relation 
\begin{equation}
\bar{M}^2_{PL} = \frac{M^3_5}{k}(1 - e ^{-2kr_c \pi}),
\end{equation}
which means that $k \sim \bar{M}_5 \sim \bar{M}_{PL}$.
There are two 3-dimensional branes in the model 
with equal and opposite tension 
localised at the point $y = \pi r_c$ (so called the TeV brane) 
and at $y =0$ (referred to as the Planck brane). All the SM fields are 
constrained to the TeV brane, while gravity propagates in the additional 
dimension.
Using a linear expansion of the metric
\begin{equation}
g_{\mu\nu} = e^{-2ky}(\eta_{\mu\nu} + \frac{2}{M^{3/2}_5} h_{\mu\nu})
\end{equation}
one can obtain the interaction of the gravitons with the SM fields
\begin{equation}
L = - \frac{1}{\bar{M}_{PL}}T^{\mu\nu}h^{(0)}_{\mu\nu}(x) - 
\frac{1}{\Lambda_{\pi}}\sum_n T^{\mu\nu}h^{(n)}_{\mu\nu}(x),
\end{equation}
where $\Lambda_{\pi} \sim \bar{M}_{PL}e ^{-kr_c\pi}$.
  The couplings of massive 
states are suppressed  by $\Lambda^{-1}_{\pi}$, while the zero mode 
couples with usual strength, $\bar{M}^{-1}_{PL}$. The physical scale on the 
$TeV$ brane, $\Lambda_{\pi}$, is of the order of $1~TeV$ for $kr_c \sim 12$.
 The masses of the graviton KK excitations are given by 
\begin{equation}
m_n = kx_ne^{-kr_c\pi},
\end{equation}
where $x_n$  are the roots of the Bessel function $J_1(x)$.
In RS model  \cite{RS} the first graviton excitation 
has a mass $O(1)~TeV $ and it decays into jets, leptons or photons. 
The most promising  mode for the graviton 
resonance detection at the LHC  is the use of the 
lepton decay modes. The  signature  
$q\bar{q},gg \rightarrow G_{res1} \rightarrow l^{+}l^{-}$ 
has been studied in refs.\cite{GRAVITON}. The signal is 
visible for $M_{G,res1} \leq 2~TeV $. Moreover, for 
$M_{G,res1} \leq 1.5~TeV $ from the 
measurement of lepton angular distribution 
it is possible to confirm that the resonance is spin-2 particle. 

Other nontrivial prediction of the RS model is the existence of 
relatively light scalar particle    called radion and denoted as 
$\Phi$. Radion is characterised by its mass $m_{\Phi}$,  some scale 
$\Lambda_{{\Phi}}$ and mixing parameter $\zeta$ with Higgs boson.
The interactions of the radion with the SM particles are given by 
\begin{equation}
L_{int} = \frac{\Phi}{\Lambda_{\Phi}}T^{\mu}_{\mu}(SM),
\end{equation}
where $\Lambda_{\Phi} = <\Phi> \sim O(1)~TeV$ and
\begin{equation}
T^{\mu}_{\mu}(SM) = \sum_{f} m_f\bar{f} f  - 2m^2_W W^+_{\mu}W^{\mu} - m^2_Z
Z_{\mu}Z^{\mu} + m^2_H H^2 + ... 
\end{equation}
The radion interactions are very similar to those of the SM Higgs boson.
Note that  the radion has anomalous couplings from the trace anomaly 
to a pair of gluons(photons), in addition to the loop diagrams 
with the top-quark 
\begin{equation}
T^{\mu}_{\mu}(SM)^{anom} = \sum_a \frac{\beta_a(g_a)}{2g_a}F^a_{\mu\nu}
F^{a\mu\nu},
\end{equation}
where $\beta_{QCD}/2g_s = -\frac{\alpha_s}{8\pi}(11- 2n_f/3)$ and 
$\beta_{QED}/2e = - \frac{11}{3}(\alpha/8\pi)$. Because of the anomalous 
coupling of the radion to gluons, the gluon fusion will be the most important 
production  channel for the radion in hadronic collisions.   
In general the Higgs boson and radion should mix due to nonzero mixing term 
\begin{equation}
S_{\xi} = \xi \int d^4x \sqrt{g_{vis}} R(g_{vis})H^{+}H,
\end{equation}
where $R(g_{vis})$ is the Ricci tensor for the induced metric 
on the visible brane.  

At the LHC the radion production subprocesses are $gg \rightarrow \Phi$
(the dominant channel), $qq^{'} \rightarrow W \Phi, q \bar{q} \rightarrow 
Z\Phi, q q^{'} \rightarrow qq^{'}\Phi$ and $q\bar{q} \rightarrow 
t\bar{t}\Phi$.  The most interesting radion decay 
modes which could be used for its 
discovery are: $ \Phi \rightarrow \gamma \gamma, ZZ, hh$. 
For heavy radion ($m_{\Phi} \geq 2M_Z$) the cleanest signature is 
\begin{equation}
gg \rightarrow \Phi \rightarrow ZZ \rightarrow 4l \,.
\end{equation} 
The radion LHC 
discovery limit depends on its mass and lies between $\Lambda_{\Phi} =
1~TeV$ and $\Lambda_{\Phi} = 10~TeV$ \cite{ATCOL}.     

In the ADD and RS models all of the SM particles are confined to a brane while 
gravitons are free to move in the extra dimensions. However, there are 
no deep reasons why the SM particles have to be  confined  on a brane. 
In ref. \cite{MAT} the scenario where all particles are free to move to all 
dimensions has been considered (braneless scenario)
\footnote{Similar model has been proposed in refs.\cite{ANT},\cite{APPEL}}. 
For the simplest case of 
one extra dimension the momentum conservation in the fifth 
dimension  leads after the compactification 
to the conservation of the KK numbers.
Because of the KK number conservation KK states are pair produced at 
the LHC in close analogy with the case of supersymmetry with 
$R$-parity conservation. Therefore the LHC phenomenology is 
determined by  the pair 
production of KK quarks and KK gluons \cite{ANT}, \cite{MAT}
\begin{equation}
qq^{'} \rightarrow q^{(1)}q^{'(1)} \,,
\end{equation}
\begin{equation}
q\bar{q} \rightarrow q^{(1)}\bar{q}^{(1)} \,,
\end{equation}
\begin{equation}
gg \rightarrow g^{(1)}g^{(1)} \,,
\end{equation}
\begin{equation}
gg, q\bar{q} \rightarrow q^{(1)}\bar{q}^{(1)} \,.
\end{equation}
Each KK quark $q^{(1)}$ decays into jet and KK 
photon $\gamma^{(1)}$. 
Therefore the LHC signature will be jets with missing energy 
as in the case of supersymmetry. Also very interesting decay chain is 
$q^{(1)}$ decays into $W^{(1)}$ and $Z^{(1)}$ with the subsequent decays 
of $W^{(1)}, Z^{(1)}$ into leptons leading to the signature with 
isolated leptons, jets and missing energy again in close 
analogy with the supersymmetry case. The LHC will be able to discover 
KK quarks and gluons with the masses up to $1.5~TeV$ \cite{MAT}.

Note that there are mixed scenarios where some SM particles live on a 
brane while others can propagate in additional dimensions. For instance, in 
$5DSM$-model \cite{5DSM} the 5-th dimension $y$ is compactified on the orbifold
$S^1/Z_2$ which has two fixed points at $y = 0$ and $y = \pi R_c$. 
The SM gauge fields propagate in the $5D$ bulk, while the chiral matter is 
localised on the $4D$ boundaries \cite{5DSM}. 
In this model the first excitation
 of the  gauge bosons can be directly produced in Drell-Yan processes 
mediated by the first KK modes of the gauge bosons, 
$pp \rightarrow Z^{(1)} \rightarrow l^+l^-$. 
The LHC will be able to discover the KK gauge bosons with the masses 
up to $6~TeV$ \cite{5DSM1}.
       
In the ADD scenario the Planck scale at which the gravity becomes strong is 
$M_D \sim 1~TeV$ for $d = 10$. In this scenario a production of black holes 
should be possible at $\sqrt{s} \gg 1~TeV$. Black hole intermediate states 
are expected to dominate $s$-channel scattering since in the 
string theory the number of such states grows with black hole mass 
faster than the number of perturbative states \cite{GIDD}.  The Schwartzchild 
radius of a $(4 + d)$-dimensional black hole with the mass $M_{BH}$ 
for spin $J = 0$ has the form \cite{Mey}
\begin{equation}
R_S(M_{BH}) = \frac{1}{M_D}(\frac{M_{BH}}{M_D})^{1/(1 + d)}
\end{equation}

The black hole production cross section of two partons a and b is 
taken in a simple geometrical form  \cite{GIDD}
\begin{equation}
\sigma_{ab \rightarrow BH}(s) \approx \pi R^2_S(s)
\end{equation}
The cross section (194) has no small coupling constant 
and it rises rapidly with an energy. With TeV scale gravity the production of 
the black holes should be dominant process at the LHC.
The experimental consequences of 
the black hole decays in ADD model are very distinctive \cite{KIS}:

a. flavor-blind(thermal) decays,

b. hard prompt charged leptons and photons (with energy $E \geq 100~GeV$), 

c. the ratio of hadronic to leptonic activity is close to $5 : 1$,

d. complete cut-off of hadronic jets with $p_{T} > R^{-1}_S$,

e. small missing energy.

These signatures have almost vanishing background. The LHC black 
hole discovery potential is maximal one in $e/\mu + X $ channel and 
the scales up to $M_D \leq 9~TeV$ can be reached \cite{DIMLAN}. Note that 
the described scenario could be too crude and too optimistic 
(see refs. \cite{PES}). To our mind the situation with 
the possibility of black hole detection at the LHC is 
not very clear and a lot of further work is required.

\begin{figure}[hbt]

\vspace*{0.5cm}
\hspace*{0.0cm}
\epsfxsize=15cm \epsfbox{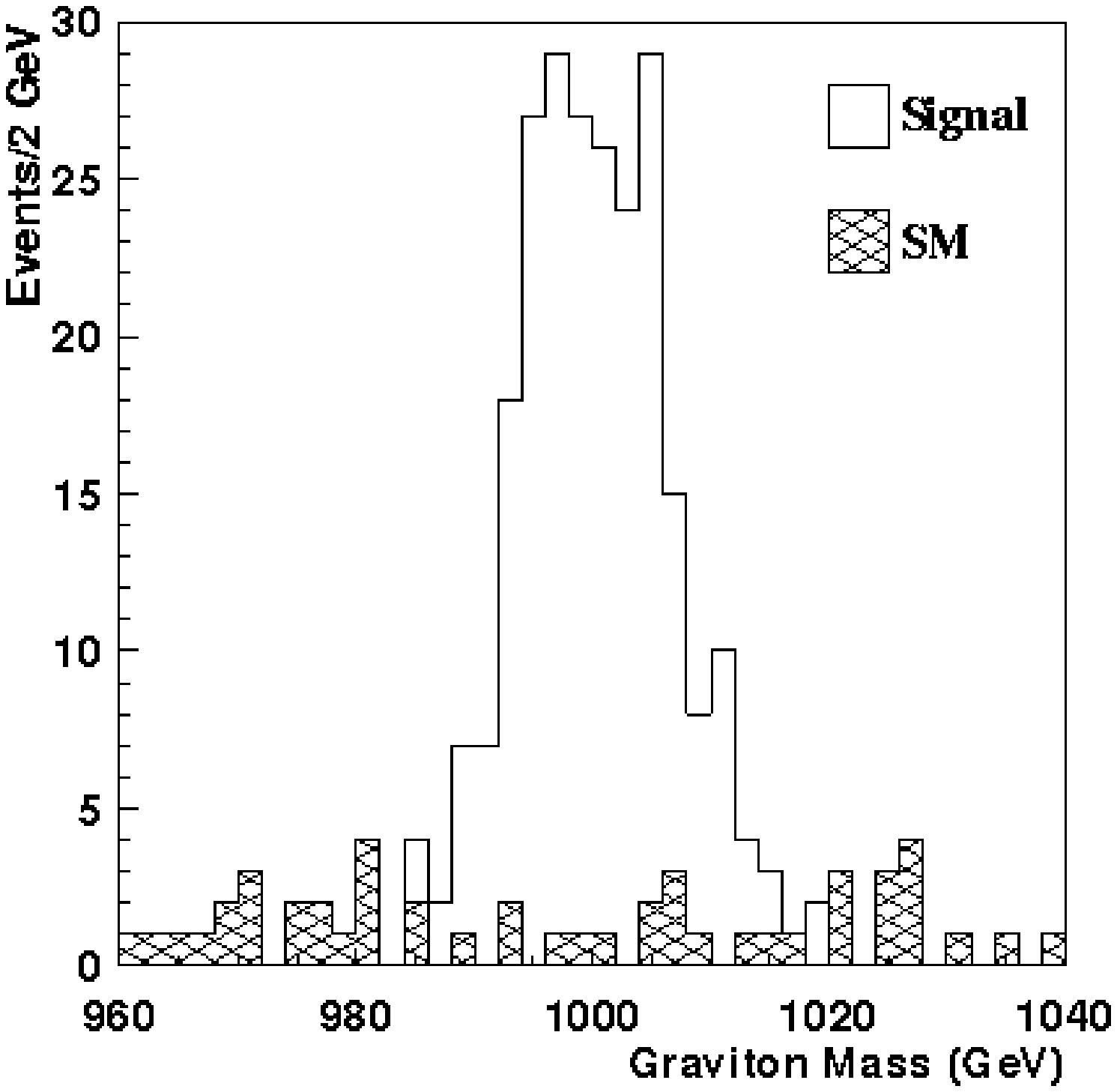}
\vspace*{0.0cm}

\caption[] {\label{fg:kun2} \it Distributions of the 
$e^+e^-$ invariant mass in signal from a graviton resonance  of mass 
$1~TeV$ and in background after event selection for $L_t = 100~fb^{-1}$ }
\end{figure}


\subsection{Extra gauge bosons}

Many string inspired supersymmetric electroweak models and grand unified 
models based on extended gauge groups $(SO(10), E_6, ..)$ predict the 
existence of new relatively light neutral $Z^{'}$-bosons 
and charged $W^{'}$-bosons \cite{Zprime}. 
The LHC 
$Z^{'}$-boson discovery potential depends on the couplings of 
$Z^{'}$-boson with quarks and leptons.
The Lagrangian describing a single $Z^{'}$ and its interactions 
with the SM fields has the form \cite{Zprime} 
\begin{equation}
L_{Z^{'}} = - \frac{1}{4}F^{'}_{\mu\nu}F^{'\mu\nu} - 
\frac{\sin \chi}{2}F^{'}_{\mu\nu}F^{\mu\nu} + 
\frac{1}{2}M^2_{Z^{'}}Z^{'}_{\mu}Z^{'\mu} 
+\delta M^2 Z^{'}_{\mu}Z^{\mu} - \frac{e}{2c_Ws_W}\sum_i 
\bar{\psi}_i \gamma^{\mu}
(f^i_V - g^i_A\gamma_{5}) \psi_i Z^{'}_{\mu},
\end{equation}
where $c_W = \cos \theta_W, s_W = \sin \theta_W$, $F_{\mu\nu}$,
  $F^{'}_{\mu\nu}$ are the field strength tensors for the 
hypercharge and the $Z^{'}$ boson respectively, $\psi_i$ are the matter 
fields with the $Z^{'}$ vector and axial charges $f^i_V$ and $f^i_A$
and $Z_{\mu}$ is the electroweak $Z$-boson. The mixing angle between 
$Z$- and $Z^{'}$-bosons is
\begin{equation}
\xi \approx \frac{\delta M^2}{M^2_Z - M^2_{Z^{'}}}
\end{equation}
If the $Z^{'}$ charges are generation-dependent, tree-level 
flavor-changing neutral currents will generally arise. There exist 
severe constraints in the first two generations coming from precision 
measurements such as the $K_L - K_S$ mass splitting and 
$Br(\mu \rightarrow 3e)$. If the $Z^{'}$ 
interactions commute with the SM gauge group, 
then per generation, there are 5 independent $Z^{'} {\bar{\psi}} 
\psi$ couplings 
; one can choose them in the form $f^u_V,f^u_A, f^d_V, f^e_V, f^e_A$.

Two $Z^{'}$ models are usually considered. 
In the first model an effective $SU_L(2) \otimes U_Y(1) \otimes 
U_{Y^{`}}(1)$ gauge group originates from the breaking of the exceptional 
$E_6$ gauge group 

$E_6 \rightarrow SO(10) \otimes U(1)_{\psi} \rightarrow 
SU(5) \otimes U_{\chi}(1) \otimes U_{\psi}(1) \rightarrow 
SU_c(3) \otimes SU_L(2) \otimes U_Y(1) \otimes U_{Y^{`}}(1)$
The lightest new $Z^{`}$-boson is defined as 
\begin{equation}
Z^{`} = Z^{`}_{\chi} cos \beta + Z^{`}_{\psi} sin \beta \\,
\end{equation}
where $\beta$ is the mixing parameter. In the second model new $Z^{`}$ boson 
arises in  $SU_L(2) \otimes SU_R(2) \otimes U_{B-L}(1)$ left-right symmetric 
models. The $Z^{`}$ boson in such model couples to a linear combination 
of the right-handed and $B-L$ currents. Sometimes as a test example the 
non-realistic case  of $Z^{`}$ boson with the same fermion couplings 
as the SM $Z$ boson is considered. 

The $Z^{`}$ decay width into a massless fermion-antifermion pair 
reads
\begin{equation}
\Gamma^f_{Z^{`}} = N_{c}\frac{\alpha M_{Z^{`}}}{12 c^2_W}
[(f^i_V)^2 + (g^i_A)^2] \,,
\end{equation}
where $N_c$ is the colour factor and $\alpha$ is the effective 
electromagnetic coupling constant to be evaluated at the scale 
$M_{Z^{`}}$ leading to $\alpha \sim 1/128$. Typically in the 
considered models $Z^{`}$ boson is rather narrow \cite{Zprime} 
with the total decay width $ \Gamma_{t}(Z^{`}) \sim O(10^{-2})M_{Z^{`}}$ 
and with $Br(Z^{`} \rightarrow e^{+}e^{-}) \sim 0.05$.   

The main mechanism for the production of such new neutral vector
bosons is the quark-antiquark fusion. The cross section is
given by the standard formula
\begin{eqnarray}
&&\sigma (pp \rightarrow Z^{'} + ...) = \sum_{i}
\frac{12\pi^{2}\Gamma(Z^{'} \rightarrow \bar{q}_iq_i)}{9sM_{Z^{'}}}
\int_{M^2_{Z^{'}}/s}^{1} \frac{dx}{x} \\ \nonumber
&&[\bar{q}_{pi}(x,\mu)
q_{pi}(x^{-1}M^2_{Z^{'}}s^{-1}, \mu)
 +q_{pi}(x,\mu)\bar{q}_{pi}(x^{-1}M^2_{Z^{'}}s^{-1}, \mu)] 
\end{eqnarray}
Here $\bar{q}_{pi}(x, \mu )$ and $q_{pi}(x,\mu )$
are the parton distributions of the antiquark $\bar{q}_i$ and
quark $q_i$ in the proton at the normalisation
point $\mu \sim M_{Z^{'}}$ and
$\Gamma(Z^{'} \rightarrow \bar{q}_{i}q_i)$ is the hadronic decay width of
the $Z^{'}$ boson into quark-antiquark pair with a flavour $i$.

The best way to detect $Z^{'}$-bosons is to use the 
$Z^{'} \rightarrow e^{+}e^{-}, \mu^{+}\mu^{-}, jet~jet$ decay modes.
The study of the angular distribution of lepton pairs allows  
to obtain nontrivial information on $Z^{'}$-boson coupling constants 
with quarks and leptons and confirm that $Z^{'}$-boson is spin-1 particle. 
For considered $Z^{'}$-boson  models 
new $Z^{`}$ bosons can be observed in the reaction 
$pp \rightarrow Z^{`} \rightarrow l^{+}l^{-}$, up to masses 
about $5~TeV$ for an integrated luminosity of $100~fb^{-1}$ \cite{CMS},
 \cite{ATCOL}, \cite{dit1}. 
The measurements of  the forward-backward lepton charge asymmetry, 
both on $Z^{`}$ peak and in the interference region plus the 
measurement of the $Z^{`}$ rapidity distribution allow to discriminate 
between different $Z^{`}$ models for $Z^{`}$ masses up to $2 - 2.5~TeV$ 
for total luminosity $L_t = 100~fb^{-1}$ \cite{dit1}.

The most attractive candidate for $W^{'}$ is the $W_R$ gauge boson associated 
with the left-right symmetric  models \cite{Leftright}. These models provide 
a spontaneous origin for parity violation in weak interactions. 
The gauge group of left-right symmetric model is $SU_c(3) \otimes 
SU_L(2) \otimes SU_R(2) \otimes  U(1)_{B-L}$ with the SM hypercharge 
identified as $Y = T_{3R} + \frac{1}{2}(B-L)$, $T_{3R}$ being the third 
component of $SU_R(2)$. The fermions transform under the gauge group as 
$q_L(3,2,1,1/3) + q_R(3,1,2,1/3)   $ for quarks and  
$l_L(1,2,1,-1) + l_R(1,1,2,-1)$ for leptons. The model requires the 
introduction of right-handed neutrino $\nu_R$ which is  the essential 
ingredient for the see-saw mechanism for explaining the smallness 
of the ordinary neutrino masses. A Higgs bidoublet $\Phi(1,2,3,0)$ is 
usually introduced to generate fermion masses. 
    
The main production mechanism 
for the $W^{'}$-boson is the quark-antiquark fusion similar to the case of 
$Z^{'}$-boson production. If right-handed neutrino $\nu_R$ is heavier 
than $W_R$  the decay mode  $W_R \rightarrow \nu_R +l$  is forbidden 
kinematically and the dominant decay of $W_R$ will be into dijets. 
If $\nu_R$ is lighter than $W_R$ the decay $W_R \rightarrow l_R \nu_R$ 
is allowed. The decay of $\nu_R \rightarrow e_R q \bar{q}^{'}$ leads to 
the $e~e~jet~jet$ signature. The use of the signature 
$pp \rightarrow W_R \rightarrow e \nu_R \rightarrow 
e e q\bar{q}$ allows to discover $W_R$ boson up to masses of 
$4.6~TeV$ for $L_t = 30~fb^{-1}$ and $m_{\nu_{R}} \leq 
2.8~TeV$ \cite{ATCOL}.
 
For the $W^{'}$ boson with coupling constants 
to the SM fermions equal to the
ordinary $W$-boson coupling constants 
the best way to look for $W^{'}$-boson is 
through its leptonic decay mode $W^{'} \rightarrow l\nu$. For 
such model  it would be possible to discover 
the $W^{'}$-boson through its leptonic mode with a mass up to 
$6~TeV$ \cite{ATCOL}, \cite{CMS}. 
By the measurement of the $W^{'}$-boson transverse mass 
distribution it is possible to determine its mass with the   
accuracy $(50-100)~GeV$ \cite{ATCOL}.

\begin{figure}[hbt]

\vspace*{0.5cm}
\hspace*{0.0cm}
\epsfxsize=10cm \epsfbox{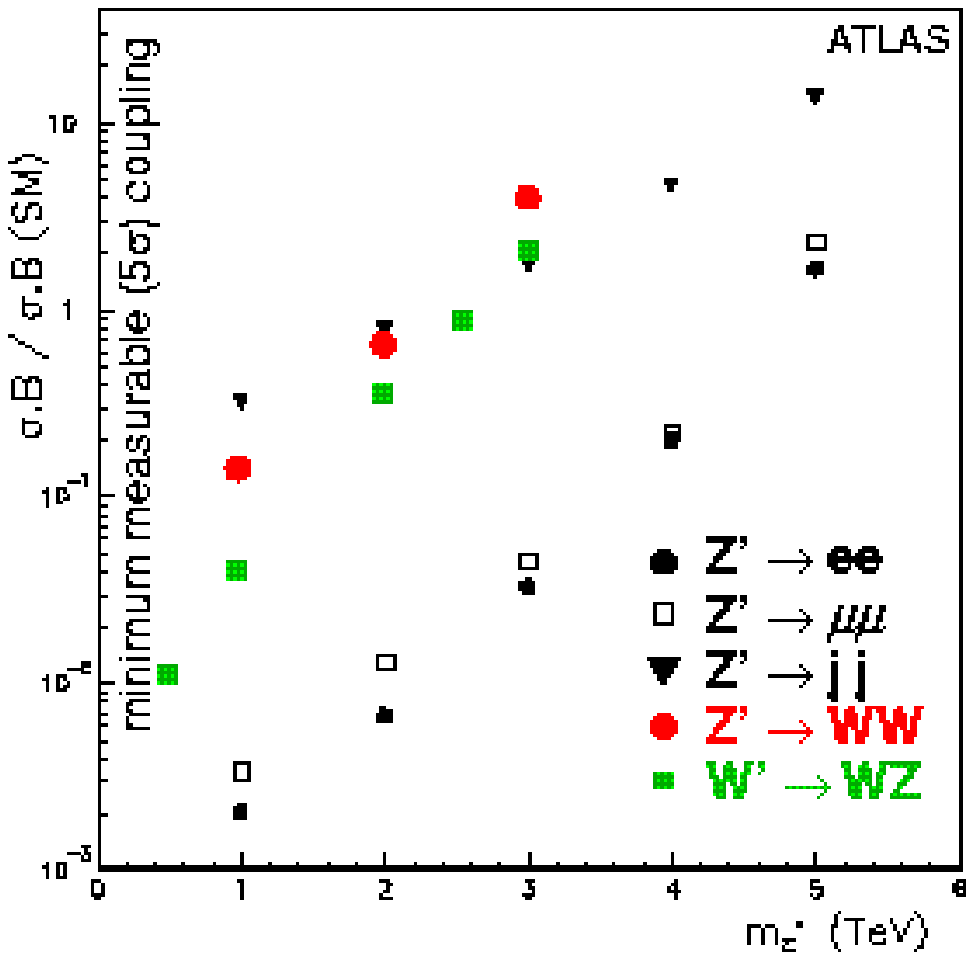}
\vspace*{0.0cm}

\caption[] {\label{fg:kun2} \it $5\sigma$ limits on $W^{`}$ and $Z^{`}$ 
coupling for fermionic $(100~fb^{-1}$ and bosonic $(300~fb^{-1}$ 
modes }

\end{figure}


\begin{figure}[hbt]

\vspace*{0.5cm}
\hspace*{0.0cm}
\epsfxsize=10cm \epsfbox{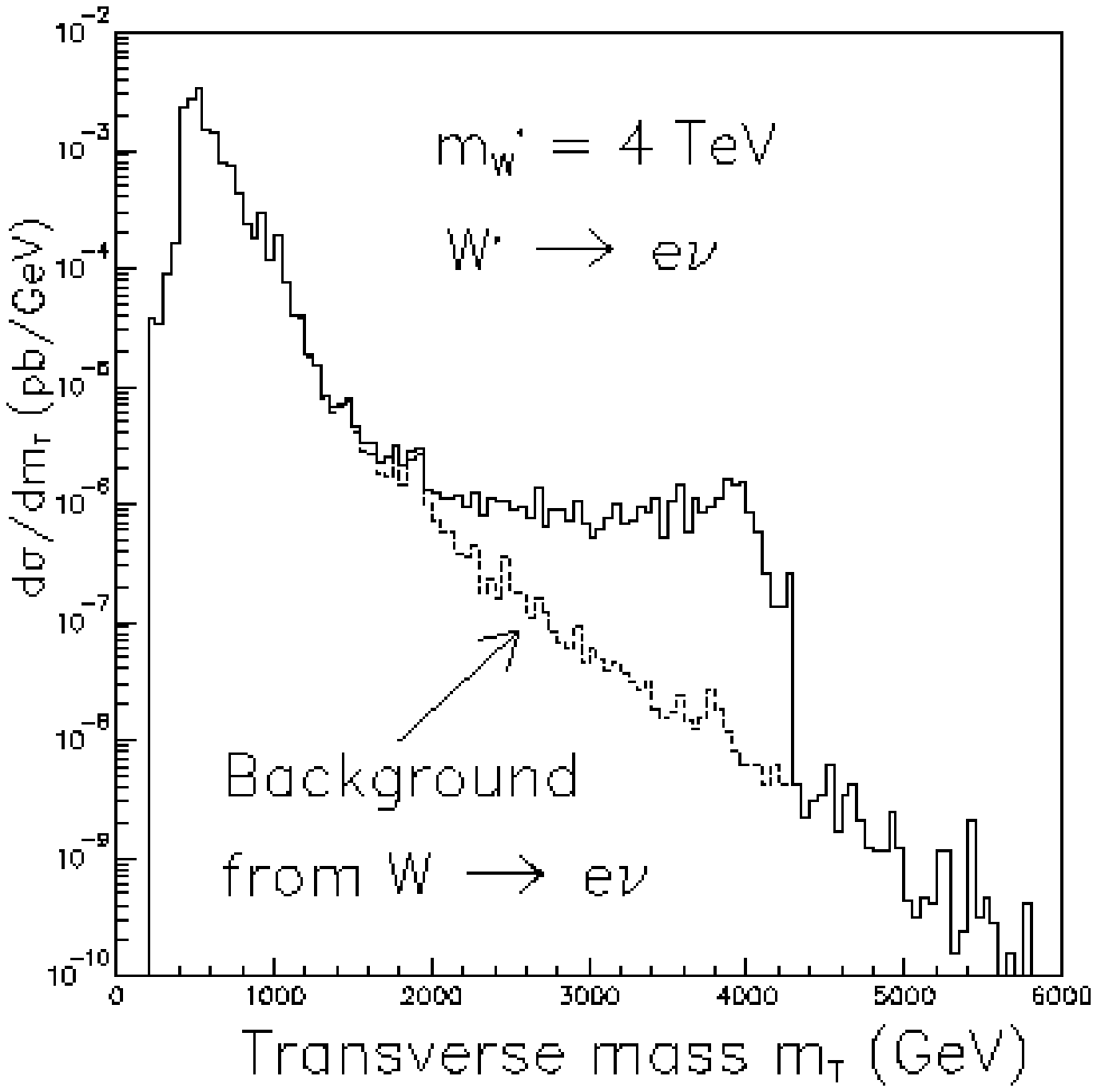}
\vspace*{0.0cm}

\caption[] {\label{fg: kun3} \it Expected electron-neutrino transverse 
mass distribution in ATLAS for $W^{`} \rightarrow e\nu$ decays }
\end{figure}


\subsection{Heavy neutrino }

Left-right symmetric models 
\cite{Leftright} based on $SU_c(3) \otimes SU_L(2) \otimes 
SU_R(2) \otimes U(1)$ gauge group predict the existence of heavy Majorana 
neutrinos $\nu_{R,e,\mu,\tau}$. 
For $m_{\nu_R} < M_{W_R}$   it is possible to look for heavy 
Majorana neutrinos in heavy 
right-handed $W_R$-boson decay using the signature

$pp \rightarrow W_R + ... \rightarrow l(\nu_{R,l} \rightarrow ljj)+...$

Due to Majorana nature of neutrino 
halph of events will be with the same sign leptons plus $\geq 2$ 
jets from $\nu_{R,l}\rightarrow ljj $ decay that makes the signature 
with the same sign leptons the most promising one for both the 
ATLAS \cite{ATCOL} and CMS \cite{GNI}. 
For  $L_t = 30~fb^{-1}$ it is possible to detect heavy neutrino with 
a mass up to $2.8~TeV$.

\subsection{Sgoldstinos}

It is well known, that exist models of supergravity breaking with 
relatively light sgoldstinos (scalar $S$ and pseudoscalar $P$ particles
---  superpartners of goldstino $\psi$). 
Such pattern emerges in a number of non-minimal 
supergravity models~\cite{ellis} and also in gauge mediation
models if supersymmetry is broken via non-trivial superpotential (see,
Ref.~\cite{gmm} and references therein).
To the leading order in $1/F$, where
$F$ is the parameter of supersymmetry breaking, and to 
zero order in MSSM gauge
and Yukawa coupling constants, the interactions between 
the component fields of
goldstino supermultiplet and MSSM fields have been derived in
 Ref.~\cite{Gorbunov:2001pd}. 
They correspond to the most attractive for collider studies 
processes where only one of these {\it new} particles appears in a
 final state. In this case light gravitino behaves exactly
as goldstino. For sgoldstinos, as they are R-even, 
only sgoldstino couplings to goldstino and 
sgoldstino couplings to SM fields have been included as the
 most interesting phenomenologically. 

All relevant sgoldstino coupling constants 
presented in Ref.~\cite{Gorbunov:2001pd} are
completely determined by the MSSM soft terms and the parameter of
supersymmetry breaking $F$, but sgoldstino masses ($m_S,m_P$) remain
free. If sgoldstino masses are of the order of electroweak scale and 
$\sqrt{F}\sim1$ ~TeV --- sgoldstino may be detected 
in collisions of high energy particles at 
supercolliders~\cite{Perazzi:2000id,Perazzi:2000ty}.    

There are flavor-conserving and flavor-violating interactions of
sgoldstino fields.
As concerns flavor-conserving interactions, the strongest bounds arise
from astrophysics and cosmology, that is $\sqrt{F}\geq
10^6$~GeV~\cite{Nowakowski:1994ag,Gorbunov:2000th}, or
$m_{3/2}>600$~eV, for models with $m_{S(P)}<10$~keV and MSSM soft
flavor-conserving terms being of the order of electroweak scale.  For
the intermediate sgoldstino masses (up to a few MeV) constraints from
the study of SN explosions and reactor experiments lead to
$\sqrt{F}\geq 
300$~TeV~\cite{Gorbunov:2000th}.  For heavier
sgoldstinos, low energy processes (such as rare decays of mesons)
provide limits at the level of
$\sqrt{F}\geq 500$~GeV~\cite{Gorbunov:2000th}.

The collider experiments exhibit the same level of sensitivity to
light sgoldstinos as rare meson decays.  Indeed the
studies~\cite{Dicus:1989gg,Dicus:1990su,Dicus:1990dy,Dicus:1996ua} of
the light sgoldstino ($m_{S,P}\leq a~few$~MeV) phenomenology based
on the effective low-energy Lagrangian derived from N=1 linear
supergravity yield the bounds: $\sqrt{F}\geq 500$~GeV (combined
bound on $Z\to S\bar{f}f,P\bar{f}f$~\cite{Dicus:1990dy}; combined
bound on $e^+e^-\to\gamma S,\gamma P$~\cite{Dicus:1990su}) at
$M_{soft}\sim 100$~GeV, $\sqrt{F}\geq 1$~TeV~\cite{Dicus:1996ua}
(combined bound on $p\bar{p}\to gS,gP$) at gluino mass $M_3\sim 500$~GeV.
Searches for heavier sgoldstinos at colliders, though exploiting
another technique, results in similar bounds on the scale of
supersymmetry breaking.  Most powerful among the operating machines,
LEP and Tevatron, give a constraint of the order of 1~TeV on
supersymmetry breaking scale in models with light sgoldstinos. Indeed,
the analysis carried out by DELPHI Collaboration~\cite{Abreu:2000ij}
yields the limit $\sqrt{F}>500\div200$~GeV at sgoldstino masses
$m_{S,P}=10\div150$~GeV and $M_{soft}\sim100$~GeV.  The constraint
depends on the MSSM soft breaking parameters. In particular, it is
stronger by about a hundred GeV in the model with degenerate gauginos.
At Tevatron, a few events in $p\bar{p}\to S\gamma(Z)$ channel, and
about $10^4$ events in $p\bar{p}\to S$ channel would be produced at
$\sqrt{F}=1$~TeV and $M_{soft}\sim100$~GeV for integrated luminosity
${\cal L}=100$~pb$^{-1}$ and sgoldstino mass of the order of
100~GeV~\cite{Perazzi:2000ty}.  This gives rise to a possibility to
detect sgoldstino, if it decays inside the detector into photons and
$\sqrt{F}$ is not larger than $1.5\div2$~TeV.

In terms of $SU(3)_c\times SU(2)_L\times U(1)_Y$ fields the sgoldstino
effective Lagrangian reads~\cite{Gorbunov:2001pd}:

\begin{eqnarray}
&&L_{S} = -\sum_{all~gauge\atop fields}
{M_\alpha\over2\sqrt{2}F}S\cdot F_{a~\mu\nu}^\alpha F_a^{\alpha~\mu\nu}
-{{\cal A}^L_{ab}\over\sqrt{2} F}y^L_{ab}\cdot S \\ \nonumber
&&\bigl(\epsilon_{ij} l_a^je_b^c h_D^i +h.c.\bigr) -
{{\cal A}_{ab}^D\over\sqrt{2} F}y_{ab}^D\cdot S
\bigl( \epsilon_{ij} q_a^jd_b^c h_D^i+h.c.\bigr) \\ \nonumber
&&-{{\cal A}_{ab}^U\over\sqrt{2} F}y_{ab}^U\cdot S
\bigl( \epsilon_{ij} q_a^iu_b^ch_U^j+h.c.\bigr)\;,
\end{eqnarray}

\begin{eqnarray}
&&L_{P} = \sum_{all~gauge\atop fields}
{M_\alpha\over 4\sqrt{2}F}P\cdot F_{a~\mu\nu}^\alpha 
\epsilon^{\mu\nu\lambda\rho}F_{a~\lambda\rho}^\alpha \\ \nonumber
&&-i{{\cal A}^L_{ab}\over\sqrt{2} F}y^L_{ab}\cdot P
\bigl(\epsilon_{ij} l_a^je_b^c h_D^i -h.c.\bigr) -  
i{{\cal A}_{ab}^D\over\sqrt{2} F}y_{ab}^D\cdot P \\ \nonumber 
&&\bigl( \epsilon_{ij} q_a^jd_b^c h_D^i-h.c.\bigr)
-i{{\cal A}_{ab}^U\over\sqrt{2} F}y_{ab}^U\cdot P
\bigl( \epsilon_{ij} q_a^iu_b^ch_U^j-h.c.\bigr)\;.
\end{eqnarray}

\begin{eqnarray}
&&L_{\psi,S,P} =i\partial_{\mu}\bar{\psi}\bar{\sigma}^\mu\psi
+\partial_{\mu} S\partial^{\mu} S \\ \nonumber
&&-\frac{1}{2} m_S^2S^2+ 
\partial_{\mu} P\partial^{\mu} P-\frac{1}{2} m_P^2P^2 \\ \nonumber
&& +\frac{m_S^2}{2\sqrt{2}F}S (\psi\psi+\bar{\psi}\bar{\psi})
-i\frac{m^2_S2}{\sqrt{2}F}P(\psi\psi-\bar{\psi}\bar{\psi})\;.
\end{eqnarray}
where $M_\alpha$ are
gaugino masses and $A_{\alpha\beta}$, $y_{\alpha\beta}$ are soft trilinear
coupling constants. Usually the case when  ${\cal A}_{ab}=A$ and Yukawas
$y_{ab}\propto\delta_{ab}$ is considered for numerical estimates. 

At hadron colliders sgoldstinos will be produced mostly by gluon
resonant scattering $gg\to S(P)$~\cite{Perazzi:2000ty}. The associated
production $gg\to S(P)g$ has several times smaller cross section than
resonant production and the corresponding discovery potential (in
analogy with SM Higgs boson case) is expected to be weaker than the
discovery potential for the resonant mode $gg\to S(P)$. 

One has to consider the
subsequent decay of the sgoldstino inside the detector.
Indeed, for the range of parameters that are relevant for this study,
sgoldstinos are expected to decay inside the detector, not
far from the collision point. Then, assuming that the
supersymmetric partners (others than the gravitino $\tilde G$) are too
heavy to be relevant for the sgoldstino decays, the main decay channels are:
\[
S(P)\to gg, \gamma\gamma, \tilde{G}\tilde{G}, f \bar f, \gamma Z, WW, ZZ.
\]
The corresponding widths have been calculated 
in Refs.~\cite{Perazzi:2000id,Perazzi:2000ty}.
                                    
For   sgoldstinos decaying into pairs of massless gauge bosons, one has
\[
\Gamma(S(P)\rightarrow \gamma\gamma)=
{M_{\gamma\gamma}^2m_{S(P)}^3\over 32\pi F^2}\;,
~~~~
\Gamma(S(P)\to gg)={M_3^2m_{S(P)}^3\over 4\pi F^2}\;,
\]
where $M_{\gamma\gamma}=M_1\cos^2\theta_W+M_2\sin^2\theta_W$,
and $\theta_W$ is the electroweak mixing angle. Note that for
$M_{\gamma\gamma}\sim M_3$ gluonic mode dominates over the photonic 
one due to the color factor enhancement.
  
For the values of $\sqrt{F}$ which are interesting, gravitino is very light,
with mass in the range $m_{\tilde{G}}=\sqrt{8\pi/3}\;
F/M_{Pl}\simeq10^{-3}\div 10^{-1}$~eV. Then, the sgoldstinos decay rates
into two gravitinos are given by
\[
\Gamma(S(P)\to\tilde{G}\tilde{G})={m_{S(P)}^5\over 32\pi F^2}\;,
\]
and become comparable with the rate into two photons for heavy
sgoldstinos, such that $m_{S(P)}\sim M_{\gamma\gamma}$.

\begin{table}[htb]
\begin{center}
\vspace{2mm}  
\begin{tabular}{|c|c|c|c|c|}     
\hline
Model&$M_1$&$M_2$&$M_3$&$A$\\  
\hline
I&100~GeV&300~GeV&500~GeV&300~GeV\\
\hline
II&300~GeV&300~GeV&300~GeV&300~GeV\\
\hline
\end{tabular}
\caption{The sets of parameters  which the LHC sensitivity is presented
for.
\label{sets}}
\end{center}
\end{table}

In ref. \cite{GORBK} the LHC sgoldstino discovery potential has been studied. 
Two sets of the MSSM soft SUSY breaking parameters have been considered
(see table 1).  The most reliable signatures with $\gamma\gamma$ 
and $ZZ$ in a final state have been studied.
The main result is that it would be possible to discover sgoldstino at the LHC 
with  $\sqrt{F} \leq (2 - 8)~TeV$ (see Figs. 28, 29).

\begin{figure}[hbt]

\vspace*{0.5cm}
\hspace*{0.0cm}
\epsfxsize=15cm \epsfbox{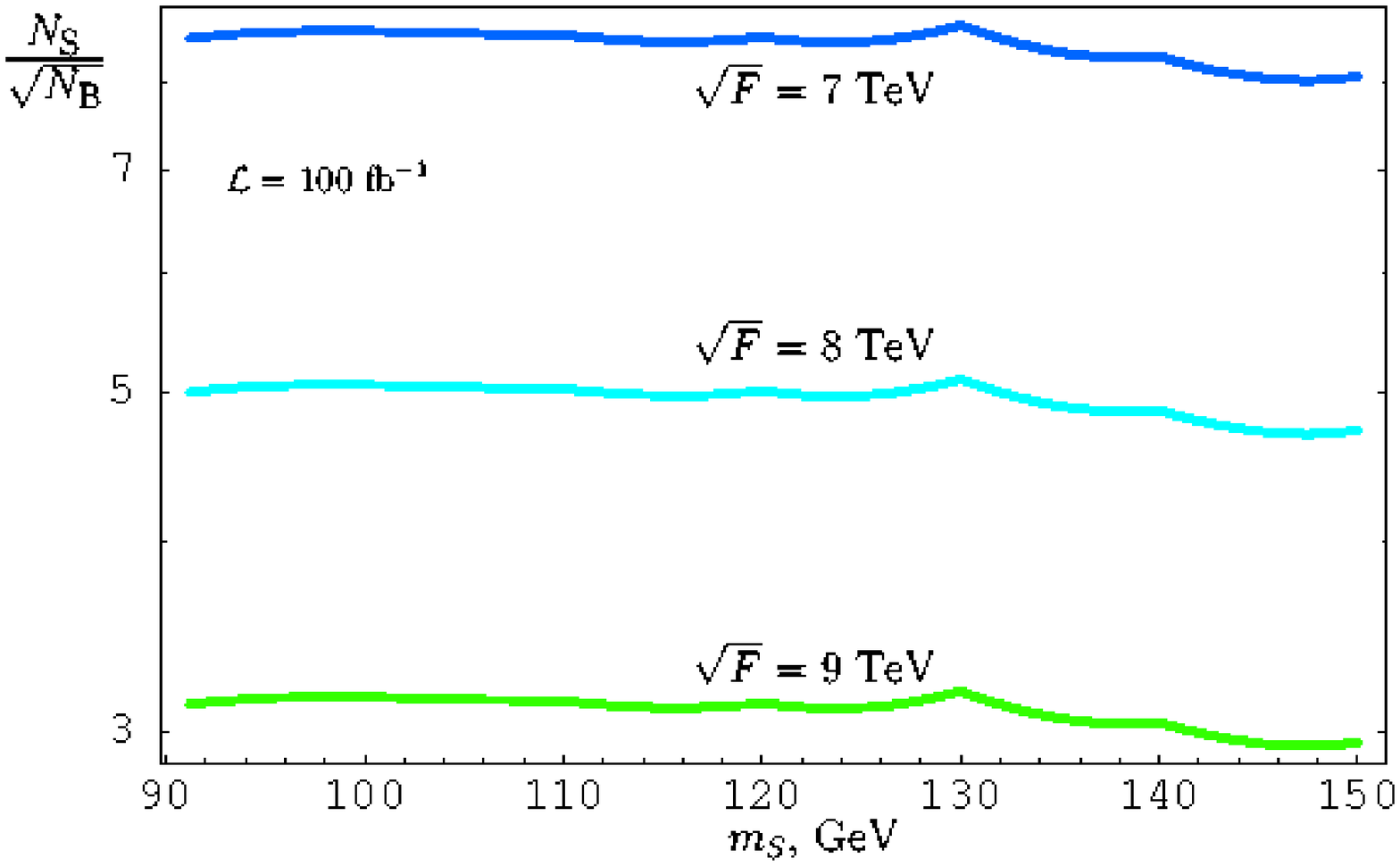}
\vspace*{0.0cm}

\caption[] {\label{fg:kun2} \it Signal significance of $\gamma \gamma$    
channel as a function of sgoldstino mass $m_S$ for the model II.    }

\end{figure}


\begin{figure}[hbt]

\vspace*{0.5cm}
\hspace*{0.0cm}
\epsfxsize=15cm \epsfbox{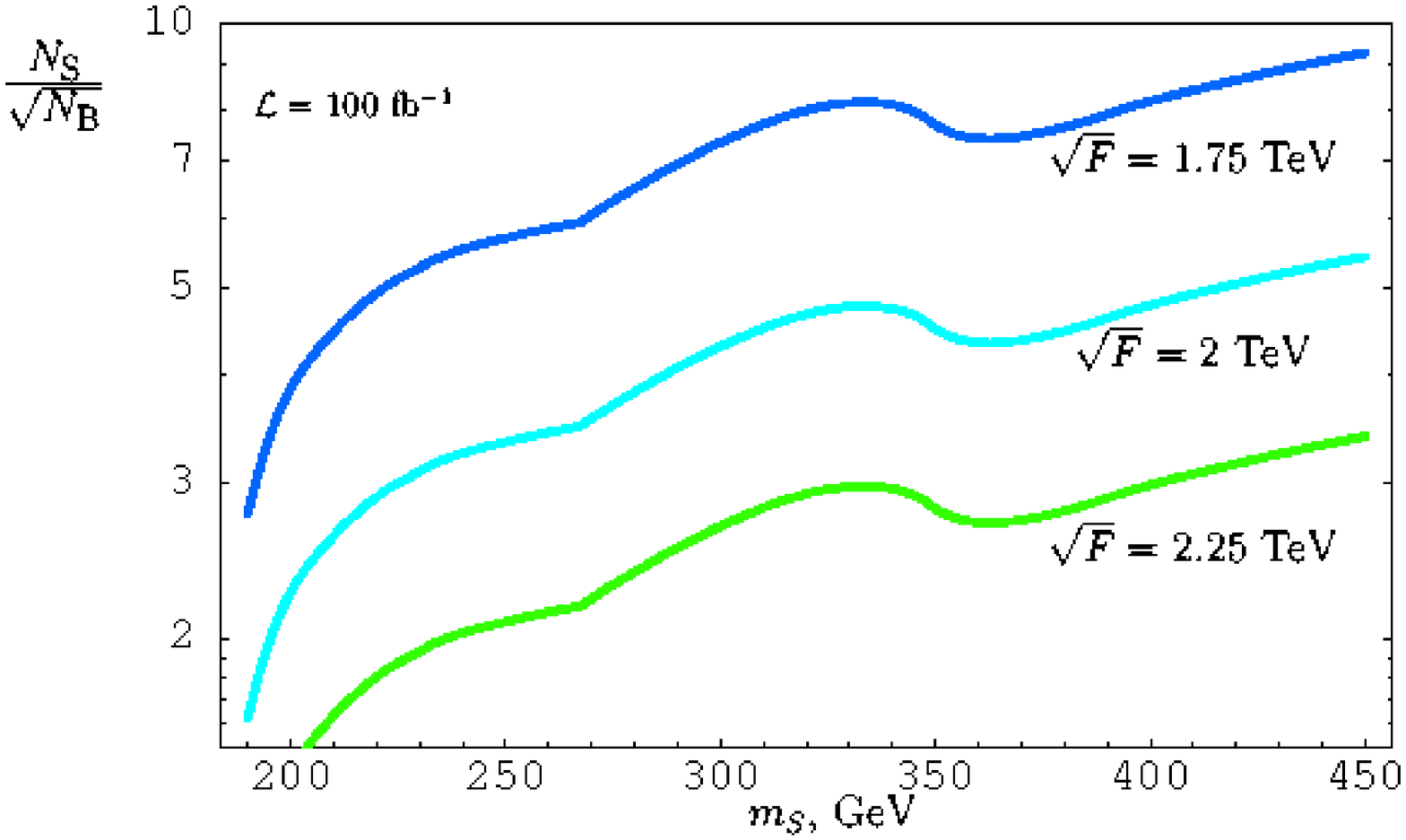}
\vspace*{0.0cm}

\caption[] {\label{fg:kun2} \it Signal significance of $Z Z$               
channel as a function of sgoldstino mass $m_S$ for the model I.     }

\end{figure}


\subsection{Scalar leptoquarks}

Scalar leptoquarks (LQ) are particles having both lepton and 
baryon numbers different 
from zero. They are predicted by many models \cite{LEPTOQUARK} 
with gauge symmetry larger than $SU_c(3) \otimes SU_L(2) \otimes U(1)$ 
of the SM. 
Leptoquarks decay mainly into quark and lepton. 
At LHC  both pair and single leptoquark production mechanisms  are possible:
\begin{equation}
q + g \rightarrow LQ +l \rightarrow 2l +j \,,
\end{equation}
\begin{equation}
g + g \rightarrow LQ +LQ \rightarrow 2l +2j \,.
\end{equation}
Single leptoquark production cross section depends on unknown 
Yukawa coupling constant of leptoquark with quark and lepton. 
Pair leptoquark production cross section depends mainly on 
leptoquark mass. Leptoquark pair production mechanism has been studied 
in refs. \cite{WRO1}, \cite{ATCOL} 
for leptoquark detection at LHC. The main signature here are 
the events with 2 jets and 2 isolated leptons arising from 
leptoquark decays  with the invariant jet-lepton mass equal to 
leptoquark mass. For the first and second generation leptoquarks 
it would be possible to discover them with the masses up to 
$1.6~TeV$ at the LHC for $L_t = 100~fb^{-1}$ \cite{WRO1}, \cite{ATCOL}.

\begin{figure}[hbt]

\vspace*{0.5cm}
\hspace*{0.0cm}
\epsfxsize=15cm \epsfbox{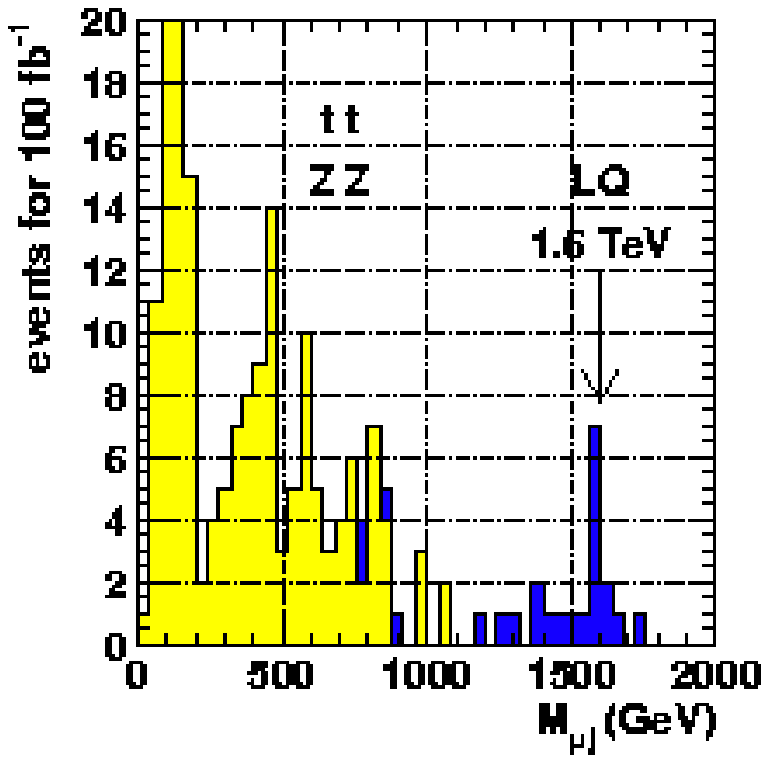}
\vspace*{0.0cm}

\caption[] {\label{fg:kun2} \it Mass distribution for scalar, $2^{nd}$  
generation leptoquarks with $M_{LQ2} = 1.6~TeV$ produced in pairs }

\end{figure}


\subsection{Compositeness}

In the SM  quarks and leptons are the fundamental point-like particles. 
But the proliferation of quarks and leptons has inspired the 
speculations that they are composite structures, bound states of 
more fundamental constituents often called preons. 
If quarks have substructure, it will be revealed in the deviation of the jet 
cross-section from that predicted by QCD. The deviation can be 
parametrised by an 
interaction of the form 
\begin{equation}
\delta L = \frac{4\pi}{\Lambda^{2}}\bar{q}\gamma^{\mu}q\bar{q}\gamma_{\mu}q
\end{equation}
which is strong at a scale $\Lambda$. Comparing the QCD predictions 
for the jet cross section at high $p_T$ with data it would be possible 
to restrict the value of $\Lambda$.  
The main conclusion is that the LHC(ATLAS) at full luminosity 
$300~fb^{-1}$ will be able to probe the  
compositeness with $\Lambda \leq 20~TeV$ 
provided the  systematic uncertainties are smaller than the 
statistical ones \cite{ATCOL}.

\begin{figure}[hbt]

\vspace*{0.5cm}
\hspace*{0.0cm}
\epsfxsize=15cm \epsfbox{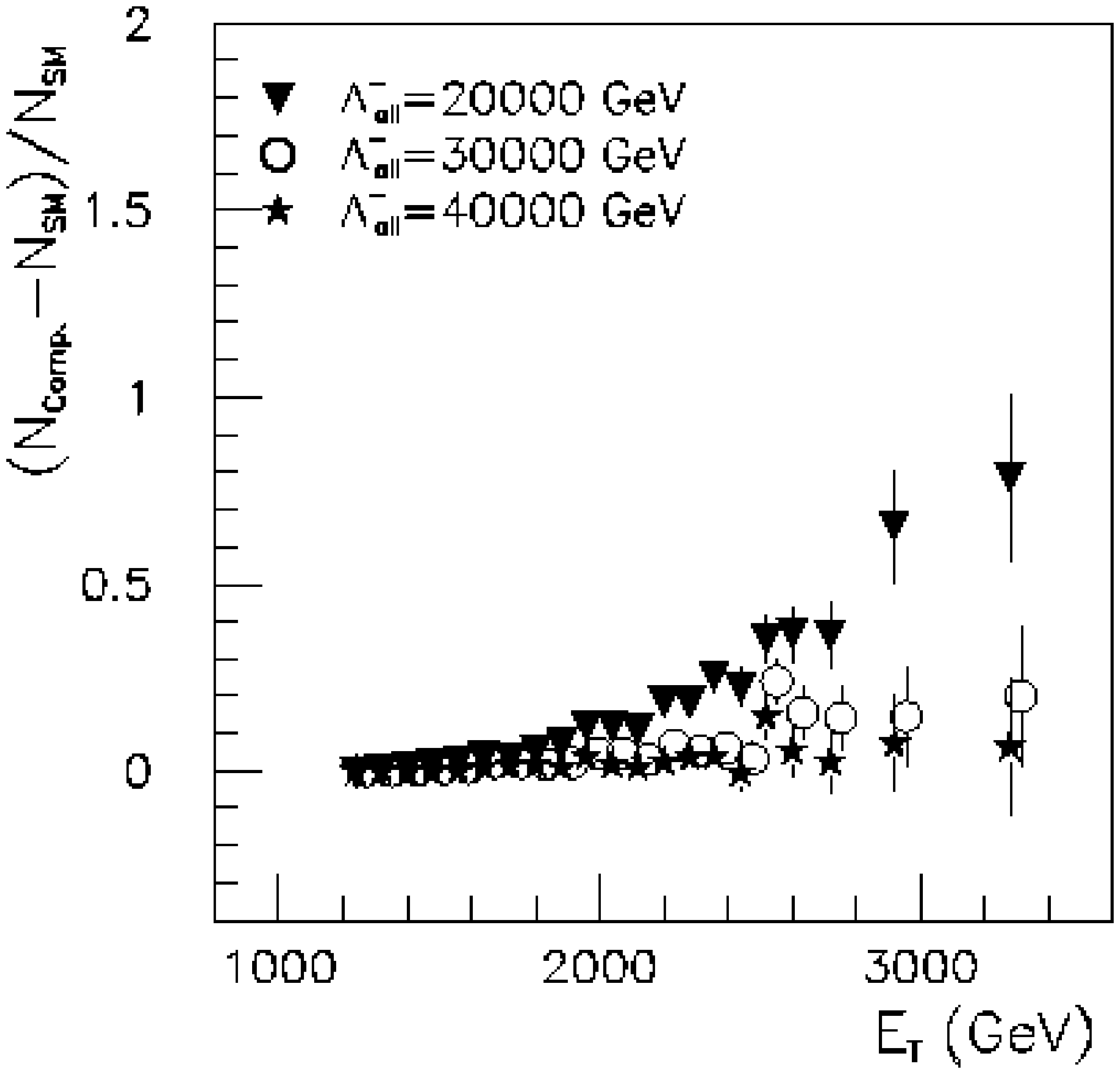}
\vspace*{0.0cm}

\caption[] {\label{fg:kun2} \it Difference of the SM prediction and the 
effect of compositeness on the jet $E_T$ distribution, normalised to the SM 
rate. The errors correspond to $L_t = 300~fb^{-1}$ 
for various values of the compositeness scale $\Lambda$} 

\end{figure}


In ref.\cite{GUPTA} the possibility to 
search for quark-lepton contact interaction 

\begin{equation}
\delta L = \frac{4\pi}{\Lambda_{ql}^{2}}\bar{l}\gamma^{\mu}l\bar{q}
\gamma_{\mu}q
\end{equation}
at LHC(CMS) has been studied. The Drell-Yan process 
$pp \rightarrow q\bar{q} \rightarrow 
\gamma^{*}/Z^{*} \rightarrow l^{+}l^{-}$ has been considered. 
The interaction (206) modifies SM predictions
for Drell-Yan cross section in the high dielectron mass 
region. The main conclusion is that for $L_t = 100~fb^{-1}$ 
LHC(CMS)  will be able to obtain lower bound  $\Lambda_{ql} \geq 35~TeV$ at 
$95~\%$   $C.L.$ level.

\subsection{$R$-parity violation}

Most of supersymmetric phenomenology assumes the MSSM which
conserves R-parity. As a consequence of R-parity conservation
supersymmetric particles can only be produced in pairs and a
supersymmetric state cannot decay into conventional states.
This has untrivial consequence for the search for
supersymmetric particles at supercolliders; in particular
most of experimental searches of SUSY rely on pair production and
on missing transverse momentum $p_T^{miss}$ as a signal for the
production of the LSP, which must be stable and electrically neutral.
However, at present there are no deep theoretical
motivations in favour of R-parity conservation.  The phenomenology
of the models with explicit R-parity violation at hadron colliders
has been studied in refs. \cite{DIM1}.
The terms in the superpotential (75) violate baryon and lepton 
number and, if present in the Lagrangian of 
the MSSM, they generate an unacceptably large amplitude
for proton decay suppressed only by the inverse squark mass squared.
The R-parity prohibits the dangerous terms (75).  
However, R-parity conservation 
is not the single way to construct
a minimal supersymmetric extension of the SM.  It is easy to
write down alternative to R-parity symmetries which allow for a
different set of couplings. For example, under the transformation
\begin{equation}
(Q,\bar{U}, \bar{D}) \rightarrow -(Q,\bar{U}, \bar{D}),\;\;
(L,\bar{E}, H_{1,2}) \rightarrow +(L,\bar{E}, H_{1,2}) \,,
\end{equation}
only the quark superfields change sign. If the superpotential (75) 
 is invariant under transformations (207) then only the last, baryon number
violating term 
$\bar{U} \bar{D} \bar{D}$  is forbidden. This gives a new model in which
a single supersymmetric state can couple to standard model states
breaking R-parity. Similarly, there are analogous transformations
forbidding the lepton number violating terms.

In the direct search for supersymmetric particles the phenomenology
is altered considerably when including R-parity violating terms in the
superpotential. In general both the production mechanisms and the
decay patterns  can change. Other than the standard supersymmetric pair
production of particles there is now the possibility of production
of R-odd final states as well. Also, if all supersymmetric particles
decay in the detector, we will no longer have the standard $p_t^{miss}$
signal and the decay patterns will all be altered. In particular, the LSP
will decay mainly into three-body final states \cite {DIM1}.
However, except for LSP, which now decays, all particles
predominantly decay as in the MSSM.
Consider the case when LSP decays within the detector. 
If lepton number is 
violated, the SUSY signal will contain multiple leptons from 
$\tilde{\chi}^0_1 \rightarrow l^+l^-\nu, l q\bar{q}$ \cite{DIM1}. If baryon 
number is conserved, the LSP will decay into jets, $\tilde{\chi}^0_1 
\rightarrow qqq$, giving events with  high jet multiplicity and without 
missing transverse energy. It is not easy to extract such signal since 
the QCD background is huge. It is possible to detect SUSY events using 
cascade decays involving leptons, for instance, 
$\tilde{\chi}^0_2 \rightarrow \tilde{l}^{\pm}l^{\mp} 
\rightarrow qqql^+l^-$.     

Note that it is possible to construct a model \cite{KR4} 
with supersmall $R$-parity 
violation and with relatively long-lived $t \sim (10^{-1} - 10^{-9}) ~sec$ 
charged $\tilde{\tau}_R$ slepton playing the role of the LSP. The 
phenomenology of  such model is the similar to the GMSB (gauge mediated 
supersymmetry breaking)   model \cite{GRDGOR} phenomenology 
with the NLSP $\tilde{\tau}$. \footnote{Remember that in GMSB model 
gravitino  $\tilde{G}$ 
becomes LSP (lightest stable superparticle). Neutralino $\chi^0_1$ or 
stau $\tilde{\tau}$ can be the ``Next to LSP'' and it can be stable or 
long lived . Such $\tilde{\tau}$ would look like  a ``heavy muon'' 
traversing detector with velocity significantly lower than the velocity 
of light. One can measure its time of flight and hence calculate 
the mass $m_{\tilde{\tau}}$ \cite{WROCH}, \cite{ATCOL}.}.

\subsection{Additional Higgs bosons with big Yukawa coupling constants}

Many Higgs doublet model where each Higgs doublet couples with its
own quark with relatively big Yukawa coupling constant has been
considered in ref. \cite{KRAS3}. For non small Yukawa coupling constants
the main reaction for the production of the Higgs doublets
corresponding to the first and the second generations is quark-antiquark
fusion. The phenomenology of the Higgs doublets corresponding to
the third generation is very similar to the phenomenology of the
model with two Higgs doublets. The cross section for the quark-antiquark
fusion in quark-parton model in the approximation of the infinitely
narrow resonances is given by the standard formula
\begin{eqnarray}
&&\sigma(AB \rightarrow H_{q_iq_j} + X) =
\frac{4\pi^{2}\Gamma(H_{q_iq_j} \rightarrow \bar{q}_iq_j)}{9sM_H}
\int^1_{\frac{M^2_H}{s}}\frac{dx}{x} \\ \nonumber
&&[\bar{q}_{Ai}(x,\mu)q_{Bj}(x^{-1}M^2_Hs^{-1}, \mu) +
q_{Aj}(x,\mu)\bar{q}_{Bj}(x^{-1}M^2_Hs^{-1}, \mu)] \,.
\end{eqnarray}
Here $\bar{q}_{Ai}(x,\mu)$ and $q_{Aj}(x,\mu)$ are parton distributions
of the antiquark $\bar{q}_i$ and quark $q_j$ in hadron A at the
normalisation point $\mu \sim M_H$ and the  $\Gamma(H_{q_iq_j}
\rightarrow \bar{q}_iq_j)$ is the hadronic decay width of the Higgs
boson into quark-antiquark pair. For the Yukawa Lagrangian
\begin{equation}
L_Y = h_{q_iq_j}\bar{q}_{Li}q_{Rj}H_{q_iq_j} + h.c.\,,
\end{equation}
the hadronic decay width for massless quarks is
\begin{equation}
\Gamma(H_{q_iq_j} \rightarrow \bar{q}_iq_j) =
\frac{3M_H h^2_{q_iq_j}}{16\pi}\,.
\end{equation}

The value of the renormalization point $\mu$ has been chosen equal to
the mass $M_H$ of the corresponding Higgs boson. The variation of the
renormalization point $\mu$ in the interval $0.5M_H - 2M_H$ leads to
the variation of cross section less than 50 percent. In considered
model there are Higgs bosons which couple both with
down quarks and leptons so the best signature
is the search for  electrically neutral Higgs boson decays
into $e^+e^-$ or $\mu^{+} \mu^{-}$ pairs. For  charged Higgs bosons
the best way to detect them is to look for their decays into charged
leptons and neutrino. The Higgs doublets which couple with up quarks
in model with massless neutrino do not couple with leptons so the only
way to detect them is the search for the resonance type structure
in the distribution of the  dijet cross section on the dijet invariant
mass as in the case of all Higgs bosons, since in the considered models all
Higgs bosons decay mainly into quark-antiquark pairs that leads at the
hadron level to additional dijet events. However, the accuracy of the
determination of the dijet invariant mass  is $O(10)$ percent, so it would
be not so easy to find stringent bound on the Higgs boson mass by the
measurement of dijet differential cross section at LHC.  In considered 
model, due to the smallness of the vacuum expectation
values of the Higgs doublets corresponding to the u, d, s and c quarks,
after electroweak symmetry breaking the mass splitting inside the Higgs
doublets is small, so in such models the search for neutral Higgs boson
decaying into lepton pair is in fact the search for the corresponding
Higgs isodoublet. The main background in the search for neutral
Higgs bosons through their decays into lepton pair is the
Drell-Yan process which is under control. The main conclusion of
the ref.\cite{KRAS3} is that at LHC for $L_t = 100$ $fb^{-1}$ and
for the  Yukawa coupling constant $h_{Y} = 1$
it would be possible to detect
such Higgs bosons with the masses up to $4.5 - 5$ TeV.

\subsection{Astroparticle applications}

Let us briefly mention interesting proposal for astroparticle 
physics application of the CMS detector \footnote{S.N.Gninenko, 
private communication, to be published}.
One of the main design features of the  CMS detector  is 
the presence of  a large volume of high magnetic field 
instrumented all around its surface with the electromagnetic
calorimeter (ECAL) and surrounded by almost 4$\pi$ hermetic 
 hadronic calorimeter (HCAL).
This feature provides unique 
opportunity for the most sensitive search for cosmic scalar or 
pseudoscalar particles (e.g. such as axion) with two-photon
interaction vertex in the energy region from several GeV 
up to ultra-high values, which might be relevant to the GZK cutoff.
New particles, if they exist, 
 would  penetrate the CMS HCAL detector, used as efficient VETO against 
cosmic ray interactions, and would be 
observed in the CMS ECAL through the Primakoff effect, i.e. through
their conversion into real high energy photons in the process of 
interactions with the virtual photons from the magnetic field of the CMS 
superconducting solenoid. 
For high energies, axion-photon conversion is  coherent throughout the CMS 
detector volume, thus enhancing the  signal  and 
 allowing a substantial 
increase in the sensitivity to axion masses up to 
 $0 \leq  m_a \leq 32\times \sqrt{E_a[GeV]}$,
where $m_a$ is  the axion mass in $e$V.
A preliminary constraint on the product of the 
 coupling to two photons and integral flux of particle through the detector  
$\Phi_a \times g_{a\gamma\gamma}$ could be  set
in the case of zero signal.

\section{Conclusion}

There are no doubts that at present the supergoal number one 
of the experimental high energy physics is the 
search for the Higgs boson - the last non discovered 
cornerstone of the Standard Model. The LHC will be 
able to 
discover the Higgs boson and to check 
its basic properties. The experimental Higgs boson discovery will be 
the triumph of the idea of the renormalizability (in some sense it 
will be the ``experimental proof'' of the renormalizability of the 
electroweak interactions). The LHC will be able also to 
discover the low energy 
broken supersymmetry with the squark and gluino masses up to 2.5 TeV. 
Also there is nonzero probability to find something new beyond the SM or  the 
MSSM (extra dimensions, $Z^{'}$-bosons, $W^{'}$-bosons, compositeness, ...). 
 At any rate after the LHC we will know the mechanism 
of the electroweak symmetry breaking (the Higgs boson or something more 
exotic?) and the basic elements of the matter structure at TeV scale.

We thank  our colleagues from INR theoretical department
for useful discussions. The work of N.V.Krasnikov 
has been supported by RFFI grant No 03-02-16933.

\end{document}